\documentclass[rmp, nofootinbib, reprint]{revtex4-2}

\usepackage{amsfonts,amsmath,amssymb,ascmac,bm,tensor, mathtools}
\usepackage{fnpct} 
\usepackage{comment}
\usepackage{ifpdf}
\usepackage{slashed}
\usepackage{color, subcaption}
\usepackage[mathscr]{eucal}
\usepackage[utf8]{inputenc}
\usepackage{physics}
\usepackage{cancel}
\usepackage{soul, multirow}
\usepackage{simpler-wick}

\usepackage{ragged2e}
\justifying

\usepackage[font=small]{caption}

\ifpdf
  \usepackage{graphicx}     %   usepackage without driver option
  \usepackage[bookmarksopen,colorlinks=true,linkcolor=bblue,citecolor=bblue,urlcolor=ppink]{hyperref}
\else     % For (p)LaTeX + dvipdfmx
\fi

\hypersetup{
           breaklinks=true,   % splits links across lines
           colorlinks=true,   % displays links as colored text instead of blocks
           linkcolor=bblue,
           citecolor=bblue,
           urlcolor=ppink
        }

\definecolor{red}{rgb}{1,0,0}
\definecolor{darkred}{rgb}{0.6,0,0}
\definecolor{darkgreen}{rgb}{0.992447,0.623778,0.034597}
\definecolor{ppink}{rgb}{1,0.4,0.4} 
\definecolor{bblue}{rgb}{0.284602,0.317763,0.963947}
\definecolor{purple}{rgb}{0.5 ,0, 0.7}
 
\newcommand{\rref}[1]{\textcite{#1}} %RMP does not use Ref.

\renewcommand{\vev}[1]{ \left< {#1} \right> }

\newcommand{\NL}{\text{NL} }
\newcommand{\R}{\text{R} }
\newcommand{\GW}{\text{GW} }
\newcommand{\Pl}{\text{Pl} }
\newcommand{\tmax}{\text{max} }

\newcommand{\ee}{\text{e}}
\newcommand{\ii}{\text{i}}

\newcommand{\MeV}{\text{MeV}}

\newcommand{\Mpc}{\text{Mpc}}
\newcommand{\eq}{\text{eq}}

\newcommand{\PBH}{\text{PBH}}
\newcommand{\e}{\text{e}}
\newcommand{\n}{\text{n}}
\newcommand{\kk}{\text{kin}}
\renewcommand{\ss}{\text{s}}

\newcommand{\eva}{\text{eva}}
\newcommand{\obs}{\text{obs}}
\newcommand{\tot}{\text{tot}}

\newcommand{\bfk}{\mathbf{k}}
\newcommand{\bfq}{\mathbf{q}}
\newcommand{\bfx}{\mathbf{x}}

\newcommand{\eqf}{{\text{eq},1}}
\newcommand{\eqs}{{\text{eq},2}}

\newcommand{\mm}{\text{m}}
\newcommand{\rr}{\text{r}}
\newcommand{\cc}{\text{c}}

\newcommand{\UV}{\text{UV}}

\newcommand{\HGF}[4]{{}_2 F_1 \ensuremath{\left(#1, #2; #3; #4 \right)} }

\definecolor{brown(web)}{rgb}{0.65, 0.16, 0.16}
\newcommand{\para}[1]{\phantom{.} \\ \noindent \textsf{\color{brown(web)}{#1.}}}
\definecolor{purpleTT}{rgb}{0.5,0, 0.7}

\makeatletter
\newcommand\footnoteref[1]{\protected@xdef\@thefnmark{\ref{#1}}\@footnotemark}
\makeatother

\allowdisplaybreaks[1]

\begin{document}
 
%%%%%%%%%%%%%%%%%%%%%%%%%%%
%%%%%%%%%%% Title %%%%%%%%%%%
%%%%%%%%%%%%%%%%%%%%%%%%%%%
 
\title{
The poltergeist mechanism - Enhancement of scalar-induced gravitational waves with early matter-dominated era -
}

\author{Keisuke Inomata}
\affiliation{William H. Miller III Department of Physics and Astronomy, Johns Hopkins University, 3400 N. Charles Street, Baltimore, Maryland, 21218, USA}

\author{Kazunori Kohri}
\affiliation{Division of Science, National Astronomical Observatory of Japan and The Graduate University for Advanced Studies (SOKENDAI), Mitaka, Tokyo 181-8588, Japan}
\affiliation{Department of Astronomy, School of Science, The University of Tokyo, Bunkyo-ku, Tokyo 113-0033, Japan}
\affiliation{Theory Center, IPNS, KEK, 1-1 Oho, Tsukuba, Ibaraki 305-0801, Japan}
\affiliation{Kavli IPMU (WPI), UTIAS, The University of Tokyo, Kashiwa, Chiba 277-8583, Japan}

\author{Takahiro Terada}
\affiliation{Kobayashi-Maskawa Institute for the Origin of Particles and the Universe, Nagoya University, Tokai National Higher Education and Research System, Furo-cho, Chikusa-ku, Nagoya 464-8602, Japan}

\begin{abstract} 
\noindent
Gravitational waves induced by primordial density perturbations provide a powerful probe of the Universe’s thermal history, which may include an early matter-dominated (eMD) era predicted by well-motivated particle-physics models.
The induced GWs can be significantly enhanced when the Universe undergoes a sudden transition from an eMD era to an era with pressure, such as a radiation or kination era. 
This enhancement arises from the growth of density perturbations during the eMD era and their rapid oscillations during the era with pressure.
This phenomenon is called the poltergeist mechanism. 
In this review, we explain the essence of the poltergeist mechanism and explore concrete scenarios in which such an enhancement can occur.
\end{abstract}

\date{\today}
\maketitle

\tableofcontents

%%%%%%%%%%%%%%%%%%%%%%%%%%%%%%%%
\section{Introduction}
%%%%%%%%%%%%%%%%%%%%%%%%%%%%%%%%

Over the last decade, gravitational waves (GWs) have been directly measured with the ground-based interferometers~\cite{Abbott:2016blz}. 
Recently, pulsar timing array (PTA) experiments~\cite{NANOGrav:2023gor,EPTA:2023fyk,Reardon:2023gzh,Xu:2023wog} have also reported evidence of the GW signal in a different (nanohertz) frequency range. 
GWs provide a new observational window into the Universe, revealing phenomena inaccessible to electromagnetic wave observations.
In particular, GWs are expected to serve as powerful tools for studying the early Universe. 
Since gravity couples to all forms of matter and energy, GWs—whether their amplitudes are large or small—are inevitably generated throughout cosmic evolution.
Moreover, once produced, GWs experience little damping or diffusion due to the weakness of the gravitational interactions.
As a result, GWs generated in the early Universe can carry information from their production epoch all the way to the present day.

In the early Universe, GWs can arise from various sources: quantum fluctuations of tensor perturbation during inflation~\cite{Rubakov:1982df,Fabbri:1983us,Abbott:1984fp}, bubble collisions associated with first-order phase transition~\cite{Witten:1984rs,Hogan:1986qda,Kosowsky:1991ua,Kosowsky:1992rz,Kosowsky:1992vn,Kamionkowski:1993fg}, topological defects such as cosmic strings and domain walls~\cite{Vilenkin:1981bx,Vilenkin:1981zs,Gelmini:1988sf,Accetta:1988bg}, and scalar perturbations. 
In this review, we focus on GWs induced by scalar perturbations as a probe of the early Universe.

Scalar perturbations originate from the quantum fluctuations of scalar fields during inflation. 
In particular, the fluctuations of the inflaton field, a field that dominates the Universe during inflation, generate density perturbations that seed the cosmic microwave background (CMB) anisotropies and the large-scale structure (LSS) observed today. 
These first-order scalar (density) perturbations induce GWs at second order in perturbations.
Such GWs are commonly referred to as induced GWs, secondary GWs, or scalar-induced GWs (SIGWs).

To the best of our knowledge, the earliest study on SIGWs was made by \rref{10.1143/PTP.37.831}, who studied nonlinear perturbation theory of pressureless matter and first pointed out that density perturbations can generate GWs.
The production of GWs from density perturbations was later discussed in \rref{Matarrese:1993zf,Matarrese:1997ay} in a similar context (the nonlinear perturbation theory of pressureless matter).
The phenomenology of SIGWs began with \rref{Mollerach:2003nq}, which examined how secondary tensor and vector perturbations sourced by density perturbations contribute to the CMB B-mode polarization.
\rref{Ananda:2006af} were the first to use SIGWs as a probe of small-scale perturbations, calculating the SIGWs generated during a radiation-dominated (RD) era and deriving constraints on the power spectrum of curvature perturbations on small scales.
\rref{Baumann:2007zm} extended this analysis by including the transition from the RD era to the dark-matter-dominated era (around redshift $z \sim 3400$) and accounting for the effects of neutrino anisotropic stress.

The connection between SIGWs and primordial black holes (PBHs) was highlighted by \rref{Saito:2008jc,Saito:2009jt}.
PBHs—black holes formed from the collapse of overdense regions in the early Universe before star formation~\cite{Hawking:1971ei,Carr:1974nx,Carr:1975qj}—can account for dark matter~\cite{Chapline:1975ojl,Carr:2016drx}.
For PBHs to form in significant abundance, the curvature power spectrum must be enhanced on small scales, which inevitably leads to strong SIGWs. 
Using this connection, \rref{Saito:2008jc,Saito:2009jt} showed that the PBH abundance can be constrained by stochastic GW observations.
The relation between SIGWs and PBHs was also studied by \rref{Bugaev:2009zh,Bugaev:2009kq,Bugaev:2010bb}. 
\rref{Assadullahi:2009jc} further refined the analysis of SIGW constraints on the small-scale power spectrum using the results of \rref{Saito:2008jc,Saito:2009jt}.

Following the first direct GW detection by LIGO in 2015~\cite{Abbott:2016blz}, interest in SIGWs intensified because PBHs can explain the BHs that produced the detected GWs~\cite{Bird:2016dcv,Clesse:2016vqa,Sasaki:2016jop}.
PBHs with $\mathcal O(10)\,M_\odot$, corresponding to the LIGO BHs, predict large SIGWs around the PTA frequencies ($f\sim 10^{-9}-10^{-8}$\,Hz).
This connection between the LIGO observations and SIGWs was discussed in early works such as~\cite{Inomata:2016rbd,Orlofsky:2016vbd,Nakama:2016gzw,Inomata:2017okj,Garcia-Bellido:2017aan,Di:2017ndc}.\footnote{
  The GWs can be induced by enhanced vector modes during axion (monodromy) inflation. This kind of induced GWs was also studied in the context of the LIGO results~\cite{Cheng:2016qzb,Garcia-Bellido:2016dkw,Domcke:2017fix}.
}
Furthermore, the evidence of stochastic GWs in the PTA experiments in 2023~\cite{NANOGrav:2023gor,EPTA:2023fyk,Reardon:2023gzh,Xu:2023wog} boosted interest, as SIGWs could potentially explain the observed stochastic GWs~\cite{NANOGrav:2023hvm,EPTA:2023xxk}, though the most conservative interpretation attributes it to mergers of supermassive black hole binaries.

Historically, SIGWs have drawn attention as probes of the primordial power spectrum enhanced on small scales, which is inaccessible to CMB and LSS observations.
In this review, however, we focus on another aspect of SIGWs: their sensitivity to transitions between different cosmic eras in the early Universe.
In particular, we discuss the enhancement of the SIGWs associated with an early matter-dominated (eMD) era, which can arise even if the primordial power spectrum is nearly scale-invariant even on small scales. 
The eMD era is an era without pressure that ends before the big-bang nucleosynthesis (BBN), which occurs during the standard RD era. 
The existence of an eMD era is predicted in many high-energy physics models, and understanding how its imprint appears in SIGWs is of great importance. We review the theoretical motivations for an eMD era in detail in Sec.~\ref{sec:emd}.

The phenomenology of SIGWs in the existence of an eMD era was first studied in \rref{Assadullahi:2009nf}. 
They showed that the subhorizon density perturbations during an eMD era can source large-amplitude tensor perturbations because density perturbations grow proportionally to the scale factor during that era.
\rref{Alabidi:2012ex,Alabidi:2013wtp} further explored SIGWs in specific inflation models with and without an eMD era.
These works focused on the SIGWs during the eMD era and assumed that their power spectrum remains unchanged after the eMD era, except for redshift effects.
Later, \rref{Kohri:2018awv} (see also \rref{Terada:2025cto}) revisited the calculation of SIGWs, taking into account the transition from a RD era to a matter-dominated (MD) era and vice versa. 
However, there was a missing effect or contribution to the SIGWs for the transition from an eMD era to a RD era, which was discovered by \rref{Inomata:2019zqy,Inomata:2019ivs}. 
The key finding of these works is that the final SIGWs after the transition sensitively depend on the transition timescale.

In particular, \rref{Inomata:2019zqy} showed that if the transition is gradual—as expected for a constant decay rate of the matter dominating the eMD era—the final SIGWs are smaller than those during the eMD phase.
The exponential suppression of the amplitude of the scalar perturbations during the transition leads to a corresponding suppression of SIGWs, in contrast to earlier assumptions \cite{Assadullahi:2009nf,Alabidi:2012ex,Alabidi:2013wtp} that the SIGW amplitude would remain unaffected once produced during the eMD era.

Meanwhile, \rref{Inomata:2019ivs} showed that if the transition is sudden, the final SIGWs can be much larger than the prediction of the earlier works.
This enhancement of SIGWs arises from the following behavior of density perturbations. 
The density perturbations first grow during an eMD era inside the horizon and, after the transition, they start to oscillate due to the pressure of radiation.
If the transition is sudden, the density perturbations start oscillating with the large amplitude that grows during the eMD era.
In addition, the oscillation timescale for the subhorizon perturbations is much shorter than the Hubble timescale at the transition. 
These rapid oscillations of the large-amplitude density perturbations induce strong GWs. 
This enhancement mechanism is called the poltergeist mechanism~\cite{Inomata:2020lmk}.

The main goal of this review is to elucidate the essence of the poltergeist mechanism, which provides a novel way to probe the early Universe through the timescale of the transition from an eMD era.
\rref{Inomata:2020lmk} illuminated the potential power of the poltergeist mechanism by showing that it can generate strong SIGWs if an eMD era is realized through the domination of PBHs.
This idea has since inspired numerous studies.
The poltergeist mechanism has also been applied to other compact objects, such as Q-balls and oscillons, and it may also work in scenarios involving axion rotation.
We will also review these concrete examples in this review.

The remainder of this review is organized as follows. 
In Sec.~\ref{sec:emd}, we review the theoretical motivation for an eMD era. 
Section~\ref{sec:scalar_evo} summarizes the evolution of the background quantities and the first-order perturbations in the presence of an eMD era. 
Following that, we review basic formulas for SIGWs in Sec.~\ref{sec:SIGW}.
In Sec.~\ref{sec:gradual}, we discuss the suppression of SIGWs in the case of a gradual transition from an eMD era to a RD era.
Then, we see the enhancement of SIGWs in the case of a sudden transition (the poltergeist mechanism) in Sec.~\ref{sec:sudden}, where we explain the mechanism intuitively, highlight the essence with a instantaneous-limit transition, and show concrete examples. 
We devote Sec.~\ref{sec:conclusions} to conclusions.
In Appendix~\ref{app:einstein_em_tensor}, we summarize explicit expressions of the Einstein and energy-momentum tensors up to first order in perturbations. 
Appendix~\ref{app:perturbation_during_trans} contains the derivation of the equation of motion for the scalar perturbations during the transition from an eMD era to a RD era. 
Analytic formulas for the integration kernel and the GW spectrum for the power-law power spectrum of the curvature perturbations are summarized in Appendix~\ref{app:analytic_formulas} in the context of the poltergeist mechanism.

Several reviews on SIGWs already exist.
A review by \cite{Domenech:2021ztg} covers general topics of SIGW with an emphasis on the SIGW production during an epoch with general equation of state. 
Other reviews by the same author are the one focusing on the GWs induced by (initially) isocurvature perturbations~\cite{Domenech:2023jve} and the one focusing on the PBH-associated SIGWs that may be tested by terrestrial GW detectors~\cite{Domenech:2024cjn}.
The theme of the review by~\cite{Yuan:2021qgz} is the relation to PBHs, and it also contains discussions on the next-to-leading (third) order effect in the cosmological perturbation theory. 
In contrast, our review focuses on the impact of an eMD era on SIGWs: the poltergeist mechanism in sudden transition scenarios, as well as the basic case without the poltergeist enhancement in gradual/slow transition scenarios, which we include for comparison.

Throughout this review, we use the Greek characters, such as $\mu$, $\nu$, $\rho$, and $\sigma$, for the space-time coordinates, running over the set $\{0, 1, 2, 3\}$, and the Latin characters, such as $i$, $j$, $k$, and $l$, for the space coordinates, running over the set $\{1, 2, 3\}$. We adopt the natural units with $c = k_\text{B} = \hbar=1$.

%%%%%%%%%%%%%%%%%%%%%%%%%%%%%%%%
\section{Early matter-dominated era in particle physics and cosmology}
\label{sec:emd}
%%%%%%%%%%%%%%%%%%%%%%%%%%%%%%%%

In this section, we introduce the theoretical motivation for an early matter-dominated (eMD) era, which is a necessary and essential ingredient of the poltergeist mechanism for GW production. A reader familiar with an eMD era can skip this section. 

In the concordance cosmological model, the $\Lambda$CDM model, the Universe experiences the RD era that accommodates the BBN, followed by the MD era during which the CMB radiation is last scattered. More than $80\%$ of matter is in the form of cold dark matter (CDM)~\cite{Planck:2018vyg}.  In the late-time Universe, the energy budget is dominated by dark energy, whose simplest candidate is the cosmological constant $\Lambda$.  

There are various reasons to consider cosmology beyond the $\Lambda$CDM model. Within the $\Lambda$CDM model, there are tensions such as the Hubble tension and $S_8$ tension~\cite{H0DN:2025lyy, DiValentino:2021izs, Abdalla:2022yfr}.  While these tensions may be due to lack of understanding of systematics in observations, there are possibilities that new physics is the cause of these tensions.  Apart from the cosmological tensions, the very ingredients of the $\Lambda$CDM, namely, dark energy and dark matter are not well understood in particle physics. While dark energy can be a cosmological constant,\footnote{
Recent analyses by the DESI collaborations indicate a preference for dynamical dark energy~\cite{DESI:2024mwx, DESI:2024hhd, DESI:2025zgx, DESI:2025fii}. See also a reanalysis of Dark Energy Survey~\cite{DES:2025sig}, which affects the significance of the dynamical dark energy signal.
} dark matter most likely requires new physics beyond the Standard Model (BSM) of particle physics. The dark matter may be a new particle species or a macroscopic object. In well-motivated BSM scenarios, not only a single dark matter candidate but also a bunch of new particles is usually introduced for various reasons.\footnote{
For example, a set of particles may be needed for anomaly cancellation. Or, supersymmetry requires a superpartner for each Standard-Model particle. If the dark matter is a composite particle like a dark baryon, there are excited dark baryonic particles and possibly dark mesonic particles too. Further, the completeness conjecture even states that there must exist particles with any possible gauge charges in the presence of gravity~\cite{Polchinski:2003bq}.
} Then, it is natural to consider cosmology involving these BSM particles. 

In the BSM cosmology, one often encounters new massive and unstable particles. Since their energy density redshifts like $a^{-3}$ while the radiation energy density redshifts like $a^{-4}$, where $a$ is the scale factor of the Universe, the fraction of the massive and unstable particles grows even if the initial abundance is small.  Therefore, if the initial abundance is sufficiently large and the lifetime is sufficiently long, they can eventually and tentatively dominate the energy density of the early Universe. The resultant MD era is called an eMD era. 

To name a few, there are candidates like moduli, dilaton, axion and axion-like particles in superstring theory, their superpartners like modulino, dilatino, axino, and also gravitino as well as inflaton.    Heavy particles can be composite particles and even be macroscopic compact objects such as PBHs~\cite{Hawking:1971ei, Carr:1974nx} and Q-balls~\cite{Coleman:1985ki}.  Moreover, new heavy particles or macroscopic objects are not the only reason for the eMD era.  For example, coherent oscillations of a scalar field $\phi(t)$ with its scalar potential $V(\phi)\sim \phi^{2n}$ behave like perfect fluid with the equation-of-state parameter $w \equiv P/\rho = (n-1)/(n+1)$ after the oscillation average~\cite{Turner:1983he, Johnson:2008se}, where $\rho$ and $P$ are the energy density and pressure. A generic scalar potential $V(\phi)$ can be expanded around its minimum, and the leading term in the expansion is the quadratic term $(n=1)$, leading to the dust behavior $w = 0$~\cite{Preskill:1982cy, Abbott:1982af, Dine:1982ah}. This can be viewed as a Bose-Einstein condensate of nonrelativistic $\phi$ particles.

Strictly speaking, the perturbations of the coherently oscillating scalar field in a quadratic potential behave differently from the dust-like fluid for the wavenumber in $k \gtrsim \sqrt{a m \mathcal H}$ ($m$: mass of the oscillating field)~\cite{Cembranos:2015oya} (see also \rref{Kodama:1984ziu,Hamazaki:1996ir,Kodama:1996jh}). 
However, for the perturbations that enter the horizon during an eMD era, which are the main source of the poltergeist mechanism, we can regard the fluid of the oscillating field as the dust-like fluid even at the perturbation level.
The reason is as follows. 
The comoving horizon scale at the beginning of the coherent oscillations is given as $\sim (a m)^{-1}$ at that time because a scalar field starts to coherently oscillate when $\mathcal H \sim a m$.
The eMD era begins soon or later after the coherent oscillations begin.
Then, the wavenumber of the perturbation entering the horizon during the eMD era always satisfies $k < m a|_{t = t_\text{eMD}} < \sqrt{a(t) m \mathcal H(t)}$ ($t_\text{eMD}$: the start time of the eMD era), considering the fact that $\sqrt{am\mathcal H}$ is proportional to $a^{1/4}$ during the eMD era.
Therefore, for our purpose, we can regard the coherently-oscillating field fluid as the dust-like fluid even at the perturbation level.

One of the candidates for such a coherently oscillating scalar field is inflaton.  After slow-roll inflation, it is supposed to oscillate around the minimum of the potential.\footnote{
It is also possible to consider models without inflaton oscillations with runaway-type potential. The epoch of kinetic-energy domination is called kination~\cite{Spokoiny:1993kt, Joyce:1996cp, Ferreira:1997hj}.  In this case, one should make sure that the reheating successfully occurs. See, \textit{e.g.}, \cite{Chun:2009yu, Dimopoulos:2018wfg, Opferkuch:2019zbd, Lankinen:2019ifa, Kamada:2019ewe,Bettoni:2021zhq,Laverda:2023uqv}.
} The inflaton coherent oscillation is an example of the eMD era. 
Depending on the (self-)coupling of the inflaton itself and other fields, there can be a stage of ``nonperturbative'' particle production called preheating~\cite{Dolgov:1989us, Traschen:1990sw, Shtanov:1994ce, Kofman:1994rk, Kofman:1997yn}, where the effects of the preexisting particles cannot be neglected. Let us discuss whether and how preheating affects the eMD era. The typical phenomena in preheating include parametric resonance and tachyonic instability. These are efficient mechanisms for the production of (semi)relativistic particles, and the equation of state approaches that of radiation. However, preheating cannot complete the process of energy transfer from the inflaton condensate, and one has to assume the standard perturbative decay process to complete the reheating of the Universe~\cite{Kofman:1997yn}. In fact, the preheating process is shut off when the backreaction from the produced particles to the homogeneous background (inflaton coherent oscillation) becomes important. Sooner or later, the momenta of the produced particles are redshifted by cosmic expansion. This means that even if preheating occurs, it is a tentative process and the equation-of-state will sooner or later be back to matter-like ($w=0$) until the inflaton condensate finally decays. For recent development on processes toward thermalization, see \rref{Harigaya:2013vwa, Mukaida:2015ria,Drees:2021lbm, Passaglia:2021upk,Drees:2022vvn, Mukaida:2024jiz, Fujita:2025zoa}.  

A similar but somewhat different example is a complex scalar field rotating on the body of its potential.  Again, the potential is generically quadratic expanded around its minimum.  Since a complex scalar field can be regarded as a collection of two real scalar fields, one can interpret the rotation as a combination of oscillations.  
This case too can behave as a MD era, and we will discuss it in Sec.~\ref{sssec:axion_poltergeist}. 

Let us discuss the fate of the eMD era. The most common end of the eMD era is decay of the particles/condensates dominating the eMD era.  This leads to emission of relativistic daughter particles. Then, the eMD era is followed by the RD era. In this case, there can be an early RD era, followed by the eMD era and the standard RD era.  By the decay of matter, entropy is produced and preexisting particles in the radiation are diluted. This should occur sufficiently before the BBN, as otherwise the light element abundance is affected.  On the other hand, entropy production is useful for diluting unwanted long-lived particles such as moduli and gravitinos, but it may also dilute, \textit{e.g.}, the baryon asymmetry of the Universe depending on the time of its production in the early Universe.  On the other hand, the eMD era can be followed by another cosmological epoch.  An example is the axion-rotation case where the equation-of-state parameter $w$ of the axion fluid morphs from matter-like $(w=0)$ to kinetic-energy-like $(w=1)$~\cite{Co:2019wyp, Co:2019jts} as we discuss more in Sec.~\ref{sssec:axion_poltergeist}. 

As we have seen above, an eMD era is a not-so-exotic extension of the standard cosmology and often well-motivated from particle physics.  For further discussions on an eMD era, see the review on nonstandard cosmological epochs~\cite{Allahverdi:2020bys}. 

GW signals related to an eMD era are valuable probes of the early Universe and high-energy physics. Before we discuss the induced GWs in the presence of an eMD era, we review the dynamics of the scalar component of the gravitational perturbations, which is the source of the induced GWs, in the next section.

%%%%%%%%%%%%%%%%%%%%%%%%%%%%%%%%
\section{Transition between matter and radiation-dominated eras}
\label{sec:scalar_evo}
%%%%%%%%%%%%%%%%%%%%%%%%%%%%%%%%

In this section, we summarize the evolution of the Universe through the transition between a MD era and a RD era in a pedagogical fashion. 
Specifically, we consider its evolution both in transitions eMD $\to$ RD and RD $\to$ eMD.
This is because, while the poltergeist mechanism occurs soon after an eMD $\to$ RD transition, we need to care about a RD $\to$ eMD transition that arises in the RD $\to$ eMD $\to$ RD transition, which we consider as concrete examples in Sec.~\ref{sec:sudden}.
A reader familiar with the first-order cosmological perturbations involving multiple cosmic epochs may skip this section and use this section as a reference for equations when necessary.  
Specifically, we summarize the evolution of the background and the first-order perturbations that experience the transition, using the Einstein equation,
\begin{align}
  G_{\mu\nu} = \frac{T_{\mu\nu}}{M_\Pl^2},
\end{align}
where $G_{\mu\nu}$ is the Einstein tensor, $T_{\mu\nu}$ is the energy-momentum tensor, and $1/M_\Pl^2 \equiv 8 \pi G$ with $G$ being the Newton constant.

In the conformal Newtonian gauge,\footnote{The gauge (in)dependence of SIGWs is discussed, \textit{e.g.}, in \rref{Nakamura:2004rm,Arroja:2009sh,Hwang:2017oxa,Gong:2019mui,Tomikawa:2019tvi,DeLuca:2019ufz,Inomata:2019yww,Yuan:2019fwv,Lu:2020diy,Ali:2020sfw,Chang:2020iji,Domenech:2020xin,Ota:2021fdv}.
The key point is that, within appropriate gauges where the scalar perturbation sources decouple from tensor modes on deeply subhorizon scales, SIGWs finally become gauge-independent.
Such appropriate gauges exist in the era with $w \neq 0$, when the poltergeist mechanism produces GWs, and the conformal Newtonian gauge is one of them.
} the metric is given by 
%%%%%%
\begin{align}
\dd s^2 &= g_{\mu\nu} \dd x^\mu \dd x^\nu \nonumber \\
&= a^2(\eta) \left[ -(1+2\Phi)\dd \eta^2 
\phantom{ \left\{ \frac{1}{2} \right\} }  \right. \nonumber \\
& \quad \left. + \left\{ (1-2\Psi)\delta_{ij} + \frac{1}{2} h_{ij} \right\} \dd x^i \dd x^j \right],
\end{align}
%%%%%%
where the gravitational potential $\Phi$ and $\Psi$, related to the curvature perturbations, are the first-order scalar perturbations, and $h_{ij}$ is the second-order tensor perturbations. 
We have neglected the vector perturbations\footnote{
The first-order vector metric perturbations in a perfect fluid always decay proportionally to $a^{-2}$ regardless of the transition of the equation-of-state parameter~\cite{Mukhanov:991646}.
The second-order vector perturbations induced by the other perturbations are out of the scope of this review.
} and first-order tensor perturbations to focus on the GWs induced by scalar perturbations throughout this review.
GWs induced through scalar-tensor or tensor-tensor interactions have recently been focused on in several studies~\cite{Gong:2019mui,Chang:2022vlv,Chang:2022aqk,Yu:2023lmo,Bari:2023rcw,Picard:2023sbz,Picard:2024ekd}.

For the energy-momentum tensor, we consider the perfect fluid, where anisotropic stress is absent, for the matter and the radiation fluid as 
\begin{align}
  T_{\mu \nu} = (\rho + P) u_\mu u_\nu + P g_{\mu \nu},
  \label{eq:t_munu}
\end{align}
where $\rho$ is the energy density, $P$ is the pressure, and $u_\mu$ is the four velocity.
We here explain why this perfect fluid assumption is reasonable in our setup. 
Throughout this review, we consider the case where there are two fluids: matter (dust) fluid and radiation fluid.
For matter fluid, there is no anisotropic stress within the perturbation theory because the dust by definition consists of particles with a negligible velocity dispersion, which leads to negligible pressure and anisotropic stress.
This argument applies even for the perturbation level unless the perturbation theory breaks down due to the shell crossing.
For radiation fluid, the anisotropic stress can be negligible when the radiation sufficiently interacts with each other~\cite{Weinberg:2008zzc}.
Since the eMD era is followed by the RD era where the BBN occurs, the radiation produced at the end of the eMD era consists of the standard model particles. 
The standard model particles sufficiently interact with each other in the early Universe, especially before the neutrino decoupling (the temperature of the Universe is of $\sim \mathcal O(1)\,\MeV$)~\cite{Jeong:2014gna}, which we mainly focus on in this review. 
Given this, it is reasonable to neglect the anisotropic stress even for the radiation fluid in our setup, similar to the standard cosmology.
For these reasons, the perfect fluid is a good approximation for each of the matter and radiation fluid. 
Furthermore, although the neutrino anisotropic stress is non-negligible after the neutrino decoupling, \rref{Baumann:2007zm} finds that its effect on the SIGW spectrum is typically less than $1\,\%$.

We express the perturbations as $\rho = \bar \rho + \delta \rho$, $P = \bar P + \delta P$, and $u_\mu = \bar u_\mu + \delta u_\mu$, where the upper bars represent the background values.
Note that the four velocity satisfies $u_\mu u^\mu = -1$ by definition and the homogeneous Universe background leads to $\bar u_\mu = -a(1,0,0,0)$ and $\bar u^\mu = a^{-1}(1,0,0,0)$.
In addition, we neglect the vector perturbation contribution and express $\delta u_i = \delta u_{,i}$, where $\delta u$ is the velocity potential and the subscript $_{,i} \equiv \partial_i$.

\subsection{Background evolution}
Let us begin with the background evolution.
From the \{00\} and \{$ii$\} (trace) components of the Einstein equation, we can obtain the so-called Friedmann equations:
\begin{align}
  \label{eq:hubble_eq1}
  &3M_\Pl^2 \mathcal H^2 = a^2 \bar\rho, \\
  \label{eq:hubble_eq2}
  &M_\Pl^2 \left( 2\mathcal H' + \mathcal H^2 \right) = -a^2\bar P,
\end{align}
where $\mathcal H \equiv a'/a = a H$ is the conformal Hubble parameter, and the prime denotes the conformal time derivative.
From this, we can obtain the well-known results that $a \propto \eta$ during a RD era ($w \equiv \bar P/\bar \rho=1/3$) and $a \propto \eta^2$ during a MD era ($w = 0$).
Substituting Eq.~(\ref{eq:hubble_eq1}) into Eq.~(\ref{eq:hubble_eq2}), we can obtain another form of the Friedmann equation,
\begin{align}
  \label{eq:hubble_eq3}
  \frac{a''}{a} = \frac{1}{6M_\Pl^2} (\bar \rho - 3 \bar P) a^2,
\end{align}
which we will use in the next subsection.

In the following, we discuss the background evolution during the transition between a RD era and a MD era.

%%%%%%%%%%%%%%%%%
\subsubsection{Transition from RD era to MD era}
\label{sssec:back_rd_to_md}
%%%%%%%%%%%%%%%%%

First, we consider the transition from a RD era to a MD era, which can occur at the beginning of the eMD era.
We will take into account this transition in the analysis of our concrete examples, discussed in Sec.~\ref{sec:sudden}.
This transition occurs if the Universe is dominated by radiation, but has a sufficient amount of non-relativistic matter as a subdominant component. 
This is because the radiation energy density decays faster ($a^{-4}$) than the matter energy density ($a^{-3}$).
Also, the observations of CMB and LSS have revealed that this kind of transition occurred at $z\sim 3400$ to the dark matter-dominated era~\cite{Planck:2018vyg}.
This kind of transition also occurs at the beginning of an eMD era in the early Universe.

Through the transition from a RD era to a MD era, the scale factor depends on the conformal time as~\cite{Mukhanov:991646}
\begin{align}
  a(\eta) = a_\eq \left( \left(\frac{\eta}{\eta_*}\right)^2 + 2 \left(\frac{\eta}{\eta_*}\right) \right),
\end{align}
where $\eta_*$ is defined as
\begin{align}
  \eta_* = \left( \frac{\rho_\eq a_\eq^2}{24 M_\Pl^2} \right)^{-1/2} = \frac{\eta_\eq}{\sqrt{2}-1},
\end{align}
where the subscript `eq' represents the value at the matter-radiation equality time.
For the transition from the RD era to the dark matter-dominated era, the CMB and LSS observations have measured the comoving wavenumber of the horizon scale at the matter-radiation equality time as (from the Planck TT,TE,EE+lowE+lensing+BAO result in \rref{Planck:2018vyg})
\begin{align}
  \label{eq:keq}
  \frac{k_\eq}{\Mpc^{-1}} = 0.010339 \pm 0.000063,
\end{align}
where $k_\text{eq} \equiv a_\eq H_\eq = 2(2-\sqrt{2})/\eta_\eq$. 
Using this, we can obtain the conformal equality time as 
\begin{align}
  \frac{\eta_\eq}{\Mpc} = 113.32 \pm 0.69.
\end{align}

Since the cosmological time is often characterized by the photon temperature, it is useful to derive their relation.
From the Friedmann equation (Eq.~(\ref{eq:hubble_eq1})), we obtain
%%%%%%
\begin{align}
  \frac{aH}{a_\text{eq} H_\text{eq}} = \frac{a}{a_\text{eq}} \sqrt{\frac{\rho}{2 \rho_\text{r,eq}}}.
\end{align}
%%%%%%
During a RD era, where $aH = 1/\eta$, we get
%%%%%%
\begin{align}
  \label{eq:eta_temp_relation}
  \frac{1}{k_\text{eq} \eta} = \frac{1}{\sqrt{2}} \left( \frac{g_{s*,\text{eq}}}{g_{s*}} \right)^{1/3} \left( \frac{g_{*}}{g_{*,\text{eq}} } \right)^{1/2} \frac{T}{T_\text{eq}} \ (\text{for} \ \eta \ll \eta_\eq),
\end{align}
%%%%%%
where $g_*$ and $g_{s*}$ are the effective relativistic degrees of freedom for the energy and entropy density, respectively, and we have used the entropy conservation relation:
\begin{align}
    g_{s*,\eq} a_\eq^3 T_\eq^3 = g_{s*} a^3 T^3.
\end{align}
Note that $g_{*,\text{eq}} = 3.38$, $g_{s*,\text{eq}} = 3.93$~\cite{Saikawa:2018rcs}, and $T_\text{eq} = 8.0 \times 10^{-7}$\,MeV, where we have used $T_\eq = (1+z_\eq) T_0$ with $z_\eq = 3400$~\cite{Planck:2018vyg} and $T_0 = 2.7255\,$K~\cite{Fixsen:1996nj}.

%%%%%%%%%%%%%%%%%
\subsubsection{Transition from MD era to RD era}
\label{subsec:trans_md_to_rd}
%%%%%%%%%%%%%%%%%

After the eMD era, the standard RD era must be realized. The simplest possibility is that the latter directly follows the former. Thus, we consider a transition from a MD era to a RD era. 
For example, a coherent oscillation of a purely massive scalar field could realize a MD era preceding the RD era and the decay of the scalar field leads to the transition from a MD era to the RD era~\cite{Turner:1983he}.  
Here, we consider the case where the matter decays to radiation with a decay rate of $\Gamma$, which can be time-dependent. 
In this case, the energy-momentum conservation can be expressed as~\cite{Kodama:1996jh,Hamazaki:1996ir} 
\begin{align}
  \label{eq:t_m_munu}
  T^\mu_{\text{m}\, \nu; \mu} &= \Gamma\, T^\mu_{\text{m}\, \nu} u_{\text{m}\,\mu}, \\
  \label{eq:t_r_munu}
  T^\mu_{\text{r}\, \nu; \mu} &= -\Gamma\, T^\mu_{\text{m}\, \nu} u_{\text{m}\,\mu},
\end{align}
where the subscript ``m'' and ``r'' represent the matter and radiation components, respectively.
Note that $\Gamma$ is the decay rate of the matter per unit proper time of the matter and therefore the decay rate appears with $u_{\text{m} \, \mu}$.
The total energy-momentum tensor is given as $T^\mu_{\text{tot} \, \nu} = T^\mu_{\mm\, \nu} +  T^\mu_{\rr\, \nu}$ and the energy-momentum conservation law is satisfied for the total tensor as $T^\mu_{\text{tot} \, \nu;\mu} = 0$.

Substituting Eq.~(\ref{eq:t_munu}) into Eqs.~(\ref{eq:t_m_munu}) and (\ref{eq:t_r_munu}), we obtain the following equations:
\begin{align}
  \label{eq:rho_m_with_decay}
  \rho'_\text{m} + 3 \mathcal H \rho_\text{m} &= - a \Gamma \rho_\text{m}, \\
  \label{eq:rho_r_with_decay}
  \rho'_\text{r} + 4 \mathcal H \rho_\text{r} &=  a \Gamma \rho_\text{m}.
\end{align}
Note again $\Gamma$ can be time-dependent in general.
If $\Gamma$ is time-independent, it leads to a gradual transition from a MD era to a RD era. 
A time-independent $\Gamma$ corresponds to the case where a coherent oscillating scalar field decays to radiation with the perturbative decay~\cite{Dolgov:1982th,Abbott:1982hn}.
For a time-independent $\Gamma$, there are no analytical solutions for these equations and we need to perform numerical calculations to solve them, as we do in Sec.~\ref{sec:gradual}.
On the other hand, for a time-dependent $\Gamma$ that leads to an instantaneous transition from a MD era to a RD era, we can obtain the analytic solution for $a$.
We consider this case because there are realistic cases where a nearly instantaneous transition occurs, as we will discuss in Sec.~\ref{sec:sudden}.
In this case, the scale factor dependence of the energy density instantaneously changes from $\rho \propto a^{-3}$ to $\rho \propto a^{-4}$.
Using this and Eq.~(\ref{eq:hubble_eq1}), we can obtain
%%%%%%
\begin{align}
  \label{eq:scale_sudden}
 \frac{a(\eta)}{a(\eta_\text{R})} = \begin{cases}
 \left( \cfrac{\eta}{\eta_\R} \right)^2 & (\eta < \eta_\R) \\
 2 \cfrac{\eta}{\eta_\R} -1 & (\eta \geq \eta_\R)
 \end{cases},
\end{align}
%%%%%%
where the instantaneous transition occurs at $\eta_\R$, and we have imposed that the scale factor and its time derivative are continuous at $\eta_\R$ because the total energy density does not change through the instantaneous transition.
From Eq.~(\ref{eq:scale_sudden}), we can express the conformal Hubble parameter in this case as
%%%%%%
\begin{align}
  \label{eq:hubble_sudden}
  \mathcal H = \begin{cases}
  \cfrac{2}{\eta} & (\eta < \eta_\R) \\
  \cfrac{1}{\eta - \eta_\R/2}  & (\eta \geq \eta_\R)
  \end{cases}.
\end{align}
%%%%%%

\subsection{First-order perturbations}

In this section, we summarize the evolution of the first-order scalar perturbations.

\subsubsection{One fluid}

We first briefly review the perturbation evolution in one fluid, based on \rref{Mukhanov:991646}.
From the Einstein equation, the perturbations of the Einstein tensor and the energy-momentum tensor satisfy
\begin{align}
  \label{eq:einstein_eq_pertb}
  \delta G_{\mu\nu } = \frac{1}{M_\Pl^2} \delta T_{\mu\nu}.
\end{align}
From the $\{ 00\}$ and $\{0i\}$ components of this equation, we obtain
\begin{align}
  \label{eq:pertb_eq_00_new2}
  &\Delta \Psi - 3 \mathcal H (\Psi' + \mathcal H \Phi) = \frac{a^2}{2M_\Pl^2} \delta \rho, \\
  \label{eq:pertb_eq_0i_new2}
  &(\Psi' + \mathcal H \Phi)_{,i} = -\frac{a}{2M_\Pl^2} (\bar \rho + \bar P)\delta u_{,i}.
\end{align}
From the trace and the non-diagonal components, we obtain
\begin{align}  
  \label{eq:pertb_eq_ii_new2}
  &\Psi'' + \mathcal H (2\Psi + \Phi)' + (2\mathcal H' + \mathcal H^2)\Phi + \frac{1}{2} \Delta(\Phi - \Psi) \nonumber \\
  &\qquad \qquad \qquad \qquad \qquad \qquad \qquad \qquad 
  =  \frac{a^2}{2M_\Pl^2} \delta P, \\
  \label{eq:pertb_eq_ij_new2}  
  &(\Phi - \Psi)_{,ij} = 0.
\end{align}
From the final equation, we can obtain $\Phi = \Psi$.

In one fluid, we can relate the pressure perturbation to the energy-density perturbation as 
\begin{align}
  \delta P = c_\ss^2 \delta \rho,
\end{align}
where $c_\ss$ is the sound speed.
Substituting this equation into Eq.~(\ref{eq:pertb_eq_ii_new2}) and using Eq.~(\ref{eq:pertb_eq_00_new2}), we can derive 
\begin{align}
  \label{eq:phi_only_eq}
  \Phi'' + 3(1+c_\ss^2) \mathcal H \Phi' - c_\ss^2 \Delta \Phi + (2\mathcal H' + (1+3c_\ss^2) \mathcal H^2) \Phi = 0.
\end{align}

In the following, we discuss the evolution of the perturbations specifying the era of the Universe.

%%%%%%%%%%%%%%%%%
\para{Matter-dominated era: $c_\ss^2 = 0$}
%%%%%%%%%%%%%%%%%
During a MD era, we have $c_\ss^2 = 0$ and Eq.~(\ref{eq:phi_only_eq}) becomes 
\begin{align}
  \label{eq:phi_eq_md_era}
  \Phi'' + \frac{6}{\eta}\Phi' = 0,
\end{align}
where we have used the relation $\mathcal H = 2/\eta$. 
Solving this equation, we obtain
\begin{align}
  \Phi(\bfx) = C_1(\bfx) + \frac{C_2(\bfx)}{\eta^5},
\end{align}
where $C_1$ and $C_2$ are time-independent constants.
From this equation, we can see that if the decaying mode ($\propto 1/\eta^5$) is neglected, $\Phi$ is always constant during a MD era on either superhorizon or subhorizon scales.

Substituting this solution into Eq.~(\ref{eq:pertb_eq_00_new2}), we get the solution for the energy-density perturbation as 
\begin{align}
  \frac{\delta \rho}{\bar \rho} = \frac{1}{6} \left[ (\Delta C_1 \eta^2 - 12C_1) + (\Delta C_2 \eta^2 + 18 C_2) \frac{1}{\eta^5} \right].
\end{align}
In the Fourier space, this equation becomes
\begin{align}
  \frac{\delta \rho_{\bfk}}{\bar \rho} = -\frac{1}{6} \left[ C_{1\bfk}( x^2 + 12) + C_{2\bfk}(x^2 - 18 ) \frac{1}{\eta^5} \right],
\end{align}
where the subscript ``$\bfk$'' represents the Fourier mode of perturbation, and we have introduced the compact notation
\begin{align}
    x \equiv k \eta,
\end{align}
which should not be confused with the coordinates.
When $x$ indicates coordinates, we put the sub/superscript to it as $x^\mu$, or use the bold form as $\bfx$.\footnote{
Although not optimal, this notation is adopted for consistency with the notation widely used in the literature~~\cite{Saito:2009jt,Kohri:2018awv,Domenech:2021ztg,Mukhanov:991646}, where $x = k \eta$ and $ x = k \eta /\sqrt{3}$ depending on convention.
}
This can be interpreted either as the dimensionless wavenumber or the dimensionless time.  In particular, the superhorizon and subhorizon regimes correspond to $x \ll 1$ and $x \gg 1$, respectively.

On superhorizon scales ($k \eta \ll 1$), this can be approximated as 
\begin{align}
   \frac{\delta \rho_{\bfk}}{\bar \rho} \simeq -2C_{1\bfk} \ \ (k\eta \ll 1),
   \label{eq:delta_ini}
\end{align}
where we have neglected the decaying mode.
This shows that the energy density perturbation is constant on superhorizon scales.
On the other hand, the perturbation on subhorizon scales ($k\eta \gg 1$) can be approximated as 
\begin{align}
  \label{eq:delta_rho_m_sub}
   \frac{\delta \rho_{\bfk}}{\bar \rho} \simeq -\frac{1}{6} C_{1 \bfk} x^2 \propto a \ \ (k\eta \gg 1),
\end{align}
where we have neglected the decaying mode again.
This means that the density perturbation grows proportionally to the scale factor on subhorizon scales during a MD era.
For later convenience, we here obtain the velocity potential $\delta u$ as 
\begin{align}
  \frac{\delta u_{\bfk}}{a} \simeq -\frac{\eta}{3}\Phi_{\bfk} \ 
  ,
\end{align}
where we have used Eq.~(\ref{eq:pertb_eq_0i_new2}).

%%%%%%%%%%%%%%%%%
\para{Radiation-dominated era: $c_\ss^2 = 1/3$}
%%%%%%%%%%%%%%%%%
During a RD era ($c_\ss^2 = 1/3$), we can express Eq.~(\ref{eq:phi_only_eq}) in the Fourier space as
\begin{align}
  \label{eq:phi_eq_rd_era}
  \Phi_{\bfk}'' + \frac{4}{\eta} \Phi_{\bfk}' + \frac{k^2}{3} \Phi_{\bfk} = 0.
\end{align}
Note that $\mathcal H = 1/\eta$ during a RD era.
Solving this equation, we obtain
\begin{align}
  \label{eq:phi_evo_rd_era}
  \Phi_{\bfk} = \frac{3\sqrt{3}}{x} \left[ C_{1 \bfk} j_1(x/\sqrt{3})  + C_{2 \bfk} y_1(x/\sqrt{3}) \right],
\end{align}
where $C_{1/2,\bfk}$ is some constant and $j_1$ and $y_1$ are the spherical Bessel functions, given by 
\begin{align}
  j_1(x) &= \frac{\sin x - x \cos x}{x^2}, \\
  y_1(x) &= -\frac{\cos x + x \sin x}{x^2}.
\end{align}

Substituting Eq.~(\ref{eq:phi_evo_rd_era}) into Eq.~(\ref{eq:pertb_eq_00_new2}), we obtain
\begin{align}
  \label{eq:delta_rho_rd}
  \frac{\delta \rho_{\bfk}}{\bar \rho} =& \frac{6}{x^3} \left[ C_{1 \bfk} \left\{x (x^2 -6) \cos(x/\sqrt{3}) \right. \right. \nonumber \\
  &\left.\left.\qquad\qquad\quad
  - 2\sqrt{3} (x^2 -3) \sin(x/\sqrt{3}) \right\} \right.\nonumber \\
  &\left. + C_{2 \bfk} \left\{ x (x^2 -6) \sin(x/\sqrt{3}) \right. \right. \nonumber\\
  &\left.\left.\qquad\qquad\quad
  + 2\sqrt{3} (x^2 -3) \cos(x/\sqrt{3})\right\} \right].
\end{align}
Given $j_1(x\rightarrow 0) = x/3$ and $y_1(x\rightarrow 0) = -\infty$, we discard the $C_2$ term in Eq.~(\ref{eq:delta_rho_rd}) on the basis of the initial condition, Eq.~(\ref{eq:delta_ini}).
Then, the gravitational potential on superhorizon scales can be expressed as 
\begin{align}
  \Phi_{\bfk} \simeq C_{1 \bfk} \ \ (k \eta \ll 1).
\end{align}
On the other hand, it can be expressed on subhorizon scales as 
\begin{align}
  \Phi_{\bfk} \simeq -\frac{9 C_{1 \bfk}}{x^2}   \cos(x/\sqrt{3}) \ \ (k\eta \gg 1).
\end{align}
This equation shows that the gravitational potential oscillates and decays proportionally to $1/x^2$ on subhorizon scales during a RD era, unlike during a MD era. 
This is due to the pressure of the radiation fluid.

From Eq.~(\ref{eq:delta_rho_rd}), we can express the energy density perturbation on superhorizon scales as 
\begin{align}
  \frac{\delta \rho_{\bfk}}{\bar \rho} \simeq -2 C_{1 \bfk} \ \ (k\eta \ll 1).
\end{align}
On subhorizon scales, it is given by
\begin{align}
  \frac{\delta \rho_{\bfk}}{\bar \rho} \simeq 6 C_{1 \bfk} \cos(x/\sqrt{3}) \ \ (k\eta \gg 1).
\end{align}
The energy density perturbation does not grow on subhorizon scales during a RD era.
This is also due to the radiation pressure.
From Eq.~(\ref{eq:pertb_eq_0i_new2}), we can also obtain the expression for the velocity perturbation on superhorizon scales as
\begin{align}
  \frac{\delta u_{\bfk}}{a} \simeq -\frac{\eta}{2}\Phi_{\bfk} \ \ (k\eta \ll 1).
\end{align}
For completeness, we also record the subhorizon expression
\begin{align}
    \frac{\delta u_{\bfk}}{a} \simeq -\frac{3\sqrt{3} C_{1 \bfk}}{2 k } \sin (x/\sqrt{3}) \ \ (k\eta \gg 1).
\end{align}

\subsubsection{Two fluids}

Next, we discuss the perturbation evolution in two fluids: matter fluid and radiation fluid. 
Note that, even if we consider the adiabatic perturbation, the perturbations of matter and radiation behave differently (or the entropy perturbation appears) on subhorizon scales, and therefore the evolution cannot be described with the one-fluid formulation, given in the previous subsection.

We begin with the perturbation of Eqs.~(\ref{eq:t_m_munu}) and (\ref{eq:t_r_munu}):
\begin{align}
  \label{eq:d_t_m_munu}
  \delta(T^\mu_{\text{m}\, \nu; \mu}) &= \delta(\Gamma\, T^\mu_{\text{m}\, \nu} u_{\text{m}\,\mu}), \\
  \label{eq:d_t_r_munu}
  \delta(T^\mu_{\text{r}\, \nu; \mu}) &= -\delta(\Gamma\, T^\mu_{\text{m}\, \nu} u_{\text{m}\,\mu}).
\end{align}
After some calculation, we obtain the following equations of motion for matter and radiation perturbations~\cite{Poulin:2016nat}:
\begin{align}
\label{eq:delta_m_evo}
&\delta_\mm' + \theta_\mm -3\Psi'= -a \delta \Gamma - a\bar\Gamma \Phi, \\
\label{eq:theta_m_evo}
&\theta_\mm' + \mathcal H \theta_\mm + \Delta \Phi = 0, \\ 
\label{eq:delta_r_evo}  
&\delta_\rr' + \frac{4}{3} \theta_\rr -4 \Psi' =  a \frac{\bar \rho_\mm}{\bar \rho_\rr} \bar \Gamma \left( \frac{\delta \Gamma}{\bar \Gamma} + \delta_\mm - \delta_\rr + \Phi \right), \\ 
\label{eq:theta_r_evo}  
& \theta_\rr'  + \frac{1}{4} \Delta \delta_\rr + \Delta \Phi = a \bar \Gamma \frac{3\bar \rho_\mm}{4\bar \rho_\rr} \left( \theta_\mm - \frac{4}{3} \theta_\rr \right),
 \end{align}
where $\delta_\n \equiv \delta \rho_\n/\bar \rho_\n$ and $\theta_\n \equiv T^{0 \ \ ,i}_{\n \ i}/(\bar \rho_\n + \bar P_\n) = \Delta u_\n/a$ are measures for the density perturbation and the velocity divergence, respectively, with the component $\n \in \{\mm, \rr\}$.
We have also taken into account the perturbation of decay rate, $\Gamma = \bar \Gamma + \delta \Gamma$, for generality.
The perturbation of the decay rate can appear, \textit{e.g.}, when it depends on the scalar field value that fluctuates in space. 
See Appendix~\ref{app:perturbation_during_trans} for the detailed derivation of these equations in a general gauge.
In addition to these equations, setting $\Phi = \Psi$ (derived from Eq.~(\ref{eq:pertb_eq_ij_new2})), we can obtain the following equation of motion for $\Phi$ from Eq.~(\ref{eq:pertb_eq_00_new2}):
\begin{align}
  \Phi' = - \frac{3 \mathcal{H}^2 \Phi -\Delta \Phi + \frac{3}{2} \mathcal{H}^2\left( \frac{\bar \rho_\text{m}}{\bar \rho_\text{tot}} \delta_\text{m} + \frac{\bar \rho_\text{r}}{\bar \rho_\text{tot}} \delta_\text{r} \right)}{3\mathcal{H}},
  \label{eq:phidot_eq}
\end{align}
where $\bar \rho_\text{tot} = \bar \rho_\text{m} + \bar \rho_\text{r}$.
The equation of motion for $\delta \Gamma$ depends on the model and, in the case of the perturbative decay rate with the constant coupling constant, $\Gamma$ is constant and $\delta \Gamma = 0$.
The evolution of the perturbations during the transition from a MD era to a RD era depends on the details of the decay process of the matter.
We will concretely discuss the evolution of the perturbations during several types of transitions from a MD era to a RD era in Secs.~\ref{sec:gradual} and \ref{sec:sudden}.

While the MD $\to$ RD transition is directly related to the poltergeist mechanism, we may have to take into account the RD $\to$ MD transition because realistic situations predict the RD $\to$ MD $\to$ RD transition.
Motivated by this, we here summarize the expression of the gravitational potential that experiences the RD $\to$ MD transition on subhorizon.
For the RD $\to$ MD transition, the decay of the matter particles is irrelevant to the transition from a RD era to a MD era. 
The evolution of the perturbations through the transition from a RD era to a MD era has been studied in \rref{Bardeen:1985tr} with the background evolution summarized in Sec.~\ref{sssec:back_rd_to_md}.
The perturbation evolution Eqs.~(\ref{eq:delta_m_evo})--(\ref{eq:theta_r_evo}) can be used  with $\Gamma = 0$.
The amplitude of the gravitational potential after the transition can be numerically fitted with~\cite{Bardeen:1985tr,Dodelson:1282338}:
\begin{align}
  &\Phi_\text{plateau} (x_\eq) \equiv \Phi (x)|_{\eta_\eq \ll \eta} \nonumber \\
  &\simeq \frac{\text{ln}[1+0.146\, x_\eq]}{\left(0.146\, x_\eq \right)}\left[ 1 + 0.242\, x_\eq + \left(1.01\, x_\eq \right)^2 \right. \nonumber  \\ 
  & \left. \qquad 
+ \left(0.341\, x_\eq \right)^3 + \left(0.418\, x_\eq \right)^4 \right]^{-0.25},    \label{eq:Phi_plateau}
\end{align}
where $\eta_\eq$ is the time when $\rho_\rr = \rho_\mm$ during the transition, $x_\eq \equiv k \eta_\eq$, and $\Phi_\text{plateau}$ is normalized as $\Phi_\text{plateau}(x_\eq \rightarrow 0) \rightarrow 1$.
We will use this expression in the concrete examples discussed in Sec.~\ref{sec:sudden}.

%%%%%%%%%%%%%%%%%%%%%%%%%%%%%%%%
\section{A primer on scalar-induced gravitational waves}\label{sec:SIGW}
%%%%%%%%%%%%%%%%%%%%%%%%%%%%%%%%

In this section, we summarize the key equations for the calculation of SIGWs.
Building upon the background evolution and first-order perturbations summarized in the previous section, we now focus on the second-order tensor perturbations that arise from mode couplings of scalar fluctuations.
We outline how these tensor modes are sourced by products of scalar perturbations through the Einstein equations and describe the standard procedure to obtain the present-day GW spectrum, explored by current GW observations.

\subsection{Equation of motion for the second-order tensor perturbations}

The tensor perturbations can be expressed with the polarization tensors $\ee^\lambda_{ij}(\hat k)$ and the Fourier modes $h^\lambda_{\bfk}$ as 
%%%%%%
\begin{align}
  \label{eq:tensor_fourier}
  h_{ij}(\bfx) = \sum_{\lambda = +, \times} \int \frac{\dd^3 k}{(2\pi)^3} \ee^\lambda_{ij}(\hat k) h^\lambda_{\bfk} e^{i \bfk\cdot\bfx},
\end{align}
%%%%%%
where $\hat k \equiv \bfk/|\bfk|$ is the unit vector in the direction of $\bfk$. 
The polarization tensors are defined as 
%%%%%%
\begin{align}
    \label{eq:pol_plus}
  \ee^+_{ij}(\hat k) &\equiv \frac{1}{\sqrt{2}} \left[ \ee_i(\hat k) \ee_j(\hat k) -  \bar \ee_i(\hat k) \bar \ee_j(\hat k) \right], \\
    \label{eq:pol_cross}  
  \ee^\times_{ij}(\hat k) &\equiv \frac{1}{\sqrt{2}} \left[ \ee_i(\hat k) \bar \ee_j(\hat k) +  \bar \ee_i(\hat k) \ee_j(\hat k) \right],
\end{align}
%%%%%%
where ${\mathbf e}$ and $\bar{\mathbf e}$ are the unit orthogonal vectors satisfying ${\mathbf e} \times \bar {\mathbf e} = \hat k$.
They satisfy the following equations:
%%%%%%
\begin{align}
k^i \ee^\lambda_{ij}(\hat k) = 0, \ \ee^{\lambda \, i}_{\ \ \ i}(\hat k)=0, \ \ee^{\lambda \, ij}(\hat k) \ee^{\lambda'*}_{ij}(\hat k) = \delta^{\lambda \lambda'}.
\end{align}
%%%%%%
That is, the polarization tensors are transverse, traceless, and orthonormal. 
Note that the spatial indexes are raised or lowered by $\delta_{ij}$ throughout this review.
If we set $\hat k$ to be $z$-axis and $\ee_i$ and $\bar \ee_i$ to be the unit vectors along $x$-axis and $y$-axis respectively, we can explicitly express the polarization tensors as\footnote{
If we compare $\ee_{ij}(\hat k)$ and $\ee_{ij}(-\hat k)$ with $\ee_i$ fixed as $\ee_i(\hat k) = \ee_i(-\hat k)$, we find $\bar \ee_i(\hat k) = -\bar \ee_i(-\hat k)$ and therefore $\ee^+_{ij}(\hat k) = \ee^+_{ij}(-\hat k)$ and $\ee^\times_{ij}(\hat k) = -\ee^\times_{ij}(-\hat k)$.
} 
%%%%%%
\begin{align}
  \ee_{ij}^+ = \frac{1}{\sqrt{2}} \begin{pmatrix}
  1 & 0 & 0 \\
  0 & -1 & 0 \\
  0 & 0 & 0
  \end{pmatrix}, \quad
  \ee_{ij}^\times = \frac{1}{\sqrt{2}} \begin{pmatrix}
  0 & 1 & 0 \\
  1 & 0 & 0 \\
  0 & 0 & 0
  \end{pmatrix}.
\end{align}
%%%%%%

We here focus on the transverse-traceless (TT) part of the (perturbed) Einstein equation: 
%%%%%%
\begin{align}
\hat {\mathcal T}^\lambda_{ij}\,^{lm} \delta G_{lm} = \frac{1}{M_\Pl^2} \hat {\mathcal T}^\lambda_{ij}\,^{lm} \delta T_{lm},
\label{eq:einstein_eq_2nd}
\end{align}
%%%%%%
where $\hat {\mathcal T}^\lambda_{ij}\,^{lm}$ is the projection operator that extracts the TT part. 
Since the polarization tensors are defined in Fourier space as in Eqs.~(\ref{eq:pol_plus}) and (\ref{eq:pol_cross}), it is natural to express the projection operator also in Fourier space as
%%%%%%
\begin{align}
\hat {\mathcal T}^\lambda_{ij}\,^{lm} A_{lm}(\bfx) &= \int \frac{\dd^3 k}{(2\pi)^3} e^{i\bfk \cdot \bfx} \ee_{ij}^{\lambda}(\hat k) \ee^{\lambda \, lm}(\hat k) A_{lm}(\bfk),
\end{align}
%%%%%%
where $A_{lm}$ is an arbitrary tensor and $A_{lm}(\bfk) = \int \dd^3 x A_{lm}(\bfx) e^{-i\bfk\cdot \bfx}$.
This expression in Fourier space should be regarded as the definition of $\hat {\mathcal T}_{ij}^{\ \ lm}$. 
Since we are interested in the power spectrum of the Fourier modes, the expression of $\hat{\mathcal T}_{ij}^{\ \ lm}$ in Fourier space is sufficient for our purpose.

We hereafter impose $\Phi = \Psi$ by assuming that the anisotropic stress is negligible at least at first order in perturbation. 
Then, we can obtain the TT part of the Einstein tensor up to the second order as~\footnote{
See \rref{Yuan:2019udt, Zhou:2021vcw,Chang:2022nzu, Wang:2023sij,Zhou:2024ncc} for the calculation of SIGWs with higher-order perturbations.
} 
%%%%%%
\begin{align}
\hat {\mathcal T}^\lambda_{ij}\,^{lm} \delta G_{lm} =&  \frac{1}{4} \left( {h^\lambda_{ij}}''+ 2\mathcal H {h^\lambda_{ij}}' - 2\mathcal H^2 h^\lambda_{ij} - 4 \mathcal H' h^\lambda_{ij} -\Delta h^\lambda_{ij} \right) \nonumber \\
&+ \hat {\mathcal T}^\lambda_{ij}\,^{lm} \left( 2\Phi_{,l} \Phi_{,m} + 4 \Phi \Phi_{,lm} \right).
\label{eq:einstein_2nd_order}
\end{align}
%%%%%%
The TT part of the energy-momentum tensor up to the second order is given by 
%%%%%%
\begin{align}
\hat {\mathcal T}^\lambda_{ij}\,^{lm} \delta T_{lm} = (\bar \rho + \bar P ) \hat {\mathcal T}^\lambda_{ij}\,^{lm} \delta u_{,l} \delta u_{,m} + \frac{a^2\bar P}{2} h^\lambda_{lm},
\label{eq:emt_2nd_order}
\end{align}
%%%%%%
where we have used the fact that $h_{ij}$ is the second-order perturbations and neglected the contribution from $\delta P h_{ij}$.
Remember that we have neglected the first-order tensor perturbations because they are not enhanced through the transition. 
For clarity, we note that we do take into account the contribution of $\delta P$ to obtain the linear equation of motion of the scalar perturbations, Eqs.~(\ref{eq:delta_m_evo})-(\ref{eq:theta_r_evo}), whose derivation can be seen in Appendix~\ref{app:perturbation_during_trans}.

From Eq.~(\ref{eq:pertb_eq_0i_new2}), we can express the velocity perturbation as 
%%%%%%
\begin{align}
\delta u_{,i} = -\frac{2M_\Pl^2}{a( \bar \rho + \bar P ) } \left( \mathcal H \Phi + {\Phi}' \right)_{,i}.
\label{eq:delta_u_s4}
\end{align}
%%%%%%

Using this and Eq.~(\ref{eq:hubble_eq2}), we obtain the equation of motion for SIGWs \cite{Mollerach:2003nq,Ananda:2006af,Baumann:2007zm}:
%%%%%%
\begin{align}
\label{eq:h_ij_eom}
{h^{\lambda}_{ij}}'' + 2\mathcal H {h^{\lambda }_{ij}}' - \Delta h^{\lambda }_{ij} = -4 \hat {\mathcal T}^\lambda_{ij}\,^{lm} \mathcal S_{lm},
\end{align}
%%%%%%
where the source term $\mathcal S_{ij}$ is given by 
%%%%%%
\begin{align}
\label{eq:mathcal_s_def_real}
\mathcal S_{ij} \equiv&\  2 \Phi_{,i} \Phi_{,j} + 4\Phi \Phi_{,ij} \nonumber \\
&- \frac{4}{3(1+w) \mathcal H^2} ( \mathcal H \Phi + {\Phi}\,' )_{,i} (\mathcal H \Phi +  {\Phi} \, ' )_{,j},
\end{align}
and $w = \bar{P} / \bar{\rho}$ is the equation-of-state parameter.  
%%%%%%
In the Fourier space, Eq.~(\ref{eq:h_ij_eom}) can be written as
%%%%%%
\begin{align}
{h^\lambda_{\bfk}}'' + 2\mathcal H {h^\lambda_{\bfk}}' + k^2 h^\lambda_{\bfk} = 4\mathcal S^\lambda_\bfk.
\label{eq:tensor_eom_w_source}
\end{align}
%%%%%%
The source function $\mathcal S^\lambda_\bfk(\eta)$ is defined as
%%%%%%
\begin{align}
\label{eq:mathcal_s_def}
\mathcal S^\lambda_\bfk \equiv& -\ee^{\lambda \, ij}(\hat k) \mathcal S_{\bfk, ij} \nonumber \\
=& \int \frac{\dd^3 q}{(2\pi)^3} \ee^{\lambda\, ij}(\hat k) q_i q_j \Biggr[ 2\Phi_{\bfq} \Phi_{ \bfk - \bfq} \Biggr. \nonumber \\
&\Biggr. + \frac{4}{3(1+w)}\left( \Phi_{\bfq} + \frac{{\Phi'_{\bfq}}}{\mathcal H}\right)\left( \Phi_{\bfk - \bfq} + \frac{{\Phi'_{\bfk-\bfq}}}{\mathcal H}\right) \Biggr],
\end{align}
%%%%%%
where we have used the relation $\ee^{\lambda \, ij}(\hat k)k_{j} = 0$.

Before solving this equation, we note that Eqs.~(\ref{eq:mathcal_s_def_real}) and (\ref{eq:mathcal_s_def}) are based on the one-fluid assumption.
In two fluids of matter and radiation, an additional term arises from the velocity difference between the matter and radiation~\cite{Domenech:2020ssp,Gurian:2021rfv,Kumar:2024hsi}.
However, this velocity difference contribution is negligible before and after the transition because the Universe can be approximated as one fluid at that time~\cite{Domenech:2020ssp}.
Since the main focus of this review is on the poltergeist GW production, which occurs after the transition, we neglect the velocity difference contribution throughout this review. 
On the other hand, for the precise spectrum of the SIGWs produced in the middle of the transition (not before or after the transition), the velocity difference contribution must be taken into account.
See \rref{Kumar:2024hsi} for the discussion on the velocity difference contributions for such SIGWs.
For clarity, we note that we take into account the velocity difference \emph{for the linear perturbations}. Neglecting the velocity difference \emph{for the source term} at the second order does not change the evolution of the linear perturbations, which is used for the calculation of the poltergeist GW production occurring after the transition.

\subsection{Solving the second-order tensor perturbations}

Let us solve Eq.~(\ref{eq:tensor_eom_w_source}) with the Green function method. 
We can rewrite Eq.~(\ref{eq:tensor_eom_w_source}) as 
%%%%%%
\begin{align}
(a h_{\bfk}^\lambda)'' + \left( k^2 - \frac{a''}{a} \right) (a h_{\bfk}^\lambda) = 4a\mathcal S^\lambda(\bfk, \eta).
\end{align}
%%%%%%
The formal solution of this is given by
%%%%%%
\begin{align}
h_{\bfk}^\lambda (\eta) = \frac{4}{a} \int^\eta \dd \bar \eta G_{k}(\eta; \bar \eta) a(\bar \eta) \mathcal S^\lambda(\bfk, \bar \eta),
\label{eq:tensor_mode_time_dep}
\end{align}
%%%%%%
where $G_k$ is the causal (retarded) Green function that follows
%%%%%%
\begin{align}
G_{k}''(\eta; \bar\eta) + \left(k^2 - \frac{a''}{a} \right) G_{k}(\eta; \bar \eta) = \delta(\eta- \bar \eta).
\label{eq:green_func_eq}
\end{align}
%%%%%%
Note that the prime means the derivative with respect to $\eta$, not $\bar \eta$.
The explicit expression for the Green function depends on the cosmic era.
During a RD era, the scale factor is given by $a \propto \eta$ and the causal Green function becomes
%%%%%%
\begin{align}
  \label{eq:green_rd}
 k G^\text{RD}_{k} (\eta;\bar \eta) = \Theta(\eta- \bar \eta) \sin (x- \bar x),
\end{align}
%%%%%%
where $x=k\eta$ and $\bar x = k \bar \eta$ and $\Theta(\eta)$ is the Heaviside step function:
%%%%%%
\begin{align}
  \Theta(\eta) = \begin{cases}
  1 & (\eta \geq 0) \\
  0 & (\eta< 0)
  \end{cases}.
\end{align}
%%%%%%
During a MD era, the Green function is given by
%%%%%%
\begin{align}
  \label{eq:green_md}
 k G^\text{MD}_{k} (\eta; \bar \eta) = -\Theta(\eta- \bar \eta) x \bar x \left( j_1(x) y_1(\bar x) - j_1(\bar x) y_1(x) \right),
\end{align}
%%%%%%
where we have used the relation $a \propto \eta^2$, which is valid during a MD era.

For later convenience, we introduce the Green function for the tensor perturbations that are induced during an eMD era and experience an instantaneous transition to a RD era, which we denote by $G_k^{\text{eMD} \rightarrow \text{RD}}$.
In this case, the scale factor is given by (Eq.~(\ref{eq:scale_sudden})):
%%%%%%
\begin{align}
   \frac{a(\eta)}{a(\eta_\text{R})} = \begin{cases}
 \left( \cfrac{\eta}{\eta_\R} \right)^2 & (\eta < \eta_\R) \\
 2 \cfrac{\eta}{\eta_\R} -1 & (\eta \geq \eta_\R)
 \end{cases}, \tag{\ref{eq:scale_sudden}}
\end{align}
%%%%%%
where $\eta_\R$ is the conformal time at the transition.
In this case, the Green function satisfies
\begin{align}
  {G_k^{\text{eMD} \rightarrow \text{RD}}}''(\eta;\bar \eta) + k^2 G_k^{\text{eMD} \rightarrow \text{RD}}(\eta; \bar \eta) = 0 \ \ (\eta > \eta_\R).
  \label{eq:g_md_rd_free}
\end{align}
Note that $G_k^{\text{eMD} \rightarrow \text{RD}}$ is for the modes that are induced by the source in $\bar \eta < \eta_\R$ and evolve freely after that.
To match the Green function during the MD era, given in Eq.~(\ref{eq:green_md}), to that in $\eta > \eta_\R$, we impose the continuity of $G_k^{\text{eMD} \rightarrow \text{RD}}$ and ${G_k^{\text{eMD} \rightarrow \text{RD}}}'$ at $\eta_\R$.
We can confirm the continuity of ${G_k^{\text{eMD} \rightarrow \text{RD}}}'$ at the transition as follows. 
The source-free equation of motion for the Green function (Eq.~(\ref{eq:green_func_eq})) can be written as
%%%%%%
\begin{align}
  \left(a^2 \left(\frac{G_k}{a} \right)'\right)' = -a^2 k^2 \left( \frac{G_k}{a} \right).
\end{align}
%%%%%%
If we integrate both sides between $\eta_\R-\epsilon$ and $\eta_\R + \epsilon$ and take $\epsilon\rightarrow 0$, the right-hand side becomes zero because $G_k$ and $a$ are continuous at $\eta_\R$.
Then, we can see that $a^2 (G_k/a)' = aG_k' - a'G_k$ is continuous at $\eta_\R$.
Since $\mathcal H$ (that is, $a'$) is continuous at $\eta_\R$, $G_k'$ is also continuous at $\eta_\R$.
Imposing that $G_k$ and $G'_k$ are continuous at $\eta_\R$, we can obtain the Green function that experiences the transition from a MD era to a RD era as 
%%%%%%
\begin{align}
\label{eq:green_func_md_rd}
kG_k^{\text{eMD} \rightarrow \text{RD}}(\eta; \bar \eta) &= k G_k^\text{MD}(\eta_\R;\bar \eta) \cos(x-x_\R) \nonumber \\
&\quad + {G_k^\text{MD}}'(\eta_\R; \bar \eta) \sin(x-x_\R) \nonumber \\
& = C(x,x_\R) \bar x j_1(\bar x) + D(x,x_\R) \bar x y_1(\bar x),
\end{align}
%%%%%%
where $\bar \eta < \eta_\R$ and $C(x,x_\R)$ and $D(x,x_\R)$ are defined as 
%%%%%%
\begin{align}
  \label{eq:green_func_md_c}
  C(x,x_\R) &= \frac{\sin x - 2x_\R(\cos x + x_\R \sin x) + \sin(x-2x_\R)}{2x_\R^2},  \\
  \label{eq:green_func_md_d}
  D(x,x_\R) &= \frac{(2x_\R^2 -1)\cos x - 2x_\R \sin x + \cos(x-2x_\R)}{2x_\R^2}.
\end{align}
%%%%%%
Eq.~(\ref{eq:green_func_md_rd}) and its time derivative are continuous at $\eta_\R$ and satisfy Eq.~(\ref{eq:g_md_rd_free}).
Note that $G_k^{\text{eMD} \rightarrow \text{RD}}$ is defined only in $\eta > \eta_\R$.

For clarity, we here show the concrete expression of the Green function:
\begin{align}
k G_k(\eta;\bar \eta) = \begin{cases}
kG_k^{\text{MD}}(\eta; \bar \eta) &\text{for } \eta \leq \eta_\R \\ 
kG_k^{\text{eMD} \rightarrow \text{RD}}(\eta; \bar \eta) &\text{for } \eta > \eta_\R \text{ and } \bar \eta \leq \eta_\R  \\
kG_k^{\text{RD}}(\eta; \bar \eta) &\text{for } \eta > \eta_\R \text{ and } \bar \eta > \eta_\R 
\end{cases}.
\end{align}
We can see that $G_k$ and $G_k'$ are continuous not only at $\eta = \eta_\R$, but also at $\bar \eta = \eta_\R$ by using the relations $\lim_{\epsilon \to 0}{kG_k^{\text{MD}}}(\eta_\R;\eta_\R-\epsilon) = 0$ and $\lim_{\epsilon \to 0}{G_k^{\text{MD}}}'(\eta_\R;\eta_\R-\epsilon) = 1$.

%%%%%%%%%%%%%%%%%
\subsection{Energy density of induced GWs}
%%%%%%%%%%%%%%%%%

In this section, we summarize the expression of the energy density of GWs, which is often used to characterize their amplitude.
Under the assumption that the GW wavenumbers (frequencies) are much larger than the Hubble parameter ($k \gg \mathcal H$), the energy density of GWs is given by~\cite{Maggiore:1999vm}:
%%%%%%
\begin{align}
  \label{eq:gw_energy_dens}
  \rho_\text{GW}(\eta) = \frac{M_\Pl^2}{16a^2} \vev{ \overline {h_{ij,l} h^{ij,l}}},
\end{align}
%%%%%%
where the bracket denotes the expectation value and the overline indicates the time average over the oscillation period ($\simeq 1/k$).
We can express this with the power spectrum of tensor perturbations: 
%%%%%%
\begin{align}
\rho_{\GW} (\eta) &= \int \frac{\dd k}{k}  \left(\frac{M_\Pl^2}{8} \left( \frac{k}{a} \right)^2 \overline{ \mathcal P_h (k, \eta)} \right),
\end{align}
%%%%%%
where the dimensionless power spectrum is defined as
%%%%%%
\begin{align}
\vev{h^\lambda_{\bfk}(\eta) h^{\lambda' *}_{\bfk'}(\eta)}  = (2\pi)^3 \delta(\bfk- \bfk') \delta^{\lambda \lambda'} \frac{2\pi^2}{k^3} \mathcal P_{h}(k,\eta).
\label{eq:tensor_ensemble}
\end{align}
For later convenience, we here define the energy density of GWs per logarithmic interval of $k$ as 
%%%%%%
\begin{align}
\rho_\GW(\eta,k) &\equiv \frac{M_\Pl^2}{8} \left( \frac{k}{a} \right)^2 \overline{ \mathcal P_h (k, \eta)}.
\label{eq:rho_gw}
\end{align}
%%%%%%
Using this parameter, we also define the energy density parameter of GWs per logarithmic interval of $k$ as 
%%%%%%
\begin{align}
\Omega_\GW (\eta, k) \equiv \frac{\rho_\GW(\eta, k)}{\rho_{\text{tot}}} = \frac{1}{24} \left( \frac{k}{\mathcal H} \right)^2 \overline{\mathcal P_h (k, \eta)}.
\label{eq:gw_cosmological_para}
\end{align}
%%%%%%

%%%%%%%%%%%%%%%%%
\subsection{Formulas for GW spectrum}
%%%%%%%%%%%%%%%%%

In this subsection, we derive the formulas for $\mathcal P_h$ and $\Omega_\text{GW}$, defined in Eqs.~(\ref{eq:tensor_ensemble}) and (\ref{eq:gw_cosmological_para}).
We first define the transfer function $\Phi(k \eta)$ of the gravitational potential $\Phi_\bfk (\eta)$ as
%%%%%%
\begin{align}
\Phi_{\bfk}(\eta)&= \phi_{\bfk} \Phi(k\eta),
\end{align}
%%%%%%
where we normalize the transfer function as $\Phi(x\rightarrow 0) \rightarrow 1$ and $\phi_{\bfk}$ is the amplitude of the gravitational potential before it enters the horizon.

Substituting Eq.~(\ref{eq:tensor_mode_time_dep}) into the left-hand side of Eq.~(\ref{eq:tensor_ensemble}), we can reexpress the two-point function as 
%%%%%%
\begin{align}
\label{eq:tensor_ex_value}
&\vev{h^\lambda_{\bfk}(\eta)  h^{\lambda' *}_{\bfk'}(\eta)} = 
\frac{16}{a^2} \int^\eta \dd \eta_1 \int^\eta \dd \eta_2 \, a(\eta_1) a(\eta_2) \nonumber \\
&\quad
\times
G_{k}(\eta; \eta_1) G_k(\eta; \eta_2)\vev{\mathcal S^\lambda(\bfk, \eta_1) \mathcal S^{\lambda' *}(\bfk', \eta_2)}.
\end{align}
%%%%%%
The expectation value of the source terms can be written as
%%%%%%
\begin{align}
&\vev{\mathcal S^{\lambda}_{\bfk}(\eta_1) \mathcal S^{\lambda'*}_{\bfk'} (\eta_2)}  = \int \frac{\dd^3 \tilde k \dd^3 \tilde k'}{(2\pi)^6} \e^{\lambda \, ij}(\hat k) \tilde k_i \tilde k_j \e^{\lambda' \, lm} (\hat k') \tilde k_l' \tilde k_m' \nonumber \\
&\times f(\tilde {k}, |\bfk- \tilde{\bfk}|, \eta_1) f(\tilde{k}', |{\bfk}' - \tilde{\bfk}'|, \eta_2) \vev{\phi_{\tilde{\bfk}}  \phi_{ \bfk- \tilde{\bfk}}  \phi^*_{\tilde{\bfk}'} \phi^*_{\bfk'- \tilde{\bfk}'}},
\label{eq:ss_corr}
\end{align}
%%%%%%
where $f(k_1, k_2, \eta)$ is the source function, including the transfer functions as 
%%%%%%
\begin{align}
\label{eq:func_f_def}
&f(k_1, k_2, \eta) \equiv 2 \Phi(k_1 \eta) \Phi(k_2 \eta) \nonumber \\
&+ \frac{4}{3(1+w)} \left( \Phi(k_1 \eta) + \frac{\Phi'(k_1 \eta)}{\mathcal H} \right) \left(\Phi(k_2 \eta) + \frac{\Phi'(k_2 \eta)}{\mathcal H} \right).
\end{align}
%%%%%%
Throughout this review, we assume that $\phi_{\bfk}$ follows the Gaussian distribution for simplicity.
The power spectrum of the gravitational potential is given as 
%%%%%%
\begin{align}
  \vev{ \phi_{\bfk}  \phi^*_{\bfk'}} = (2\pi)^3 \delta(\bfk- \bfk') \frac{2\pi^2}{k^3} \mathcal P_{\Phi}(k).
\end{align}
%%%%%%

For the effects of the non-Gaussianity on the induced GWs, see \rref{Nakama:2016gzw, Garcia-Bellido:2017aan, Ando:2017veq, Cai:2018dig, Unal:2018yaa, Yuan:2020iwf, Atal:2021jyo, Adshead:2021hnm,  Garcia-Saenz:2022tzu, Abe:2022xur, Li:2023qua, Yuan:2023ofl, Perna:2024ehx, Inui:2024fgk, Iovino:2024sgs, Zeng:2025cer}.
Generically, the effects of non-Gaussianity are important because observably intense induced GWs require a large enhancement of the primordial curvature perturbations, which is typically associated with non-Gaussianity. 
In the context of the poltergeist mechanism, however, the GWs can be so significantly enhanced that the curvature perturbations do not need to be enhanced. This makes the role of non-Gaussianity less relevant.

To further calculate Eq.~\eqref{eq:ss_corr}, we express the wave vector with the spherical coordinates as 
\begin{align}
\tilde {\bfk} = (\tilde k \sin \theta \cos \varphi, \tilde k \sin \theta \sin \varphi, \tilde k\cos\theta), 
\end{align}
where $\bfk$ is taken along $z$-axis without loss of generality. 
This leads to the following expressions:
%%%%%%
\begin{align}
\e^{+\,ij}(\hat k) \tilde k_i \tilde k_j &= \frac{\tilde k^2}{\sqrt{2}} \sin^2 \theta \cos 2\varphi,\\
\e^{\times\,ij}(\hat k) \tilde k_i \tilde k_j &= \frac{\tilde k^2}{\sqrt{2}} \sin^2 \theta \sin 2\varphi.
\label{eq:polarization_kk}
\end{align}
%%%%%%
Substituting these into Eq.~(\ref{eq:ss_corr}) and integrating over the angle $\varphi$, we obtain
%%%%%%
\begin{align}
\label{eq:gw_source_ss_rr}
&\vev{\mathcal S^{\lambda}_{\bfk}(\eta_1) \mathcal S^{\lambda'*}_{\bfk'} (\eta_2)} \nonumber\\
&= (2\pi)^3 \delta(\bfk- \bfk') \delta^{\lambda \lambda'} \frac{2\pi^2}{k^3} \frac{1}{4} \int^\infty_0 \dd \tilde k \int^1_{-1} \dd\mu \frac{k^3 \tilde k^3}{|\bfk- \tilde {\bfk}|^3} (1-\mu^2)^2 \nonumber \\
&\quad \times  f(|\bfk- \tilde {\bfk}|,\tilde k, \eta_1)  f(|\bfk- \tilde {\bfk}|, \tilde k,\eta_2) \mathcal P_{\Phi}(\tilde k) \mathcal P_\Phi ( |\bfk- \tilde {\bfk}|),
\end{align}
%%%%%%
where $\mu$ is the cosine function with the angle between $\bfk$ and $\tilde {\bfk}$, defined as $\mu \equiv \bfk \cdot \tilde{\bfk}/(k \tilde k)$.
We have also used the relation $f(k_1,k_2,\eta) = f(k_2,k_1,\eta)$.
Changing the variables from $\tilde k$ and $\mu$ to $u \equiv |\bfk - \tilde {\bfk}|/k$ and $v \equiv \tilde k/k$ and using Eq.~(\ref{eq:tensor_ex_value}), we obtain the power spectrum of SIGWs as 
%%%%%%
\begin{align}
\label{eq:ph_express_v_u}
&\mathcal P_h(\eta, k) = 4  \left( \frac{3(1+w_k)}{5+3w_k} \right)^4 
\int^\infty_0 \dd v \int^{|1+v|}_{|1-v|} \dd u \nonumber \\
& \times \left[ \frac{4v^2 - (1 + v^2 - u^2 )^2}{4uv} \right]^2 I^2(u, v, x) \mathcal P_{\zeta}(u k) \mathcal P_{\zeta} ( v k ),
\end{align}
%%%%%%
where we have used $\zeta_\bfk = -(5+3w)/(3(1+w))\phi_\bfk$ on superhorizon scales~\cite{Mukhanov:1990me} and $\mu = (1+v^2 - u^2)/2v$, and note that $w_\text{k}$ represents the equation-of-state parameter when the tensor perturbation whose wavenumber is $\bfk$ enters the horizon. 
$I(u,v,x)$ is defined as
%%%%%%
\begin{align}
  \label{eq:i_vux_def}
  I(u,v,x) \equiv \int^x \dd \bar x \frac{a(\bar \eta)}{a(\eta)} k G_k(\eta; \bar \eta) f(u,v,\bar x).
\end{align}
%%%%%%
Note also that we have rewritten the arguments of $f$ as $f(u,v,\bar x) \equiv f(uk,vk, \bar \eta)$, which can be expressed as
\begin{align}
  \label{eq:f_uvx}
  f(u,v,\bar x) = &\frac{1}{3(1+w)} \left( 2(5+3w) \Phi(u \bar x)\Phi(v \bar x)\right. \nonumber \\
  & \  \left. +4  \mathcal H^{-1} (\Phi'(u \bar x)\Phi(v \bar x)  +  \Phi(u \bar x)\Phi'(v \bar x)) \right. \nonumber\\
  &\ \left.
   + 4 \mathcal H^{-2}  \Phi'(u \bar x)\Phi'(v \bar x) \right),
\end{align}
where $\mathcal H$ means $\mathcal H(\bar \eta)$ and the prime represents the derivative with respect to $\bar \eta$ (not $\bar x$).

It is a good time to review the physical meaning of these equations. The power spectrum of SIGWs, Eq.~\eqref{eq:ph_express_v_u}, is composed of several parts. The characteristic second-order feature can be seen from the dependence on the source $\mathcal{P}_\zeta(uk) \mathcal{P}_\zeta (vk)$, which contains the information of the primordial amplitude of the scalar perturbations. The dynamics of the scalar perturbation source after the horizon entry are encoded in the function $f(u, v, \bar{x})$, which depends quadratically on the transfer function $\Phi$. These perturbations source the GWs at the time $\bar \eta$, whose propagation to $\eta$ is described by the Green function $G_k (\eta; \bar \eta)$. The redshift of the GWs is taken into account in the factor $a(\bar \eta)/a(\eta)$ in Eq.~\eqref{eq:i_vux_def}. The factor in the square bracket in Eq.~\eqref{eq:ph_express_v_u} originates from the kinematic factor like the polarization tensor contracted with wavenumber vectors. Finally, the integrals over $u$ and $v$ take into account all the possible configurations of the wavenumber vectors of the scalar perturbations that can produce the GWs with the fixed size of the wavenumber $k$ under momentum conservation.  

The time integral over $\bar x$ [see Eq.~\eqref{eq:i_vux_def}] in the integration kernel $I(u, v, x)$ can be analytically computed for limited cases such as the constant equation-of-state parameter including a RD era~\cite{Kohri:2018awv, Espinosa:2018eve}, a MD era~\cite{Assadullahi:2009nf, Kohri:2018awv}, and an era with general $w$~\cite{Domenech:2019quo}. 

Here, we mention another choice of the variables, which is useful when we discuss the approximate formula of the poltergeist GW spectrum (see Appendix~\ref{app:analytic_formulas}).
If we use the variables $t \equiv u+v-1$ and $s \equiv u-v$ instead $u$ and $v$, we can rewrite Eq.~(\ref{eq:ph_express_v_u}) as 
%%%%%%
\begin{align}
  \label{eq:p_h_s_t_express}
  &\mathcal P_h(\eta, k) = 2  \left( \frac{3(1+w_k)}{5+3w_k} \right)^4 
\int^\infty_0 \dd t \int^{1}_{-1} \dd s \nonumber \\
& \times\left[ \frac{t(2+t)(s^2 -1)}{(1-s+t)(1+s+t)} \right]^2 I^2(u, v, x) \mathcal P_{\zeta}(u k) \mathcal P_{\zeta} ( v k ),
\end{align}
%%%%%%
where $u$ and $v$ are given with $s$ and $t$ as $u = (t+s+1)/2$ and $v = (t-s+1)/2$. An advantage of this change of variables is the simplification of the integration domain. Typically, $t$-dependence is more important than $s$-dependence in the sense that the former reflects the structure of the problem. For example, $\mathcal{O}(1)$ values of $t$ dominate in the integral for a sufficiently flat $\mathcal{P}_\zeta(k)$ such as the scale-invariant one, while large $t \sim \mathcal{O}(k_\text{peak}/k) \gg 1$ (the squeezed limit configuration) dominates for sufficiently low-frequency part of the spectrum when there is a sufficiently sharp peak in $\mathcal{P}_\zeta(k)$. 

Finally, we can reexpress the energy density parameter of SIGWs (Eq.~(\ref{eq:gw_cosmological_para})) as 
%%%%%%
\begin{align}
\label{eq:gw_cosmo_para_general}
&\Omega_\text{GW}(\eta,k) = \frac{1}{6}  \left( \frac{3(1+w_k)}{5+3w_k} \right)^4  \left( \frac{k}{\mathcal H} \right)^2 \int^\infty_0 \dd v \int^{|1+v|}_{|1-v|} \dd u \nonumber \\
&
\times \left[ \frac{4v^2 - (1 + v^2 - u^2 )^2}{4uv} \right]^2 \overline{I^2(u, v, x)} \mathcal P_{\zeta}(u k) \mathcal P_{\zeta} ( v k ),
\end{align}
%%%%%%
where the overline indicates the oscillation average again. 
In terms of $t$ and $s$, it is 
\begin{align}
    \label{eq:gw_cosmo_para_general_ts}
&\Omega_\text{GW}(\eta,k) = \frac{1}{12}  \left( \frac{3(1+w_k)}{5+3w_k} \right)^4  \left( \frac{k}{\mathcal H} \right)^2 \int^\infty_0 \dd t \int^{1}_{-1} \dd s \nonumber \\
& \times\left[ \frac{t(2+t)(s^2 -1)}{(1-s+t)(1+s+t)} \right]^2 \overline{I^2(u, v, x)} \mathcal P_{\zeta}(u k) \mathcal P_{\zeta} ( v k ).
\end{align}

%%%%%%%%%%%%%%%%%
\subsection{Late-time evolution}
\label{sec:late_time_evo}
%%%%%%%%%%%%%%%%%

Throughout this review, we mainly focus on the GWs induced during the RD era following the eMD era, when the poltergeist mechanism occurs.
During a RD era, the gravitational potential decays as $\Phi \propto a^{-2}$ on subhorizon scales (see Eq.~(\ref{eq:phi_evo_rd_era})) and the production of GWs stops a few e-folds after the perturbations enter the horizon.
Once the tensor perturbations decouple from the source, the energy density of the induced GWs behaves as that of radiation and therefore $\Omega_\text{GW}$ finally becomes constant during a RD era. 
We here define $\eta_\text{c}$ as the conformal time when the energy density parameter becomes constant. 
After the RD era, the Universe is dominated by dark matter and dark energy.
Through these eras, the energy density parameter of the induced GWs changes in the same way as that of radiation does even if no production of GWs occurs.
In the following, we summarize this kind of late-time evolution of $\Omega_\GW$.

Since the energy density of the induced GWs is proportional to $a^{-4}$, we can express the current energy density as 
%%%%%%
\begin{align}
  \rho_\text{GW}(\eta_0,k) = \left(\frac{a_\cc}{a_0} \right)^4 \rho_\text{GW}(\eta_\text{c},k).
\end{align}
%%%%%%
Note that the subscripts ``c'' and ``$0$'' mean the values at $\eta_\cc$ and $\eta_0$ (the present time), respectively.
Then, we can express the current value of the energy density parameter as 
%%%%%%
\begin{align}
  \Omega_\text{GW}(\eta_0,k)h^2 &= \left( \frac{a_\cc}{a_0} \right)^4 \left( \frac{H_\cc}{ H_0} \right)^2 \Omega_\text{GW}(\eta_\cc) h^2,
\end{align}
%%%%%%
where $h \equiv H_0/(100\,\text{km/s/Mpc})$.
Here, let us summarize the relation between $a$, $H$, and $\Omega_{\rr,0}$.
The ratio between $\rho_{\rr,0}$ and $\rho_{\rr,\cc}$ can be expressed as
%%%%%%
\begin{align}
 & \frac{\rho_{\rr,0}}{\rho_{\rr,\cc}} = \frac{g_{*,0}T_0^4}{g_{*,\cc} T_\cc^4} \nonumber \\
 \label{eq:omega_r0_omega_rc}
\Rightarrow \quad &  
  \frac{\Omega_{\rr,0}}{\Omega_{\rr,\cc}} = \frac{H_\cc^2 g_{*,0}T_0^4}{H_0^2 g_{*,\cc} T_\cc^4},
\end{align}
%%%%%%
where $g_{*}$ is the effective degrees of freedom for the radiation energy density and $T$ represents the temperature of the photon.
From the entropy conservation, we can relate the temperature to the scale factor as
%%%%%%
\begin{align}
    g_{s*,0} a_0^3 T_0^3 = g_{s*,\cc} a_\cc^3 T_\cc^3.
\end{align}
%%%%%%
From this equation and Eq.~(\ref{eq:omega_r0_omega_rc}), we derive
%%%%%%
\begin{align}
  \left( \frac{a_\cc}{a_0} \right)^4 \left( \frac{H_\cc}{ H_0} \right)^2 &= \frac{\Omega_{\rr,0} \, a_\cc^4 T_\cc^4 \, g_{*,\cc}}{\Omega_{\rr,\cc}\,  a_0^4 T_0^4 \, g_{*,0}} \nonumber \\
  &  = \Omega_{\rr,0} \left(\frac{g_{s*,0}}{g_{s*,\cc}}\right)^{4/3} \frac{g_{*,\cc}}{g_{*,0}},
\end{align}
%%%%%%
where $\Omega_{\rr,\cc} = 1$ because we assume that the GW production stops during the RD era here.
Since we focus on the GWs induced in the early Universe throughout this review, we take $g_{s*,\cc} = g_{*,\cc}$, which holds within a few percent above the QCD crossover, $T \gtrsim 1 \, \mathrm{GeV}$, and much more precisely for temperature higher than several hundreds of $\mathrm{GeV}$~\cite{Saikawa:2018rcs}. 
Then, we finally obtain 
%%%%%%
\begin{align}
  \label{eq:late_time_omega_gw}
  \Omega_\text{GW}(\eta_0,k)h^2 = 0.39 \left( \frac{g_{*,\cc}}{106.75} \right)^{-1/3} \Omega_{\rr,0} h^2 \Omega_\text{GW}(\eta_\cc,k),
\end{align}
%%%%%%
where we have substituted $g_{*,0}=3.38$ and $g_{s*,0}=3.93$~\cite{Saikawa:2018rcs}. 
Assuming the standard cosmology, $\Omega_{\rr, 0} h^2 \simeq 4.2 \times 10^{-5}$, with this formula, we can relate the intensity of the induced GWs in the RD era and the observable quantity at the present time.

%%%%%%%%%%%%%%%%%%%%%%%%%%%%%%%%
\section{Scalar-induced gravitational waves with a gradual matter-to-radiation transition}
\label{sec:gradual}
%%%%%%%%%%%%%%%%%%%%%%%%%%%%%%%%

With the equations derived in the previous sections, we discuss SIGW spectrum in the presence of an eMD era.

As we will see, the SIGW spectrum is sensitive to the timescale of the transition from an eMD era to a RD era. To discuss this point, let us classify a transition into the \emph{gradual} case and the \emph{sudden} case.
We compare two timescales: the oscillation timescale of the relevant gravitational-wave mode and the characteristic timescale of the change of the equation-of-state parameter of the Universe.  We call the latter simply as the timescale of the transition.  If the timescale of the transition is longer than or comparable with the timescale of the relevant mode, we call it a gradual transition. If the transition timescale is much shorter than the oscillation timescale, on the other hand, we call it a sudden (or rapid) transition. 

In this section, we discuss a gradual eMD-to-RD transition for a pedagogical purpose, as the standard eMD-to-RD transition is gradual for any subhorizon modes at the transition. The discussion will be contrasted in the next section, where we discuss a sudden transition and the associated strong SIGWs produced by the poltergeist mechanism. 

To be concrete, let us consider a constant decay rate $\Gamma$ for the dominant matter during the eMD era in this section. We stress that the qualitative results in this section do not sensitively depend on this assumption of constant $\Gamma$ as long as the transition is gradual.  With that said, the constant decay rate can be realized by perturbative decay of the matter field, which is arguably the simplest mechanism of the eMD-to-RD transition. Therefore, the analysis in this section is a proxy for the analysis of the SIGWs produced in the presence of the standard/gradual eMD-to-RD transition.

In the following, we review the results in \rref{Inomata:2019zqy}. 
Specifically, we first review how the perturbations evolve during the gradual transition, explaining the approximate formulas used to describe this evolution and to reduce the computational cost of the GW calculation. We then discuss the calculation of GWs induced during the transition, with particular emphasis on the limitations of perturbation theory.
The figures in this section are reproduced from the reference, \rref{Inomata:2019zqy}.

\subsection{Perturbation evolution during the gradual transition}
\begin{figure}%[ht] 
        \centering \includegraphics[width=0.45 \textwidth]{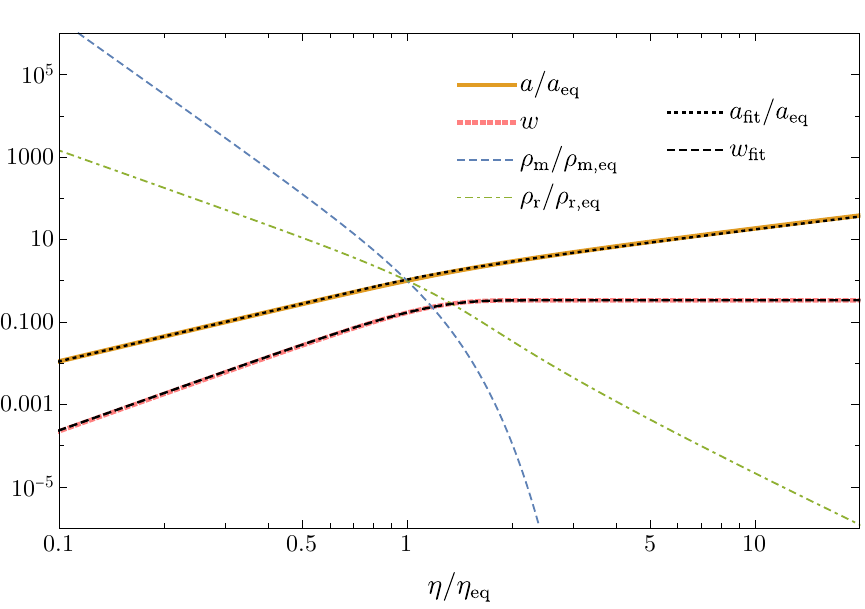}
        \caption{\justifying
        The evolution of the background quantities and the fitting functions, given by Eqs.~(\ref{eq:a_app}) and (\ref{eq:w_fit}).
        From \rref{Inomata:2019zqy}.
       }
        \label{fig:back_summary}
\end{figure}

We first numerically solve the equations of motion for background and perturbations. 
For background, we solve Eqs.~(\ref{eq:rho_m_with_decay}) and (\ref{eq:rho_r_with_decay}).
We set the initial time of the numerical calculation much before the transition from the eMD to the RD era.
Figure~\ref{fig:back_summary} shows the results. 
In this figure, we also plot the approximate formula for the scale factor:
\begin{align}
\label{eq:a_app}
\frac{a_\text{fit}(\eta)}{a(\eta_\R)} &= 
\begin{cases}
\left( \frac{\eta}{\eta_\R} \right)^2 &\ (\eta < \eta_\R), \\
2 \frac{\eta}{\eta_\R}  -1  &\ (\eta \geq \eta_\R).
\end{cases}
\end{align}
Throughout this section (for gradual transition), we take $\eta_\R=0.83 \eta_\text{eq}$ because it fits the numerical results well (see Fig.~\ref{fig:back_summary}).
We also plot the approximate formula for $w$:
\begin{align}
\label{eq:w_fit}
w_\text{fit} &= \frac{1}{3} \left( 1-\exp \left(-0.7 \left(\frac{\eta}{\eta_\text{eq}}\right)^3 \right) \right).
\end{align}
These approximate formulas fit the numerical results very well. 
In particular, Eq.~(\ref{eq:a_app}) corresponds to the situation where the background evolution changes from an eMD era to a RD era suddenly at $\eta_\R$.
This means that the sudden transition approximation works well for the evolution of the scale factor even though we consider a gradual transition.  
This fact significantly simplifies the calculation of GWs, as we will see below. 

%%%%%%%%%%%%%%%%%%
%%%%%%%%%%%%%%%%%%

On the other hand, the sudden transition approximation is not good for the perturbation evolution in the gradual transition.
For perturbations, we solve Eqs.~(\ref{eq:delta_m_evo})-(\ref{eq:phidot_eq}) with the following initial conditions~\cite{Mukhanov:991646}:
\begin{align}
&\delta_{\text{m},\text{ini}} = -2\Phi_\text{ini}, \ \delta_{\text{r},\text{ini}} = \frac{4}{3} \delta_{\text{m},\text{ini}}, \ \theta_{\text{m},\text{ini}} = \theta_{\text{r},\text{ini}} = \frac{k^2 \eta}{3} \Phi_\text{ini}.
\label{eq:ini_condition}
\end{align}

\begin{figure}%[ht] 
        \centering \includegraphics[width=0.45 \textwidth]{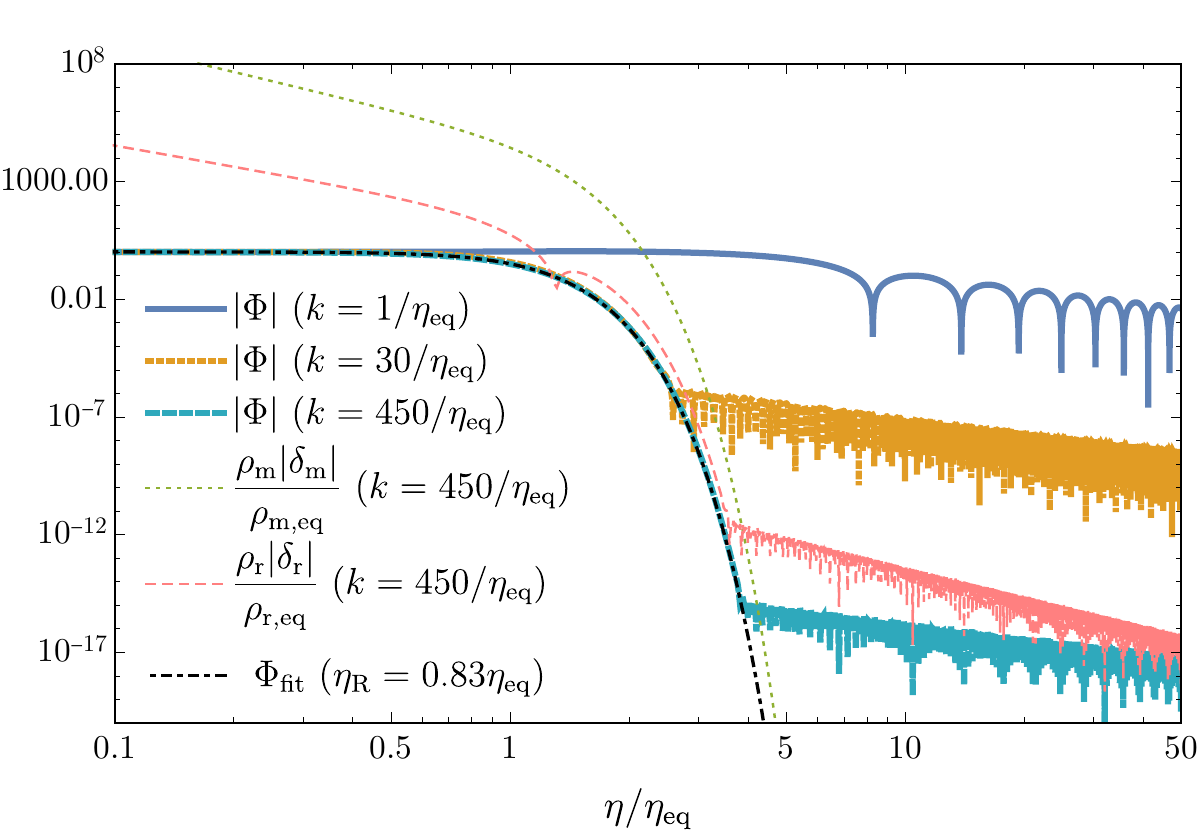}
        \caption{\justifying
        The evolution of the perturbations. $\Phi_{\text{fit}}$ is given by Eq.~\eqref{eq:phi_app_f}.
        From \rref{Inomata:2019zqy}.
        }
        \label{fig:pertb_summary}
\end{figure}

Figure~\ref{fig:pertb_summary} shows the results.
We also plot the approximate formula for the gravitational potential during the transition, $\Phi_\text{fit}$, defined as
\begin{align}
\label{eq:phi_app_f}
\Phi  &\simeq \exp \left(-\int^\eta \dd \bar \eta a(\bar \eta) \Gamma \right) \nonumber \\
& = \begin{cases} 
\exp \left( -\frac{2}{3} \left(\frac{\eta}{\eta_\R} \right)^3 \right) & (\eta < \eta_\R) \\
\exp\left( -2 \left( \left(\frac{\eta}{\eta_\R}\right)^2 - \frac{\eta}{\eta_\R} + \frac{1}{3} \right) \right) & (\eta \geq \eta_\R)
\end{cases}\\
&\equiv \Phi_\text{fit}, \nonumber
\end{align}
where we have used Eq.~(\ref{eq:a_app}) and assumed $a(\eta_\R) \Gamma = \mathcal H(\eta_\R) (= 2/\eta_\R)$.
This approximate solution is based on the intuition that the gravitational potential is approximately proportional to the matter density, whose perturbation mainly determines the gravitational potential through Poisson's equation.
In Fig.~\ref{fig:pertb_summary}, we can see that the gravitational potential that enters the horizon much before the transition significantly decays during the eMD-to-RD transition and finally oscillates. 

\begin{figure}%[ht] 
        \centering \includegraphics[width=0.45 \textwidth]{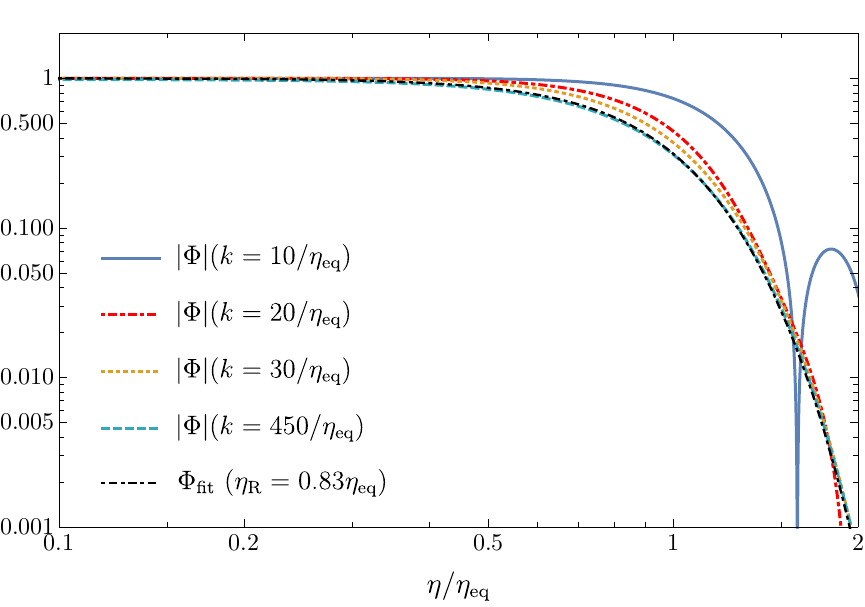}
        \caption{\justifying
        The comparison between the numerical results of $\Phi$ and the fitting function, given by Eq.~(\ref{eq:phi_app_f}).
        From \rref{Inomata:2019zqy}.
        }
        \label{fig:phi_evo}
\end{figure}

Figure~\ref{fig:phi_evo} compares the fitting function of $\Phi$, Eq.~(\ref{eq:phi_app_f}), with the numerical results with different wavenumbers. 
From this figure, we can see that the fitting function works well for $k \gtrsim 30/\eta_\eq$ but does not fit the numerical result when $k \lesssim 30/\eta_\eq$.

%%%%%%%%%%%%%%%%%%
%%%%%%%%%%%%%%%%%%

%%%%%%%%%%%%%%%%%%
%%%%%%%%%%%%%%%%%%

\subsection{Calculation of GWs induced during the gradual transition}

Next, we calculate the induced GWs.
Since we are focusing on the scalar perturbations that enter the horizon during the MD era, we can take $w_k = 0$ in Eq.~(\ref{eq:gw_cosmo_para_general}). 
Recall that $w_k$ determines the superhorizon relation between $\Phi$ and $\zeta$ and is irrelevant to the evolution during the transition. In contrast, we will use $w_\text{fit}$ for $w$ in Eq.~(\ref{eq:f_uvx}) to take into account the evolution during the transition.
Then, we obtain
\begin{align}
\label{eq:gw_cosmo_para_gradual}
&\Omega_\text{GW}(\eta,k) = \frac{27}{1250} \left( \frac{k}{\mathcal H} \right)^2 \int^\infty_0 \dd v \int^{|1+v|}_{|1-v|} \dd u \nonumber \\
&
\times \left[ \frac{4v^2 - (1 + v^2 - u^2 )^2}{4uv} \right]^2 \overline{I^2(u, v, x)} \mathcal P_{\zeta}(u k) \mathcal P_{\zeta} ( v k ).
\end{align}
We can approximately split $I$ into two parts depending on the time integral region:
\begin{align}
\label{eq:i_emd_rd_trans} 
 &I(u,v,x) \nonumber \\
 &\simeq  \int^{x_\text{R}}_0 \dd \bar{x} \left( \frac{\bar x^2/x_\R^2}{2(x/x_\R) -1} \right) k G_k^{\text{eMD}\rightarrow \text{RD}}(\eta; \bar \eta) f(u,v,\bar x) \nonumber\\
& \quad
+ \int^x_{x_\text{R}} \dd \bar x \left( \frac{2(\bar x/x_\R) -1}{2(x/x_\R) -1} \right) 
k G^{\text{RD}}_k (\eta; \bar \eta) f(u,v,\bar x) \nonumber \\ 
& \equiv I_\text{eMD}(u,v,x) + I_\text{RD}(u,v,x),
\end{align}
where $x \equiv k\eta$ and $x_\R \equiv k\eta_\R$, and we have assumed $\eta > \eta_\R$ and defined $I_\text{eMD}$ and $I_\text{RD}$ as the first and the second term after the first equality.
We have here used the fact that the scale factor can be well approximated as Eq.~(\ref{eq:a_app}), which corresponds to the sudden transition from an eMD to a RD era at $\eta_\R$.
By approximating $f$ (Eq.~(\ref{eq:f_uvx})) with the analytic fitting formulas Eqs.~(\ref{eq:w_fit}) and (\ref{eq:phi_app_f}), we can analytically express the integrands in Eq.~(\ref{eq:i_emd_rd_trans}), which significantly reduces the computational cost. 
With these approximations, we perform the time integral and the wavenumber integrals ($\int \dd v \int \dd u$) and obtain the GW spectrum. 
Note that, in this calculation, we neglect the contributions from the oscillation of the gravitational potential because the fitting function Eq.~(\ref{eq:phi_app_f}) does not cover that oscillation. 
However, for the perturbations that enter the horizon much before the transition, Figure~\ref{fig:pertb_summary} shows the oscillation of $\Phi$ begins with a very small amplitude (\textit{e.g.} $\Phi \sim \mathcal O(10^{-6})$ for $k \gtrsim 30/\eta_\eq$) and therefore we can safely neglect the contribution from the oscillation. 
This is in contrast to the sudden transition case, where the oscillation is important for the enhancement of the GW production through the poltergeist mechanism. 
We will come back to this point in the next section. 

From Eq.~(\ref{eq:i_emd_rd_trans}), we can express the square of $I$ as
\begin{align}
  \label{eq:I_split}
  \overline{I^2(u,v,x)} &=  \overline{I^2_\text{eMD}(u,v,x)} + \overline{I^2_\text{RD}(u,v,x)} \nonumber  \\ 
  &\quad + 2 \overline{I_\text{eMD}(u,v,x) I_\text{RD}(u,v,x)}.
\end{align}
Similarly, we express $\Omega_\GW$, given by Eq.~(\ref{eq:gw_cosmo_para_gradual}) as 
\begin{align}
  \label{eq:omega_gw_split}
  \Omega_\text{GW}(\eta,k) &= \Omega_\text{GW,eMD}(\eta,k) + \Omega_\text{GW,RD}(\eta,k) \nonumber \\ 
  &\quad + \Omega_\text{GW,cross}(\eta,k), 
\end{align}
where each term here corresponds to the contribution from each term in Eq.~(\ref{eq:I_split}).

To be concrete, we consider the following power spectrum of curvature perturbations:
\begin{align}
\mathcal P_\zeta = A_\ss\, \Theta(k-30/\eta_\text{eq}) \Theta(k_\text{max} -k).
\label{eq:pzeta}
\end{align} 
We have introduced the IR cutoff for wavenumber to focus on the modes that are on deeply subhorizon scales at the transition.
%that enter the horizon during the eMD era. 
We focus on those modes because, in the previous works~\cite{Assadullahi:2009nf,Alabidi:2013wtp}, the GW enhancement is realized by the modes with $k \gg 1/\eta_\eq$.
Apart from this, the fitting function Eq.~(\ref{eq:phi_app_f}) does not fit the numerical results well around $k\sim 1/\eta_\eq$. 
This makes us set $30/\eta_\eq$ as an IR cutoff to reduce computational cost.
Note that we take $k_\tmax \gg 30/\eta_\eq$, as we will see below.

On the other hand, we have introduced the UV cutoff to avoid the nonlinearity of the density perturbations. 
During an eMD era, subhorizon density perturbations grow proportionally to $a$. 
If we consider the density perturbations that enter the horizon much before the end of the eMD era, those perturbations become larger than unity ($\delta > 1$) before the transition. 
Once the density perturbations become larger than one, the perturbation theory breaks down. 
Let us here estimate the scale where the density perturbation becomes unity at the transition. 
Since the gravitational potential is constant during a MD era, Eq.~(\ref{eq:pertb_eq_00_new2}) leads to the following relation for subhorizon density perturbation:
\begin{align}
  \label{eq:delta_phi_order}
  \delta_\bfk \simeq -\frac{2}{3} k^2 \mathcal H^{-2} \Phi_\bfk \ \ (\text{for } k \gg \eta^{-1} \ \text{during MD era}),
\end{align}
where note again $\Phi_\bfk$ is constant. 
During a MD era, we have the relation $\mathcal P_\Phi = \frac{9}{25} \mathcal P_\zeta$, which leads to $\Phi_\bfk \sim - \frac{3}{5} \zeta_\bfk$. 
Using Eq.~(\ref{eq:delta_phi_order}), we can obtain the following order estimate:
\begin{align}
  |\delta| \sim \frac{1}{10}(k\eta)^2 \mathcal P_\zeta^{1/2},
\end{align}
where we have used $\mathcal H = 2/\eta$.
From this relation, we can obtain the nonlinear scale at the transition,
\begin{align}
  \label{eq:knl}
  k_\NL \sim \sqrt{10} \eta_\R^{-1} \mathcal P_\zeta^{-1/4} \sim 470/\eta_\R,
\end{align}
where $k_\NL$ is the scale where $\delta(k_\NL) = 1$ at the time $\eta_\R$ and we have substituted $\mathcal P_\zeta = 2.1\times 10^{-9}$ by simply assuming scale-invariant adiabatic perturbations.
We have also assumed that the Universe approximately behaves as a MD era until $\eta = \eta_\R$, which is somewhat justified from the expression of Eq.~(\ref{eq:a_app}).
Given this, we set $k_\tmax = 450/\eta_\R$ as a fiducial value in this section. 
For the contributions from the perturbations that enter the non-perturbative regime during an eMD era ($k>k_\NL$), see \rref{Jedamzik:2010dq,Jedamzik:2010hq,Nakama:2020kdc,Dalianis:2020gup,Eggemeier:2022gyo,Fernandez:2023ddy,Dalianis:2024kjr}. 
In particular, \rref{Fernandez:2023ddy} performed the hybrid $N$-body and lattice simulation in the case of a gradual transition and followed the GW production non-perturbatively even after the eMD era. Their results in the perturbative regime are consistent with \rref{Inomata:2019zqy}.

We here make some remarks on the limitations of the linear perturbation theory. 
If the density perturbation becomes larger than unity and the linear perturbation theory becomes unreliable at some point, we cannot obtain any reliable results after that on the perturbation evolution without the nonlinear (or nonperturbative) calculation method. 
Even if the linear perturbation theory predicts that the nonlinear density perturbations finally come back to the linear regime ($\delta < 1$), we cannot trust the perturbation evolution after the density perturbation becomes unity. 
Besides, the fact that $\Phi < 1$ is always satisfied does not mean that we can calculate the induced GW only with the linear perturbation theory. 
This is because the constancy of $\Phi$ on both super and subhorizon scales during a MD era is based on the linear perturbation theory, and, once the density perturbation becomes $\delta > 1$, the constancy of $\Phi$ is not secured.
In other words, to make $\Phi$ constant during a very long MD era, we need to consider $\delta \gg 1$, which cannot be followed with the linear perturbation theory.
We respect this logic and restrict ourselves to the parameter space where the linear perturbation theory is reliable throughout this review, including the next section, where we discuss the sudden transition case.

%%%%%%%%%%%%%%%%%%
\begin{figure}%[ht] 
        \centering \includegraphics[width=0.45 \textwidth]{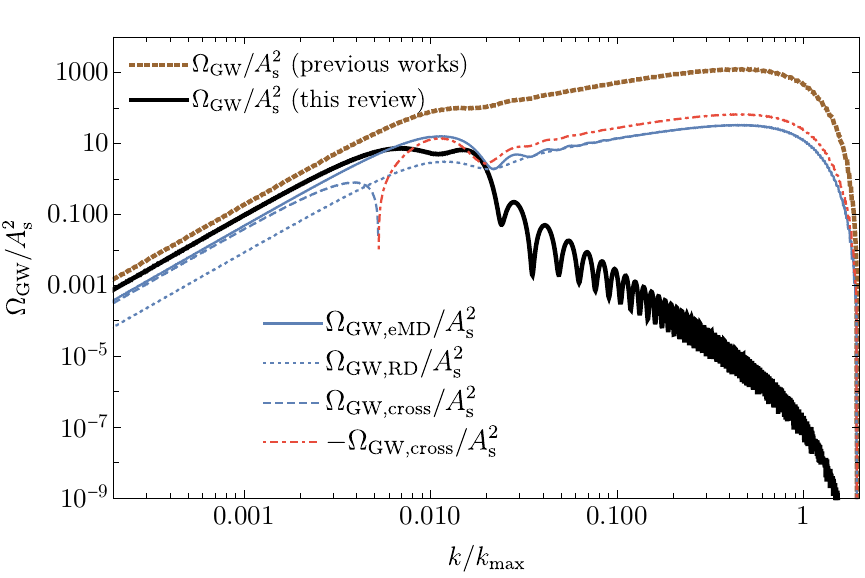}
        \caption{\justifying
        The GW spectrum at $\eta_\cc$ in the case of the gradual transition.
        The primordial curvature power spectrum is given by Eq.~(\ref{eq:pzeta}) with $k_\text{max}= 450/\eta_\R$ for all the plots.
        The black line for $\Omega_\GW$ is the sum of $\Omega_{\text{GW,eMD}}$, $\Omega_{\text{GW,RD}}$, and $\Omega_{\text{GW,cross}}$.
        For comparison, we show the GW spectrum (brown dotted line) with the implicit assumption that the previous work~\cite{Assadullahi:2009nf} makes: $\Phi=1$ until $\eta = \eta_\R$ and there is no GW production for $\eta > \eta_\R$ ($I_\text{RD} =0$).
        From \rref{Inomata:2019zqy}.
        }
        \label{fig:gw_sup_result}
\end{figure}
%%%%%%%%%%%%%%%%%%

Figure~\ref{fig:gw_sup_result} shows the GW spectrum. 
We can see that the GWs are suppressed for $k > 1/\eta_\R$. 
In the next section, we will see that the GWs are enhanced for $k > 1/\eta_\R$ if we consider a sudden transition from an eMD era to a RD era. 
The oscillation behavior of the black line in Fig.~\ref{fig:gw_sup_result} is an artifact of our piecewise approximation of the scale factor and the Green function at $\eta_\R$, Eq.~(\ref{eq:i_emd_rd_trans}).
This fact has been pointed out in \rref{Pearce:2023kxp}, where they numerically solve the scale factor and Green function.

This suppression of SIGWs is related to the unusual behavior of tensor perturbations induced by scalar perturbations during an eMD era.
Although we always call the induced tensor perturbations SIGWs throughout this review, the induced tensor perturbations during an eMD era are different from usual GWs because they do not propagate proportionally to $\sin(k\eta)$ or $\cos(k\eta)$~\cite{Inomata:2019yww}.
The induced tensor perturbations during an eMD era are coupled to the source scalar perturbations, and, once the source perturbations are suppressed during the transition, the induced tensor perturbations are also suppressed.
This is why we can see the suppression of SIGWs after the transition in Fig.~\ref{fig:gw_sup_result}.
We remark that the previous works~\cite{Assadullahi:2009nf,Kohri:2018awv} focus only on the contribution of $\Omega_{\GW,\text{eMD}}$ with the assumption $\Phi = 1$ until $\eta= \eta_\R$, so the effect of the decay of the matter and subsequent changes of the tensor mode have not been included, leading to the absence of the suppression. 
We could also alternatively interpret that the previous works assumed the instantaneous transition but missed the contribution after the transition, which leads to the poltergeist mechanism.

%%%%%%%%%%%%%%%%%%%%%%%%%%%%%%%%
\section{Scalar-induced gravitational waves with a sudden matter-to-radiation transition}
\label{sec:sudden}
%%%%%%%%%%%%%%%%%%%%%%%%%%%%%%%%

\subsection{Intuitive explanations}

\begin{table*}
    \centering
        \caption{Analogy between a pendulum and the gravitational potential in the cosmological setup}
    \label{tab:analogy}
    \setlength{\tabcolsep}{16pt}
    \begin{tabular}{ccccc} \hline
         & pendulum & & cosmology & \\ \hline \hline
        displacement angle & $\theta(t)$ & $\longleftrightarrow$ & $\Phi(k\eta)$ & gravitational potential \\ \hline
    frequency & $\omega_0 (\equiv \sqrt{g/\ell})$ & $\longleftrightarrow$ & $c_\text{s} k$ &     frequency \\ \hline 
         time-dependent friction & $\zeta(t)$ & $\longleftrightarrow$ & $\mathcal H(\eta)$ & Hubble parameter \\ \hline 
        holding force & 
        $F=\omega_0^2\theta$ & \multirow{2}{*}{$\longleftrightarrow$} & \multirow{2}{*}{$c_\text{s} = 0$} & sound speed vanishing \\
        canceling the restoring force & $\implies \omega_\text{0,eff}^2 = 0$ & & & during the eMD era \\ \hline 
       sudden release & $\dot{F}/F \gg \omega_0$ & $\longleftrightarrow$ & $c'_\text{s}/c_\text{s} \gg c_\text{s} k$  & sudden change of sound speed \\ \hline
    \end{tabular}
\end{table*}

%%%%%%%%%%%%%%%%%%
\begin{figure*}[tbhp]
\begin{center}
\subcaptionbox{damped oscillator (standard case)\label{sfig:damped_osc}}{\includegraphics[width=0.32 \textwidth]{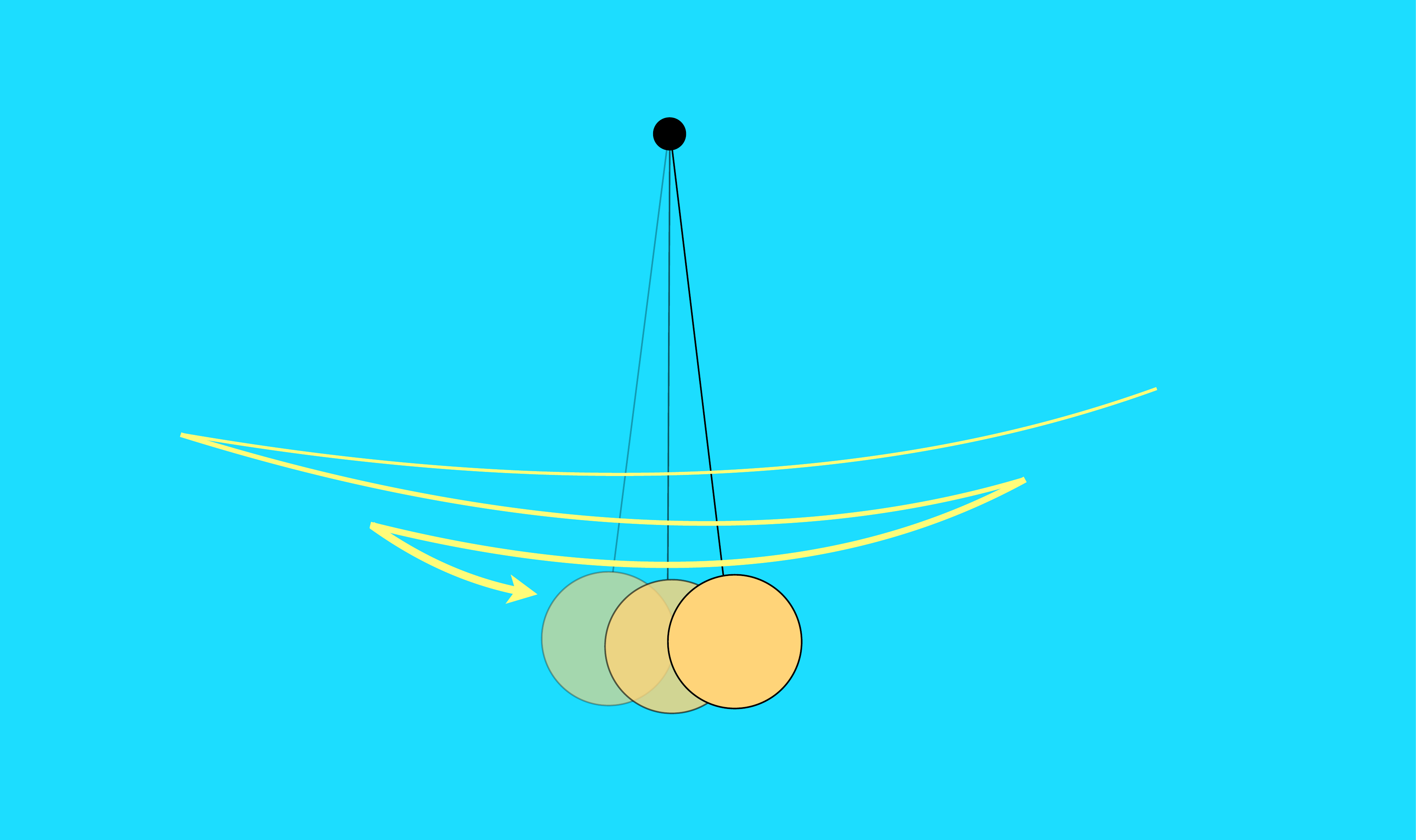}}
\subcaptionbox{fixed oscillator (eMD era) \label{sfig:fixed_osc}}{\includegraphics[width=0.32 \textwidth]{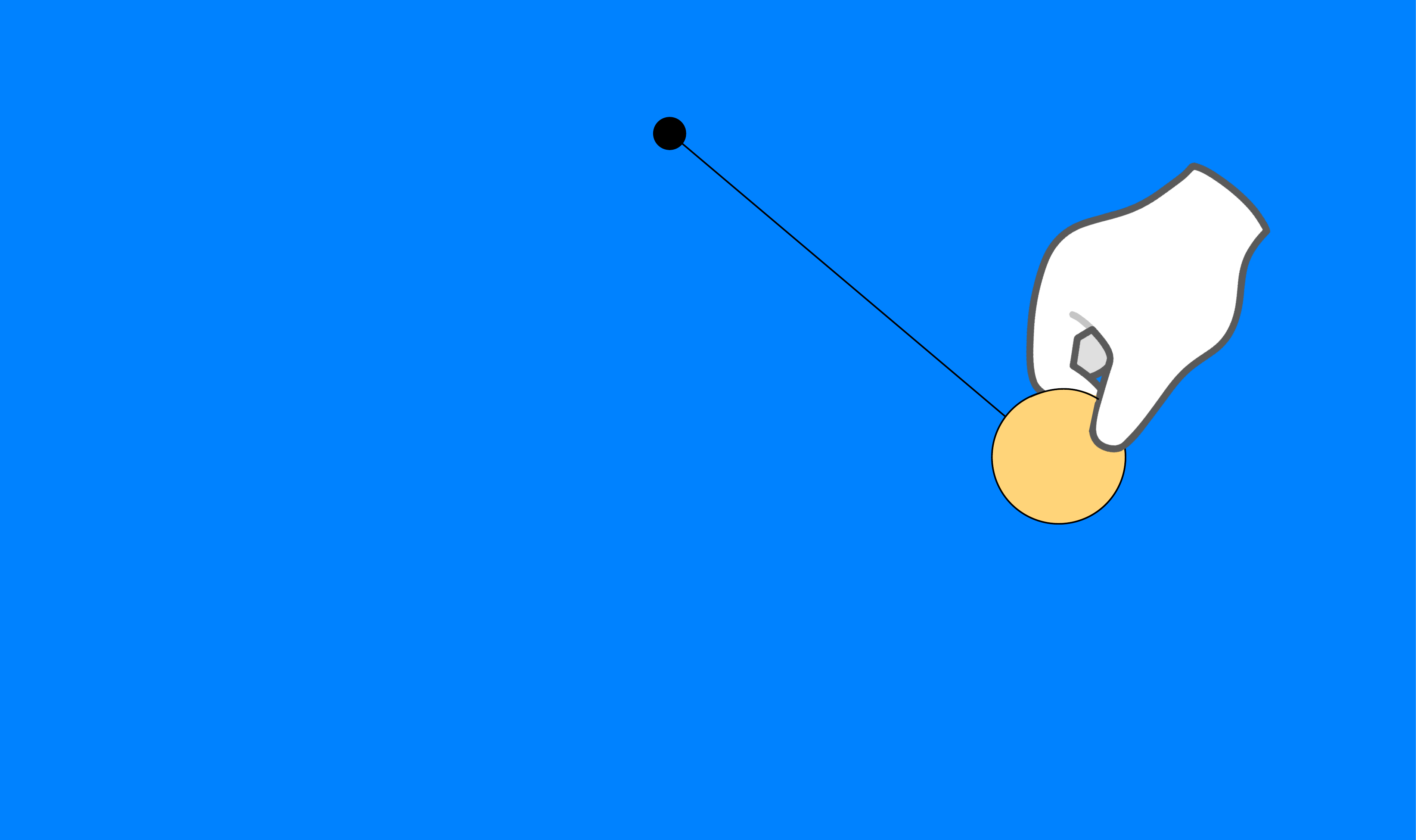}}
\subcaptionbox{released oscillator (RD era)\label{sfig:released_osc}}{
\includegraphics[width=0.32 \textwidth]{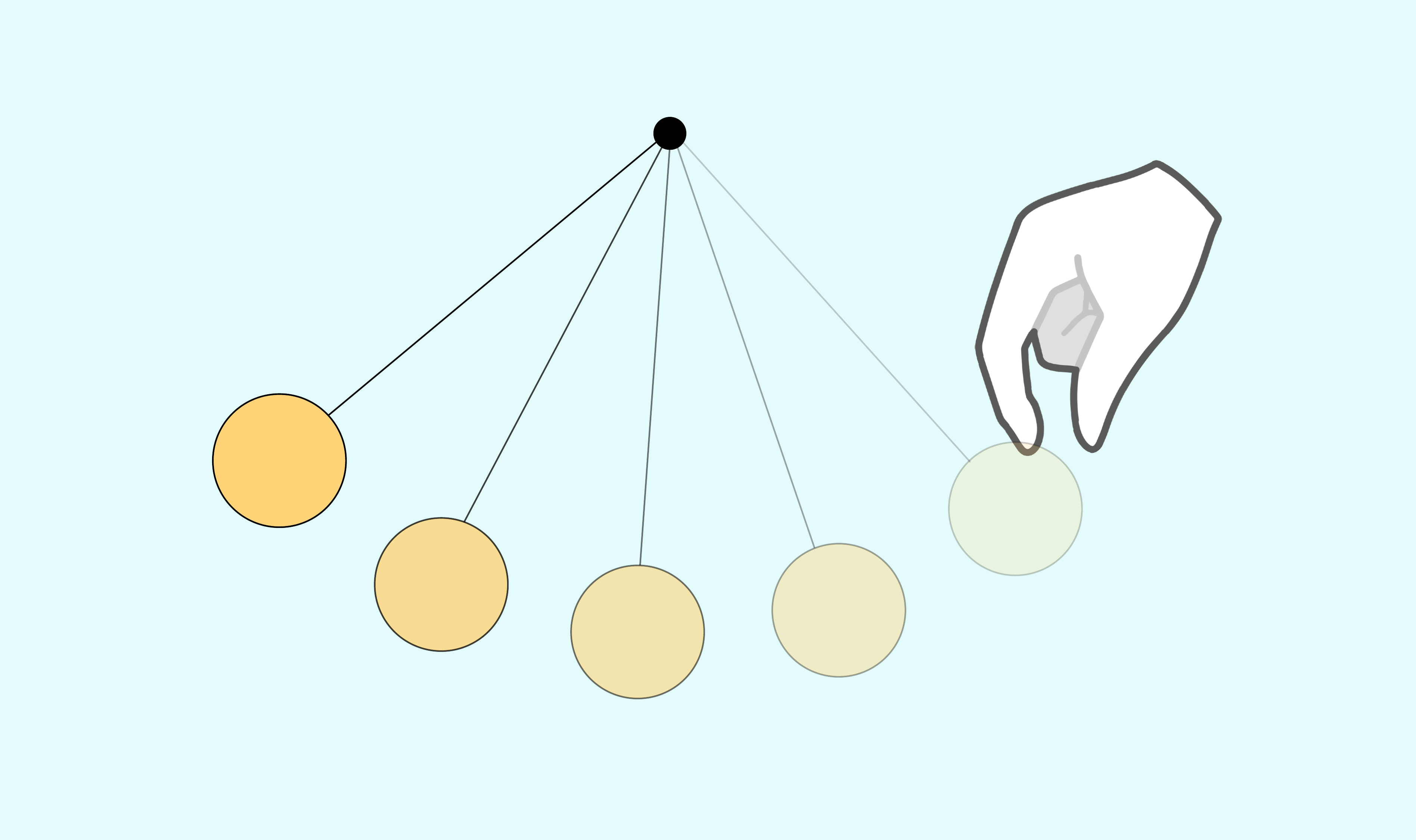}}
\end{center}
\caption{\justifying
Analogy to a pendulum in a medium with time-dependent friction. The pendulum with a fixed length corresponds to a mode of $\Phi$ with a fixed comoving wavenumber $k$. The deeper/lighter bluish background color indicates the stronger/weaker friction in analogy to the (conformal) Hubble friction, respectively. \eqref{sfig:damped_osc} The oscillation amplitude is damped by the (time-dependent) friction in analogy to the usual behavior of the gravitational potential, \textit{e.g.}, in a RD era. The yellow arrow schematically shows the damping of the oscillations. The amplitude has been damped since the horizon reentry of the mode. The induced GWs are not enhanced in this case. \eqref{sfig:fixed_osc} The position of the oscillator is kept fixed while the friction of the medium is large, so the amplitude is not damped. This is analogous to the behavior of the gravitational potential in a MD era.  \eqref{sfig:released_osc} When ``poltergeist's hand'' abruptly releases the oscillator, it begins to oscillate. By this time, the friction of the medium becomes weak. This is analogous to the behavior of the gravitational potential after a sudden transition from the eMD era to the RD era. The rapid oscillation with respect to the Hubble parameter with the undamped amplitude efficiently produces the GWs.}
\label{fig:osc_analogy}
\end{figure*}
%%%%%%%%%%%%%%%%%%

In this subsection, we intuitively explain the essence of the poltergeist mechanism of GW production. More rigorous explanations are given in the subsequent subsections. 

As we will see in more detail below, strong GWs are induced after the end of the eMD era provided that (1) the period of the eMD era is sufficiently long, and (2) the change of the equation-of-state at the end of the eMD era is sufficiently rapid.  The production is dominated just after the rapid end of the eMD era. This is because $\Phi$ decays after the eMD era so the late-time production is subdominant, while $\Phi$ oscillates only after the end of the eMD era so the strong GW production does not occur before the transition. 

There are two key elements in the mechanism. First, $\Phi$ does not decay during the eMD era even on subhorizon scales. It can oscillate after the end of the eMD era.  If the transition is sufficiently rapid, there is no time for the amplitude to decay, so it begins to oscillate with the unsuppressed amplitude.  Second, the oscillation is more rapid for more subhorizon modes. In particular, it can be hierarchically rapid compared to the cosmological timescale $H^{-1}$ for modes that entered the Hubble horizon deep (i.e., early) in the eMD epoch. This oscillation with (1) the unsuppressed amplitude and (2) the frequency larger than the Hubble scale leads to the production of strong GWs. It is significantly enhanced compared not only to the standard order-of-magnitude of $\Omega_\text{GW}$ induced in a RD era, $\mathcal{O}(\mathcal{P}_\zeta(k)^2)$, but also to the GWs induced during the eMD era.  Note that the two key elements in this mechanism require both the eMD era and the subsequent era, for which we usually consider the RD era.  Thus, the interplay between the eMD era and the subsequent era is important. 
We found the above mechanism in \rref{Inomata:2019ivs} and dubbed it as the \emph{poltergeist mechanism} in \rref{Inomata:2020lmk}.\footnote{The German word ``Poltergeist'' originates from ``poltern'', which means to make noise or to rattle, and ``Geist'', which means ghost or spirit. The poltergeist refers to supernatural phenomena in which, \textit{e.g.}, furniture spontaneously moves around and makes disruptive noises.  In our physical context of the eMD era transitioning into the RD era, the analogy is as follows.  The GWs are induced from the density perturbations or the \emph{sound waves} in the thermal bath, which is produced by the decay or \emph{death} of the matter that dominated the eMD era.  Despite the disappearance of the matter, its (``ghostly'') effect on the radiation is effective.  Therefore, it looks as if the ghost of the matter makes noises after the eMD era, and it produces the GWs.  Needless to say, our poltergeist mechanism for GW production is physical and not a supernatural effect.  This mechanism also applies to the case of the eMD era transitioning to the non-RD era such as kination.}

It may be useful to draw an analogy between the dynamics of the gravitational potential $\Phi$ and those of a pendulum (see Tab.~\ref{tab:analogy} and Fig.~\ref{fig:osc_analogy}). The second-order differential equation of motion for $\Phi$ is analogous to that of the angle $\theta$ of a driven and damped oscillator
\begin{align}
    \ddot{\theta} + 2\zeta(t) \omega_0 \dot{\theta}  + \omega_0^2 \theta = F(t),
\end{align}
where $\omega_0$ is the angular frequency in the undamped case, $\zeta$ is the damping ratio, and $F$ is the driving force. For our purposes, the analogy is qualitative. The frequency $\omega_0 = \sqrt{g/\ell}$ is given by the gravitational acceleration $g$ and the length of the pendulum $\ell$. A smaller $\ell$ (larger $\omega_0$) corresponds to a larger wavenumber $k$ in the cosmology side of the analogy. Let us focus on a fixed $\ell$ corresponding to a given mode $k$.  The friction term $2\zeta(t) \omega_0 \dot{\theta}$ is analogous to the Hubble friction term, which is also time-dependent. The strength of the friction is shown by the background color in Fig.~\ref{fig:osc_analogy}.  The deeper background color indicates the stronger friction. Suppose that the friction becomes weaker as a function of time analogously to the decrease of the Hubble friction.  
To qualitatively mimic the constancy of $\Phi$ during the eMD era, 
we imagine that the pendulum is fixed in the corresponding situation, as illustrated by the hand in Fig.~\ref{sfig:fixed_osc}.
The fixing of the amplitude of the pendulum can be implemented by the force term $F=\omega_0^2 \theta$ that counteracts the restoring force. By transposing the force term to the left-hand side, we have $\ddot{\theta} + 2\zeta(t) \omega_0 \dot{\theta}  + \omega_\text{0,eff}^2 \theta = 0$ with the vanishing effective frequency $\omega_\text{0,eff}=0$.

Let us first consider the usual case (e.g., the RD era) without the poltergeist mechanism for comparison. In the cosmology side, the amplitudes of modes decay after their horizon entry. For modes comparable to the Hubble scale, the amplitude has not yet decayed significantly, but there is no enhancement from the ratio $(k/\mathcal{H})$. For deep subhorizon modes, their frequencies are larger than the Hubble scale, but their amplitudes become too small to significantly enhance the induced GWs.

In the pendulum side of the analogy, the above usual case corresponds to the damped oscillator (Fig.~\ref{sfig:damped_osc}). 
At early time, the amplitude is large, but the friction is so large that the oscillation amplitude decays relatively rapidly.
At late time, it oscillates faster than the damping timescale due to the decrease of the friction, but the amplitude already becomes small.

Now, let us explain the poltergeist mechanism again using the analogy to the pendulum. 
The first and second halves of the dynamics in the poltergeist mechanism correspond to Figs.~\ref{sfig:fixed_osc} and \ref{sfig:released_osc}, respectively. 
This time, let us begin with the pendulum side of the analogy. 
Imagine that someone holds the pendulum to forbid the oscillation of the pendulum (Fig.~\ref{sfig:fixed_osc}). We may call the hand in the figure \textit{``poltergeist's hand''} (though matter is still alive at the stage of Fig.~\ref{sfig:fixed_osc} in the cosmology side of the analogy). The amplitude and hence the potential energy is kept fixed by the hand. Suppose that the hand abruptly releases the pendulum after waiting for the friction to become weak (Fig.~\ref{sfig:released_osc}).  Then, the pendulum can oscillate with the unsuppressed amplitude and a frequency that can be larger than (the inverse of) the damping timescale.

The poltergeist mechanism on the cosmology side is analogous. First, during the eMD era, $\Phi$ does not decay. This is true for all modes that entered the Hubble horizon during the eMD era.  Note that this effect itself was already noticed in \rref{Assadullahi:2009nf} to discuss the enhancement of the induced tensor mode during the eMD era. It turned out, however, that such a contribution is subdominant in the final GWs observed well after the end of the eMD era if the transition is rapid enough. The dominant contribution comes from the second part of the dynamics, \textit{i.e.}, after the transition into the subsequent era \cite{Inomata:2019ivs}. Again, the reason is simple; the modes oscillate rapidly compared to the Hubble scale at that time with an unsuppressed amplitude. 

In the standard not-so-sudden transition from an eMD era, for comparison, $\Phi$ decays during the transition.  On the pendulum side of the analogy, one can imagine that the holding power by the hand in Fig.~\ref{sfig:fixed_osc} becomes weaker and weaker adiabatically, so that the position (amplitude) of the oscillator adiabatically decreases. 
This case is qualitatively similar to the situation in  Fig.~\ref{sfig:damped_osc}. A concrete study of such a situation was done in \rref{Inomata:2019zqy}, reviewed in Sec.~\ref{sec:gradual}.

\subsection{Instantaneous-limit transition}

With the physical intuition explained above, let us see the concrete equations for the poltergeist mechanism. 
As the simplest example, we focus on the instantaneous-limit transition in this subsection.
We will discuss how to realize almost instantaneous transitions in the next subsection. 
Also, in this subsection, we focus on the formulation for the instantaneous transition from an eMD era to a RD era. 
The generalization to another transition (eMD $\to$ kination) will be discussed in Sec.~\ref{sssec:axion_poltergeist}.

In this limit, we can use Eq.~(\ref{eq:a_app}) for the scale factor, where $\eta_\R$ here means the time when the instantaneous transition occurs. 
Note that $\eta_\eq = \eta_\R$ in this limit. 
The main difference from the gradual transition in the previous section lies in the evolution of the perturbations.
In the following, we focus on the evolution of the gravitational potential because it determines the source terms for the induced GWs. 
We begin with the equation of motion for the gravitational potential (Eq.~(\ref{eq:phi_only_eq})) during a RD era:
\begin{align}
  \label{eq:phi_eom_rd}
    \Phi'' + 4 \mathcal H \Phi' + \frac{k^2}{3} \Phi = 0 \quad (\text{for } \eta > \eta_\R).
\end{align}
We solve this equation with the initial condition at $\eta_\R$ given by 
\begin{align}
  \Phi(k\eta_\R) = 1, \ \Phi'(k\eta_\R) = 0,
\end{align}
where we have used the fact that $\Phi$ is constant during a MD era and have imposed the continuity of $\Phi$ and $\Phi'$ at the transition. 
Their continuity is justified from the equation of motion for $\Phi$, Eq.~(\ref{eq:phi_only_eq}), because, through the transition, the value of $c_s$ remains finite and therefore the coefficients in the equation also remain finite.
The absence of infinitely large coefficients in Eq.~(\ref{eq:phi_only_eq}) leads to the continuity of $\Phi$ and $\Phi'$.
Then, we obtain 
\begin{align}
  &\Phi(x,x_\R) \nonumber \\
  &= \begin{cases}
1 & ( \text{for}\  x \leq x_\text{R}) \\
 A(x_\text{R}) \mathcal J(x,x_\R) + B(x_\R) \mathcal Y(x,x_\R)  & ( \text{for}\  x \geq x_\text{R})
\end{cases},
\label{eq:phi_formula}
\end{align}
where $\mathcal J(x,x_\R)$ and $\mathcal Y(x,x_\R)$ are the linearly independent solutions of Eq.~(\ref{eq:phi_eom_rd}), defined as
\begin{align}
\mathcal J(x,x_\R) &\equiv \frac{ 3\sqrt{3} \,  j_1\left(\frac{x-x_{\text{R}}/2}{\sqrt{3}} \right)}{x-x_{\text{R}}/2}, \\ 
\mathcal Y(x,x_\R) &\equiv \frac{3\sqrt{3} \, y_1\left( \frac{x - x_{\text{R}}/2}{\sqrt{3}} \right)}{x- x_{\text{R}}/2}, 
\label{eq:c_d_formula}
\end{align}
and the coefficients $A(x_\R)$ and $B(x_\R)$ are given by 
\begin{align}
\label{eq:a_formula}
A(x_\R) &= \left.\frac{1}{\mathcal J(x,x_\R) - \frac{\mathcal Y(x,x_\R)}{\mathcal Y'(x,x_\R)} \mathcal J'(x,x_\R)}\right|_{x \to x_\R}, \\
\label{eq:b_formula}
B(x_\R) &= \left.- \frac{\mathcal J'(x,x_\R)}{\mathcal Y'(x,x_\R)} A(x_\R)\right|_{x \to x_\R}.
\end{align}

%%%%%%%%%%%%%%%%%%
\begin{figure}%[ht] 
        \centering \includegraphics[width=0.45 \textwidth]{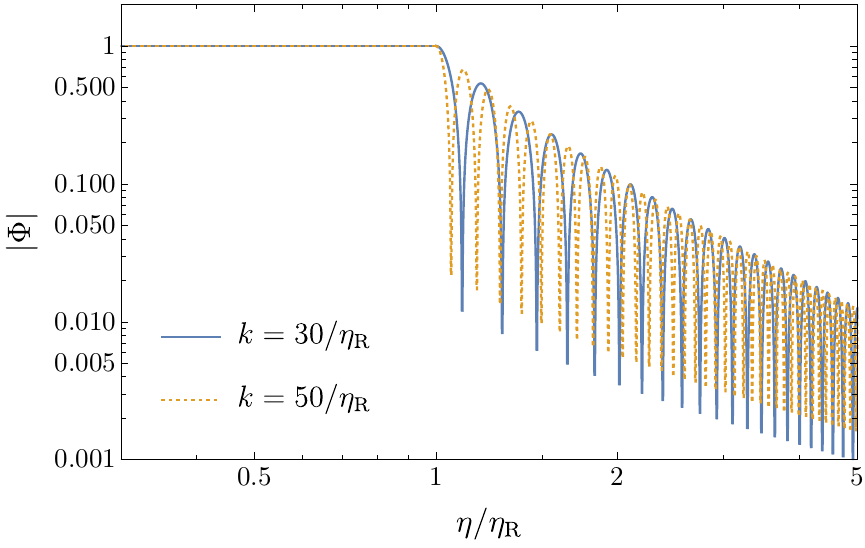}
        \caption{\justifying
        The evolution of the gravitational potential with different wave numbers in the instantaneous transition limit.
        }
        \label{fig:phi_evo_i}
\end{figure}
%%%%%%%%%%%%%%%%%%

Figure~\ref{fig:phi_evo_i} shows the evolution of the gravitational potential given by Eq.~(\ref{eq:phi_formula}).
We can see that, for the scales that enter the horizon much before the transition ($k \gg 1/\eta_\R$), the gravitational potential oscillates very rapidly in $\eta > \eta_\R$.
This rapid oscillation leads to the enhancement of the GW production. The enhancement factor originates from the time-derivative terms in Eq.~\eqref{eq:func_f_def} ($\Phi' /\mathcal{H} \sim (k/\mathcal{H}) \Phi \gg \Phi$). 

%%%%%%%%%%%%%%%%%%
\begin{figure}%[ht] 
        \centering \includegraphics[width=0.45 \textwidth]{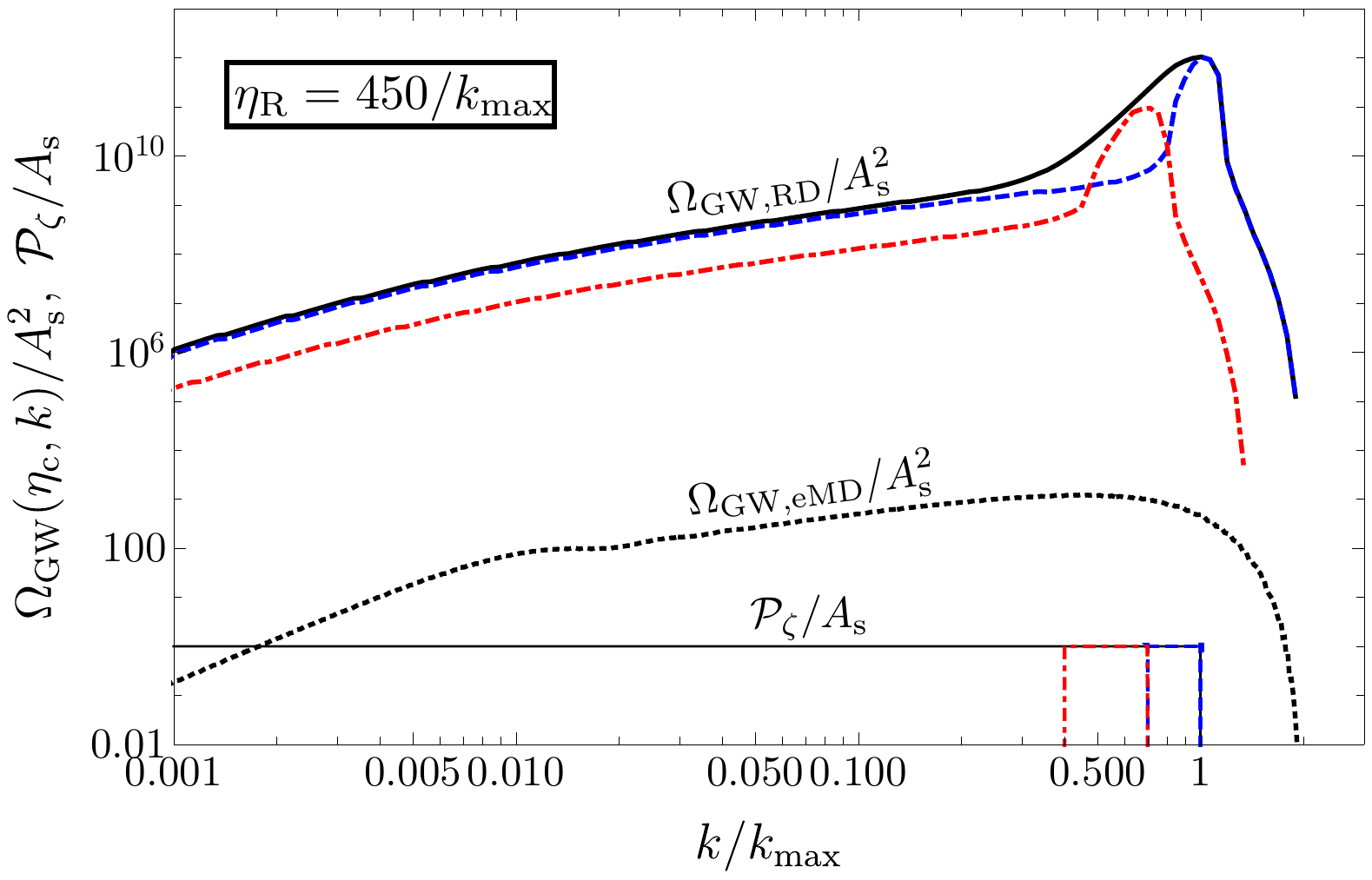}
        \caption{\justifying
        The GW spectrum produced by the poltergeist mechanism with different power spectra of curvature perturbations.
        We take $\mathcal P_\zeta = A_\ss \Theta(k_\tmax - k)$ for the black lines, $= A_\ss\Theta(k_\tmax - k)\Theta(k-0.7 k_\tmax)$ for the blue dashed line, and $= A_\ss\Theta(0.7k_\tmax - k)\Theta(k-0.4 k_\tmax)$ for the red dot-dashed line.
        $\Omega_\text{GW,RD}$ and $\Omega_\text{GW,eMD}$ are defined in Eq.~(\ref{eq:omega_gw_split}).
        From \rref{Inomata:2019ivs}.
        }
        \label{fig:gw_prof}
\end{figure}
%%%%%%%%%%%%%%%%%%

Let us calculate the GWs. 
We can use Eqs.~(\ref{eq:gw_cosmo_para_gradual}) and (\ref{eq:i_emd_rd_trans}) also in the instantaneous transition limit since they are based on the instantaneous transition for the scale factor. 
On the other hand, we substitute Eq.~(\ref{eq:phi_formula}) into Eq.~(\ref{eq:gw_cosmo_para_gradual}), instead of $\Phi_\text{fit}$, given by Eq.~(\ref{eq:phi_app_f}).
As explained in the previous subsection, the main source for the GW production in this case comes from the contributions soon after the transition because the rapid oscillations of the gravitational potential are the main source.

Figure~\ref{fig:gw_prof} shows the GW profile for the scale-invariant curvature power spectrum with a UV cutoff. 
We can see $\Omega_\text{GW,RD} \gg \Omega_\text{GW,eMD}$, each of which is defined in Eq.~(\ref{eq:omega_gw_split}).
This means $\Omega_\text{GW} \simeq \Omega_\text{GW,RD}$. 
In the figure, we also show the GW spectrum with a different power spectrum, which shows that the dominant contribution to $\Omega_\text{GW,RD}$ comes from the smallest scale of the perturbation we consider.
This is because the smaller the scale of perturbations is, the more rapidly the gravitational potential oscillates, which leads to the larger contributions to the GW production. 
In the case of $\mathcal P_\zeta (k)= A_\ss\Theta(k_\tmax-k)$, we can roughly approximate the GW spectrum as\footnote{
While we focus on the power spectrum (two-point correlation) of the poltergeist GWs throughout this review, the bispectrum (three-point correlation) of the poltergeist GWs is discussed in \rref{Hu:2025zqn}.}
\begin{widetext}
\begin{align}
&\frac{\Omega_{\text{GW}}(\eta_\cc,k)}{A_{\text{s}}^2} \simeq  
\begin{cases}
0.8 & (x_{\text{R}} \lesssim 150 x_{\text{max,R}}^{-5/3}) \\
3 \times 10^{-7} x_{\text{R}}^3 x_{\text{max,R}}^5  &   (150 x_{\text{max,R}}^{-5/3} \lesssim x_{\text{R}} \ll 1) \\
1 \times 10^{-6} x_{\text{R}} x_{\text{max,R}}^5 & (1 \ll x_{\text{R}} \lesssim x_{\text{max,R}}^{5/6}) \\
7 \times 10^{-7}   x_{\text{R}} ^7 &     (x_{\text{max,R}}^{5/6} \lesssim x_{\text{R}}   \lesssim x_{\text{max,R}})  \\
\text{(sharp drop)} & (x_{\text{max,R}} \lesssim x_{\text{R}} \leq 2 x_{\text{max,R}})
 \end{cases},  \label{eq:Omega_GW_rough_behavior}
\end{align}
\end{widetext}
where $x_{\text{max,R}} \equiv k_\tmax \eta_\R$.
We note that the IR tail of the peak is $\propto k^3$, which was later reproduced in general situations~\cite{Cai:2019cdl} (see also \rref{Yuan:2019wwo} for a logarithmic correction to $k^3$, which can also be seen in Eq.~\eqref{Omega_GW_LS_SI} but has a limitation due to dissipation~\cite{Domenech:2025bvr}). 
While we refer to Appendix~\ref{app:analytic_formulas} and \rref{Inomata:2019ivs} for the analytical derivation of this approximate form, we mention that the sharp rise with $\propto k^7$ around the peak is due to the resonant amplification of the GW production~\cite{Ananda:2006af}.
Accordingly, the peak height of $\Omega_\GW$ is proportional to $(k_\tmax \eta_\R)^7$. 

A naive interpretation of the $k^7$ dependence around the peak is as follows. 
A factor $k^2$ arises in $f$ (Eq.~(\ref{eq:func_f_def})) from the dominant term proportional to $\Phi' \Phi'$, which gives a factor $k^4$ to $\mathcal P_h$. 
Another factor arises from the integrals appearing in the expression for $\mathcal P_h$. 
In Eq.~(\ref{eq:ph_express_v_u}), the integrals over $v$ and $u$ correspond, before the change of variables, to integrals over the wavenumbers of the two scalar modes, $k_1$ and $k_2$. 
Since the resonance condition $(k_1 + k_2)/\sqrt{3} \simeq k$ holds near the peak, the integrals over $k_1$ and $k_2$ contribute an additional power of $k$ to the $k^4$ scaling, leading to $\mathcal P_h \propto k^5$. 
Furthermore, an extra factor $k^2$ arises from the relation $\rho_{\rm GW} \propto \langle h' h' \rangle \propto k^2 \mathcal P_h$ (see Eq.~(\ref{eq:rho_gw})). Combining these, we obtain $\Omega_{\rm GW} \propto k^7$ around the peak.

To be more concrete, we consider the following power spectrum of curvature perturbations:\footnote{The reconstruction of curvature power spectrum from the measurement of SIGWs with the poltergeist mechanism is discussed in \rref{LISACosmologyWorkingGroup:2025vdz,Ghaleb:2025xqn}.}
\begin{align}
  \label{eq:pzeta_inst}
  \mathcal P_\zeta = 2.1\times 10^{-9} \Theta(k_\tmax - k) \left(\frac{k}{0.05\,\Mpc^{-1}}\right)^{-0.04},
\end{align}
where we have taken the numerical values from the Planck results~\cite{Planck:2018vyg}.
Recalling the discussion below Eq.~(\ref{eq:pzeta}), we restrict ourselves to the region where $\delta < 1$ at $\eta_\R$, which leads to $k_\tmax \lesssim 450/\eta_\R$.
Note again that, if we consider $k_\tmax \gg 450/\eta_\R$, the density perturbation becomes larger than unity before $\eta_\R$ and the perturbation theory is no longer reliable. 

\begin{figure}%[ht] 
        \centering \includegraphics[width=0.45 \textwidth]{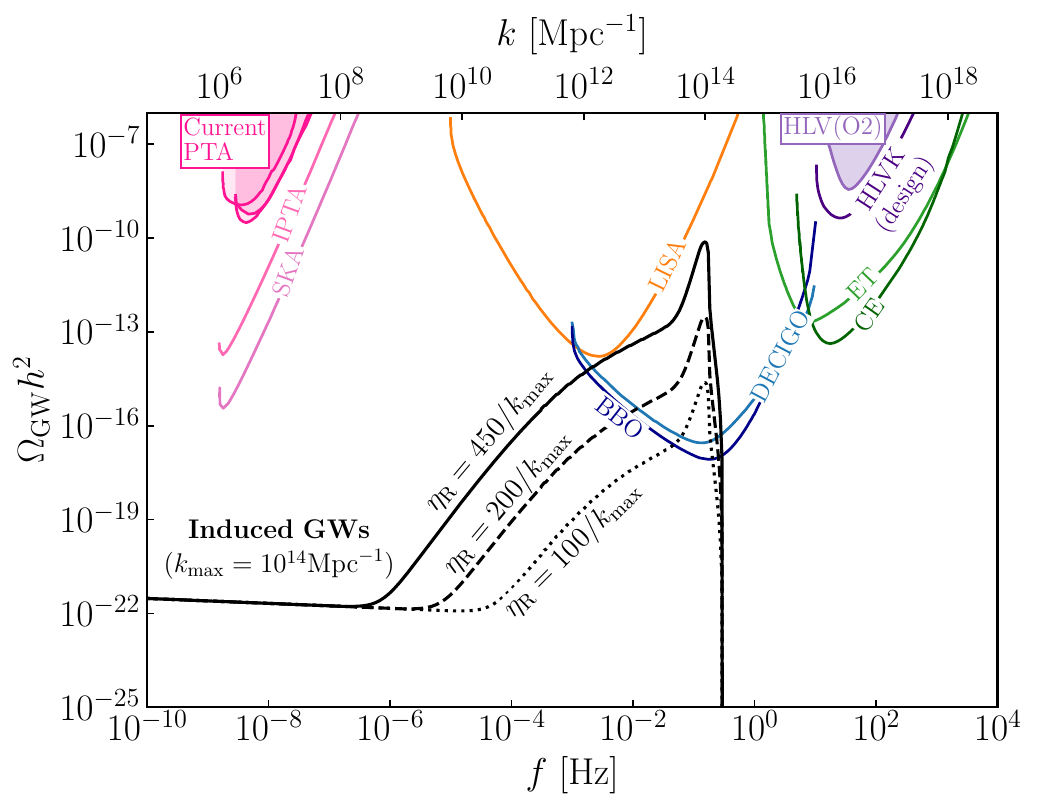}
        \caption{\justifying
        The GW spectrum produced by the poltergeist mechanism with different $\eta_\R$ with $k_\tmax$ fixed.
        The power spectrum of curvature perturbation is given by Eq.~(\ref{eq:pzeta_inst}) with $k_\tmax = 10^{14}\,\Mpc^{-1}$.
        The GW sensitivity curves are from \rref{Schmitz:2020syl}.
        The shaded regions are excluded by the current observations.
        From \rref{Inomata:2019ivs} with updated sensitivity curves.
        }
        \label{fig:gw_inst_limit}
\end{figure}

Figure~\ref{fig:gw_inst_limit} shows the GW spectrum with different $\eta_\R$ with $k_\tmax$ fixed and the sensitivities of the current and future GW observations. 
The peak scale and height of the GW spectrum are determined by $k_\tmax$ and $\eta_\R$. 
Furthermore, $\eta_\R$ is related to the reheating temperature at the end of the eMD era.

\begin{figure}%[ht] 
        \centering \includegraphics[width=0.45 \textwidth]{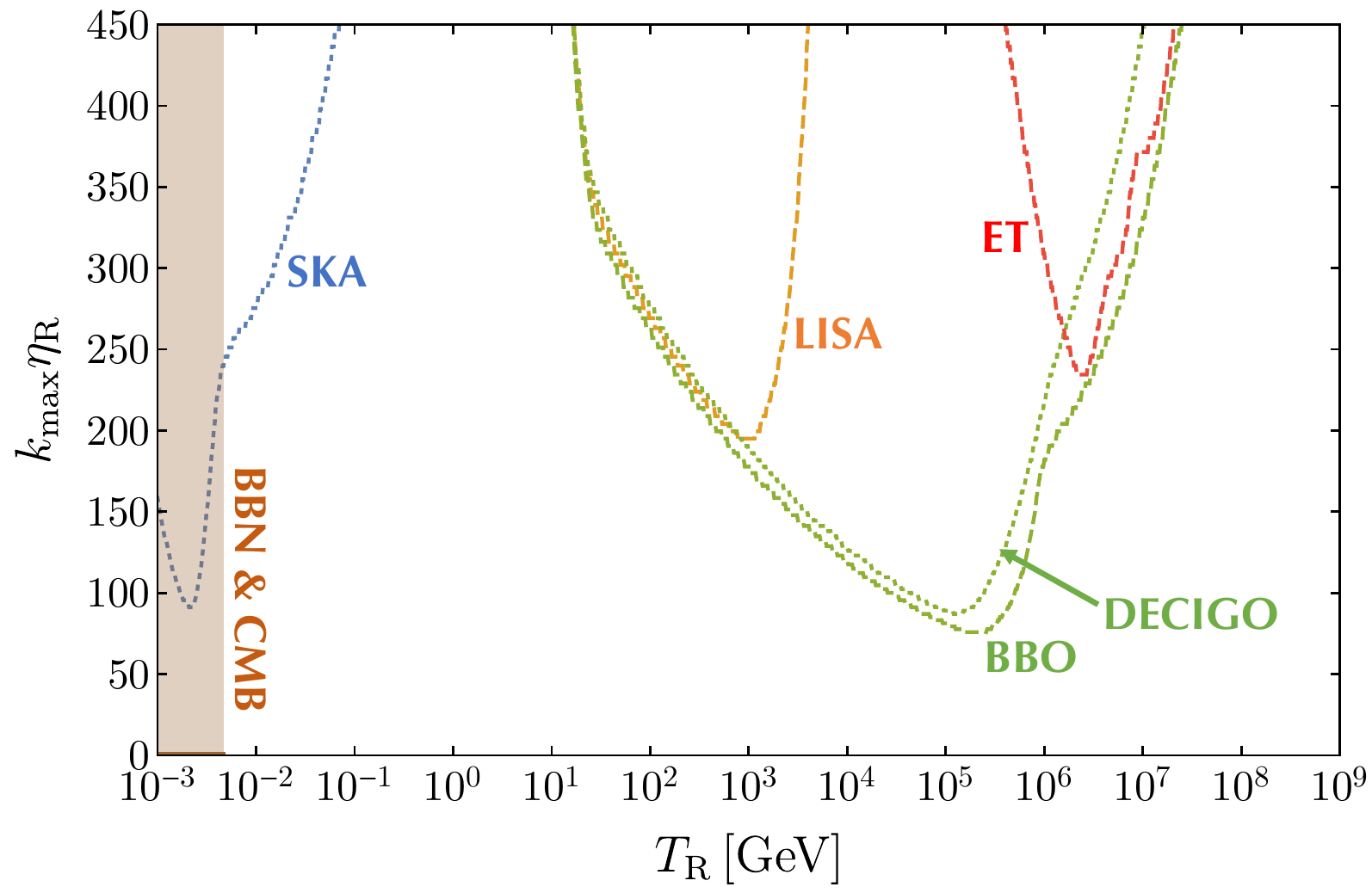}
        \caption{\justifying
        The reheating temperature and the cutoff scale that future experiments can probe.
        The observation time $T_\obs$ is $20$\,years for SKA and $1$\,year for the others.
        The dashed and dotted lines are for the signal-to-noise ratio of unity, and the regions above the lines can be probed by each experiment.
        The shaded region is excluded by the constraints on the dark radiation $N_\text{eff}$ from BBN and Planck data.
        From \rref{Inomata:2019ivs}.
        }
        \label{fig:reh_const}
\end{figure}

Figure~\ref{fig:reh_const} shows the parameter space of $k_\tmax \eta_\R$ and the reheating temperature $T_\R$ that future GW experiments can probe in the case of the instantaneous-limit transition. 
To obtain the plots in Fig.~\ref{fig:reh_const}, we calculate the signal-to-noise ratio:
\begin{align}
    \text{SNR} = \sqrt{2T_\obs} \left[ \int^{f_\tmax}_{f_\text{min}} \dd f \left( \frac{\Omega_\GW(f)}{\Omega_{\GW,\text{eff}}(f)} \right) \right],
\end{align}
where $T_\obs$ is the observation time, and $(f_\text{min},f_\tmax)$ is the observable frequency range of each project, and $\Omega_{\GW,\text{eff}}$ is the effective sensitivity curve for each project.

%%%%%%%%%%%%%%%%%%
%%%%%%%%%%%%%%%%%%

%%%%%%%%%%%%%%%%%%
%%%%%%%%%%%%%%%%%%

Before moving to concrete examples, we make three remarks here. 
First, we note that the spectrum shapes of SIGWs in Figs.~\ref{fig:gw_prof} and \ref{fig:gw_inst_limit} should be considered as a lower bound of the GW spectrum.\footnote{
Logically speaking, we cannot exclude the possibility that the contributions in $k> k_\tmax$ cancel the other contributions from $k < k_\tmax$, which leads to a smaller GW spectrum than our results in this review. However, we cannot find any physical reason to expect such cancellations between the contributions on different scales. 
} 
In particular, the sharp cutoff on the high-frequency side of the peak is due to the UV cutoff of the power spectrum introduced in Eq.~(\ref{eq:pzeta_inst}).
In realistic situations, this can be modified as follows. If the eMD era is not long enough so that the density perturbations remain linear, we should consider the effects of an era preceding the eMD era (such as an early RD era). In this case, the sharp cutoff $k_\text{max}$ is physically replaced by an effective smooth cutoff around the inverse of the horizon scale at the beginning of the eMD era. 
See, \textit{e.g.}, \rref{Kohri:2018awv, Kumar:2024hsi} for such a treatment though outside the context of the poltergeist mechanism. In this case, the GW spectrum at $k \lesssim k_\text{max} \sim k_\text{eq,1}$ will be only mildly affected by the preceding era, where $k_\text{eq,1}$ is the inverse of the horizon scale at the beginning of the eMD era.  If, on the other hand, the MD era is long enough so that the density perturbations do become nonlinear, the perturbations on $k> k_\tmax$ produce SIGWs, and modify the spectrum shape potentially significantly. 
We will come back to the nonlinear perturbation issue below Eq.~(\ref{eq:poisson_cutoff_pzeta}). 
Because of these reasons, one should not take Eq.~\eqref{eq:Omega_GW_rough_behavior} for granted in a realistic situation and for comparison with observations, though the formula is technically valid under the specified assumptions.

Second, in the above calculation, we have implicitly assumed that the density perturbations are not diffused by considering that the mean free path of relativistic particles is sufficiently smaller than the smallest perturbation scale $1/k_\tmax$ at the GW production.
If the density perturbations are diffused before they produce SIGWs, the poltergeist GWs are suppressed~\cite{Domenech:2025bvr,Yu:2025cqu}.
Within the Standard Model of particle physics, the diffusion effect can be safely neglected in the case of $T_\R > \mathcal O(1)\,\MeV$, while it could be important even in $T_\R > \mathcal O(1)\,\MeV$ in the non-Standard Model~\cite{Yu:2024xmz,Yu:2025cqu}.

Third, although we have so far discussed the gradual transition with a constant decay rate and the instantaneous transition, the intermediate transition between the gradual and the instantaneous transition was studied in \rref{Pearce:2023kxp,Padilla:2024cbq}. 
Since any rapid physical processes involve a finite time window and the magnitude of suppression of $\Phi$ sensitively depends on the transition timescale, such a quantitative study of the effect of the finite transition timescale is important.
In the next subsection, we will see the effects of the finite transition timescale in realistic situations.

\subsection{Concrete examples}

In this subsection, we introduce concrete examples where the poltergeist mechanism works. 
In particular, we discuss how the timescale of the transition from an eMD era to an era with $w \neq 0$ is determined in concrete models.\footnote{
Other applications that we do not discuss in this review can be seen in \rref{Basilakos:2023jvp,Tzerefos:2024rgb,Allahverdi:2025btx}.
}

\subsubsection{Primordial black holes}

One example of the sudden transition is the transition from a PBH-dominated era to a RD era.
A PBH is a BH produced by the large-amplitude density perturbations in the early Universe, typically before BBN~\cite{Zeldovich:1967lct,Hawking:1971ei,Carr:1974nx,Carr:1975qj}.
If many PBHs are produced during a RD era and their lifetime is long enough, they can dominate the Universe at some point. 
At the same time, if the lifetime of the dominating PBHs is shorter than the age of the present Universe, they evaporate to radiation at some point and radiation dominates the Universe again. 

In this case, there are two factors that deviate from the instantaneous transition: 1) the finite timescale of a BH evaporation and 2) the mass distribution of PBHs. 

%%%%%%%%%%%%%%%%%
\para{Timescale of a BH evaporation}
%%%%%%%%%%%%%%%%%

Even though a BH evaporation accelerates as time goes by, the evaporation is not precisely instantaneous.
We discuss here how this fact changes the prediction of the instantaneous case. 

The evolution of a PBH mass obeys~\cite{Hawking:1974sw,Hooper:2019gtx}
\begin{align}
 \frac{\dd M_\text{PBH}}{\dd t} &= - \frac{A}{M_\text{PBH}^2} \nonumber \\
 &= -7.6 \times 10^{16} \,\text{g}\, \text{s}^{-1}\, g_{\text{H}*}(T_\text{PBH}) \left( \frac{M_\text{PBH}}{10^4 \, \text{g}} \right)^{-2},
 \label{eq:m_pbh_evo}
\end{align}
where the coefficient 
\begin{align}
  A = \frac{\pi \, \mathcal G \, g_{H*}(T_\text{PBH}) M_\text{Pl}^4}{480},
\end{align}
is introduced for compact notation below. 
$\mathcal G \simeq 3.8$ is the gray-body factor and $T_\text{PBH}$ is the Hawking temperature of the PBH, given by~\cite{Hawking:1974sw} 
\begin{align}
  T_\text{PBH} = \frac{M_\text{Pl}^2}{M_\text{PBH}} \simeq 1.05 \times 10^{9} \, \text{GeV} \, \left( \frac{M_\text{PBH}}{10^4 \, \text{g}} \right)^{-1}.
  \label{eq:t_pbh_m_pbh}
\end{align}
$g_{\text{H} *}(T_\text{PBH})$ is the spin-weighted degrees of freedom for Hawking radiation with the standard model particles with $T_\text{PBH}$~\cite{Hooper:2019gtx}:
\begin{align}
  g_{\text{H}*} (T_\PBH) \simeq \begin{cases}
  108 & (T_\PBH \gg 100 \, \text{GeV} \leftrightarrow M_\PBH \ll 10^{11}\, \text{g} ) \\
  7 & ( T_\PBH \ll 1\, \text{MeV} \phantom{w} \leftrightarrow M_\PBH \gg 10^{16}\, \text{g} )
  \end{cases}.
\end{align}
The temperature dependence of $g_{\text{H}*}$ physically means that the BH evaporation cannot efficiently produce the particles heavier than the Hawking temperature ($m \gtrsim T_\text{PBH}$).
Throughout this review, we focus on tiny PBHs with $M_\text{PBH} < 10^9\, \text{g}$, which can dominate the Universe consistently with observational constraints, and fix $g_{\text{H}*} = 108$.

From Eq.~(\ref{eq:m_pbh_evo}), the PBH mass evolves as
\begin{align}
  M_\text{PBH} = \left(3A\right)^{1/3} \left(t_\text{eva} - t\right)^{1/3} \quad  \text{for}~~t \leq t_\text{eva}, 
  \label{eq:mass-time_relation}
\end{align}
where the subscript ``eva'' denotes the value at the completion of the PBH evaporation. 
We will connect $t_\text{eva}$ to temperature in Eq.~\eqref{eq:tr_pbh_mass}.
In the case of a monochromatic PBH mass function, where all PBHs have the same mass, we can define the (effective) decay rate as
\begin{align}
  \Gamma \equiv - \frac{1}{M_\text{PBH}} \frac{\dd M_\text{PBH}}{\dd t} = \frac{1}{3\left(t_\text{eva} -t \right)}.
  \label{eq:decay_rate}
\end{align}
With this decay rate, $\Gamma \rho_\PBH$ represents the energy flow from the PBHs to radiation per unit time and volume.
Note that, for a non-monochromatic PBH mass function, we cannot use Eq.~(\ref{eq:decay_rate}). We will generalize the calculation to the non-monochromatic PBH mass case below Eq.~(\ref{eq:kr_tr}). 
From Eq.~(\ref{eq:decay_rate}), we can see that the decay rate increases in time and therefore the transition is more rapid than that in the constant $\Gamma$ case, discussed in Sec.~\ref{sec:gradual}.
We here remark that the time dependence of the decay rate requires a careful treatment because, if the time is not synchronized in all space, the decay rate depends on the space. 
To simplify the calculation, we here use the expressions in synchronous gauge (see Appendix~\ref{app:perturbation_during_trans} for the equation of motion of perturbations in a general gauge).
The concrete expressions with the time-dependent decay rate in the Newtonian gauge can be seen in \rref{Pearce:2023kxp}.

The reheating (radiation) temperature at $t_\text{eva}$ is given by 
\begin{align}
  \label{eq:tr_pbh_mass}
  T_\text{R} 
   \simeq 2.8 \times& 10^4 \, \text{GeV} \left( \frac{M_{\text{PBH,i}}}{10^4 \, \text{g}} \right)^{-3/2}  \nonumber \\
& \quad  \times\left( \frac{g_{\text{H}*}(T_\text{PBH})}{108} \right)^{1/2} \left( \frac{g_{*, \text{eva}}}{106.75} \right)^{-1/4},
\end{align}
where we have used $H(t_\text{eva})= \sqrt{g_{*,\eva}T_\R^4/(90 M_\Pl^2)} = 2/(3 t_\eva)$. 
$M_{\text{PBH,i}}$ denotes the initial PBH mass at its production, which is related to $t_\eva$ as $M_\text{PBH,i} = (3A t_\eva)^{1/3}$.
Note that $g_{\ast,\eva}$ is the relativistic effective degrees of freedom at $T = T_\R$ and is different from $g_{\text{H}*}(T_\PBH)$.
We use the results of ~\rref{Planck:2018vyg, Saikawa:2018rcs} for the temperature dependence of $g_*$.
The inverse of the horizon scale at $\eta_\eva$, denoted by $k_\eva$, can be expressed with the temperature as 
\begin{align}
   k_\text{eva} = &4.7 \times 10^{11}\, \text{Mpc}^{-1} \left( \frac{g_{*,\text{eva}}}{106.75} \right)^{1/2}  \nonumber \\
   & \quad  \times \left(\frac{g_{s*,\text{eva}}}{106.75} \right)^{-1/3} \left( \frac{T_\text{R}}{2.8 \times 10^{4}\, \text{GeV} } \right),
  \label{eq:kr_tr} 
\end{align}
where $g_{*,\text{eva}}$ and $g_{s*,\text{eva}}$ represent the values at $T = T_\text{R}$.
We have obtained this equation by using $k_\bullet = a_\bullet H_\bullet$ ($\bullet = \text{eva}, \text{eq}$) and $k_\text{eva} / k_\text{eq} = a_\text{eva} H_\text{eva} / (a_\text{eq} H_\text{eq})$ with the subscript ``eq'' meaning the value at the late-time matter-radiation equality ($z_\text{eq} \simeq 3400$).

\begin{figure}%[tbh] 
        \centering \includegraphics[width=1\columnwidth]{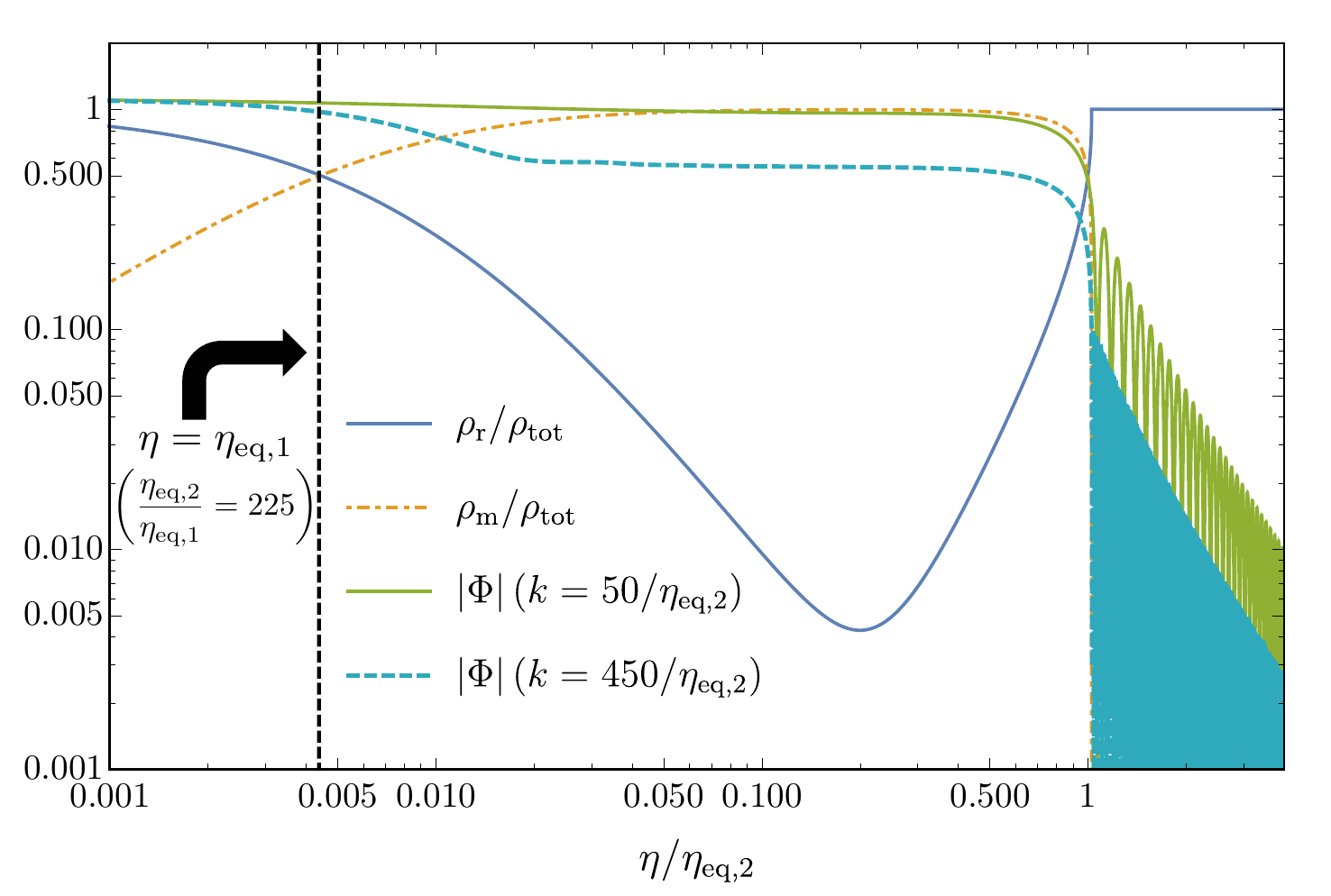}
        \caption{\justifying
      The evolution of the background energy densities and the gravitational potential.
      We take $\eta_{\eq,2}/\eta_{\eq,1} = 225$ for all the lines.
      For the latter, we normalize it to be $\Phi(x \to 0) = 10/9$ so that $\Phi \simeq 1$ for the mode that enters the horizon during the PBH-dominated era. 
      From \rref{Inomata:2020lmk}.
        }
        \label{fig:back_phi_evo}
\end{figure}

\begin{figure}%[htb] 
        \centering \includegraphics[width=1\columnwidth]{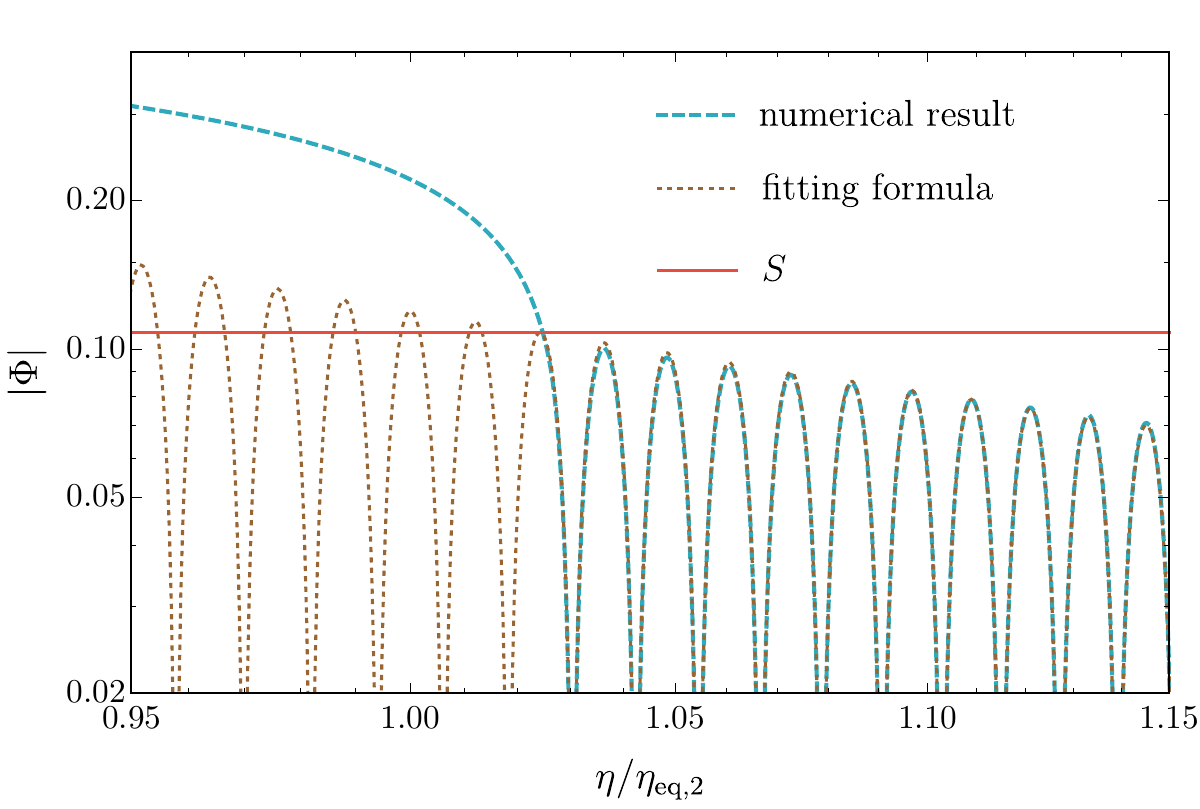}
        \caption{\justifying
        The zoom-in figure of Fig.~\ref{fig:back_phi_evo} focusing on the evolution of $\Phi$ with $k=450/\eta_\eqs$ around the transition (the cyan dashed line).
        The brown dotted line is the fitting formula Eq.~(\ref{eq:phi_osc_fit}) with $S = 0.108$ and $x_0 = 236$.
        From \rref{Inomata:2020lmk}.
        }
        \label{fig:phi_fit}
\end{figure}

Figure~\ref{fig:back_phi_evo} shows the evolution of the background and perturbations in the case of the monochromatic PBH mass.
We define $\eta_\eqf$ and $\eta_\eqs$ as the equality time of $\rho_\PBH = \rho_\rr$ at the beginning and the end of the PBH-dominated era, respectively. 
Note that we have normalized the gravitational potential as $\Phi(x \to 0) = 10/9$ during the early RD (eRD) era so that $\Phi \simeq 1$ during the PBH-dominated era for the modes that enter the horizon during that era.
We note that, if perturbations enter the horizon before the PBH-dominated era, they are suppressed until the PBH-dominated era begins (see $k= 450/\eta_\eqs$ case in Fig.~\ref{fig:back_phi_evo}). 
This suppression due to the eRD era can be caught by Eq.~(\ref{eq:Phi_plateau}). 
From Fig.~\ref{fig:back_phi_evo}, we can see that the smaller-scale gravitational potential gets more suppression during the transition.
To quantify the suppression, we define the normalization factor $S(k)$ as the amplitude of gravitational potential at the beginning of its oscillation. 
Specifically, we fit the oscillation part of $\Phi$ with Eq.~(\ref{eq:phi_formula}) modified as $x_\R \to x_0$:
\begin{align}
  \Phi_\text{osc,fit} (x,x_0) =  S(k) \left(A(x_0) \mathcal J(x,x_0) + B(x_0) \mathcal Y(x,x_0) \right),
  \label{eq:phi_osc_fit}
\end{align}
where $S$ and $x_0$ are the fitting parameters, which we determine by comparing it with the numerical result.
Figure~\ref{fig:phi_fit} compares the numerical result and Eq.~(\ref{eq:phi_osc_fit}) with the best-fit $S$ and $x_0$.
We can see that Eq.~(\ref{eq:phi_osc_fit}) fits the numerical result of the oscillation part very well.

To reduce the computational cost, let us obtain the analytical formula for the normalization factor $S$ in the following.
Similar to Eq.~(\ref{eq:phi_app_f}), we first approximate the gravitational potential during the suppression as
\begin{align}
  \frac{\Phi(t)}{\Phi_\text{plateau}} \simeq& \exp\left( - \int^t_{t_\text{i}} \dd \bar t \, \Gamma (\bar t)\right) \nonumber \\
  = & \frac{\left(3 t_\text{eva} - 3t\right)^{1/3}}{\left(3t_\text{eva} - 3 t_\text{i}\right)^{1/3}} \nonumber \\
  \simeq & \left( 1 - \frac{t}{t_\text{eva}} \right)^{1/3},
  \label{eq:phi_decay_ana}
\end{align}
where we have used $t_i/t_\text{eva} \ll 1$ and $\Phi_\text{plateau}$ is defined by Eq.~(\ref{eq:Phi_plateau}).
This approximate form of the gravitational potential is based on the assumption that the gravitational potential follows the evolution of the matter density perturbations.
We can find this assumption reasonable by looking at Eq.~(\ref{eq:phidot_eq}).
For the perturbations that enter the horizon during the eMD era, the spatial derivative terms become larger than the time derivative terms until they start to oscillate. 
This leads to the following approximate form:
\begin{align}
    \Delta \Phi \simeq \frac{3}{2} \mathcal{H}^2\left( \frac{\bar \rho_\text{m}}{\bar \rho_\text{tot}} \delta_\text{m} + \frac{\bar \rho_\text{r}}{\bar \rho_\text{tot}} \delta_\text{r} \right).
\end{align}
This is nothing but the standard Poisson equation. 
During the transition, the evolution of the gravitational potential is mainly determined by the matter density perturbations $\delta_m \bar\rho_m$ until the gravitational potential starts oscillations. This is because $\delta_m$ can grow substantially, $\delta_m \gg \delta_r$, whereas $\bar\rho_m$ and $\bar \rho_r$ are comparable with each other around the transition. Therefore, we can approximate the above equation further:
\begin{align}
    \Delta \Phi \simeq & \frac{3}{2} \mathcal{H}^2 \frac{\bar \rho_\text{m}}{\bar \rho_\text{tot}} \delta_\text{m} \nonumber \\
    \simeq & \frac{a^2 \bar \rho_m }{2 M_\text{P}^2 } \delta_m,
\end{align}
where we have used the Friedmann equation at the second equality. 
Since the transition occurs within $\sim \mathcal O(1)$ e-folds, $a$ and $\delta_m$ do not change much during the transition, and therefore the evolution of $a^2 \delta_m \bar\rho_m$ mainly comes from the time-dependence of $\bar\rho_m$, which motivates us to relate $\Phi \propto \bar\rho_m$. In terms of Fourier modes, we have
\begin{align}
    \Phi_\bfk \simeq -\frac{a^2 \bar{\rho}_{m}}{2 M_\text{P}^2 k^2} \delta_{m,\bfk}.
\end{align}
By definition, the time dependence of $\bar{\rho}_m$ is $\bar \rho_m \propto a^{-3} e^{-\int \dd \bar t \Gamma(\bar t)}$. As explained above, the time dependence of $a$ in a short time interval around the final part of evaporation can be neglected.  This leads to $\Phi \propto \bar{\rho}_m \propto e^{-\int \dd \bar t \Gamma(\bar t)}$.  Thus, we obtain Eq.~(\ref{eq:phi_decay_ana}), where the normalization follows from the initial condition.

However, the gravitational potential finally decouples from the matter density perturbations.
In other words, although we have neglected terms like $\mathcal{H}\Phi'$ and $\mathcal{H}^2\Phi$ with respect to $\Delta \Phi$ in the above discussion, that is no longer valid in the oscillating regime after the transition from the eMD era to the RD era.
Mathematically, the oscillation of gravitational potential in the RD regime is caused by the last term in Eq.~(\ref{eq:phi_eq_rd_era}). 
Given this, the necessary condition for Eq.~(\ref{eq:phi_decay_ana}) to be valid is $|\ddot \Phi| \ll \frac{k^2}{3a^2}|\Phi|$. 
From this, we can expect that $\Phi$ starts to oscillate when or before this inequality is violated:
\begin{align}
  \left| \frac{\ddot \Phi}{\Phi} \right|_{t=t_\text{dec}} \simeq  \frac{2}{9\left(t_\text{dec} - t_\text{eva}\right)^2} \lesssim \frac{k^2}{3a^2},
  \label{eq:dec_cond}
\end{align}
where $t_\text{dec}$ is the time when the gravitational potential decouples from the matter density perturbation.

Based on this expectation, we define the lower bound of the normalization factor $S(k)$ as 
\begin{align}
  S_\text{low}(k) &\equiv  \left( 1 - \frac{t_\text{dec}}{t_\text{eva}} \right)^{1/3} \Phi_\text{plateau}(x_\eqf) \nonumber \\
&\simeq \left( \frac{\sqrt{6}}{k \eta_\text{eva} } \right)^{1/3} \Phi_\text{plateau}(x_\eqf) ,  \label{eq:s_low} 
\end{align}
where we have used Eqs.~(\ref{eq:phi_decay_ana}) and (\ref{eq:dec_cond}), and the relation $\eta a = 3t$, valid during the PBH-dominated era.

\begin{figure}[htb] 
  \centering \includegraphics[width=1\columnwidth]{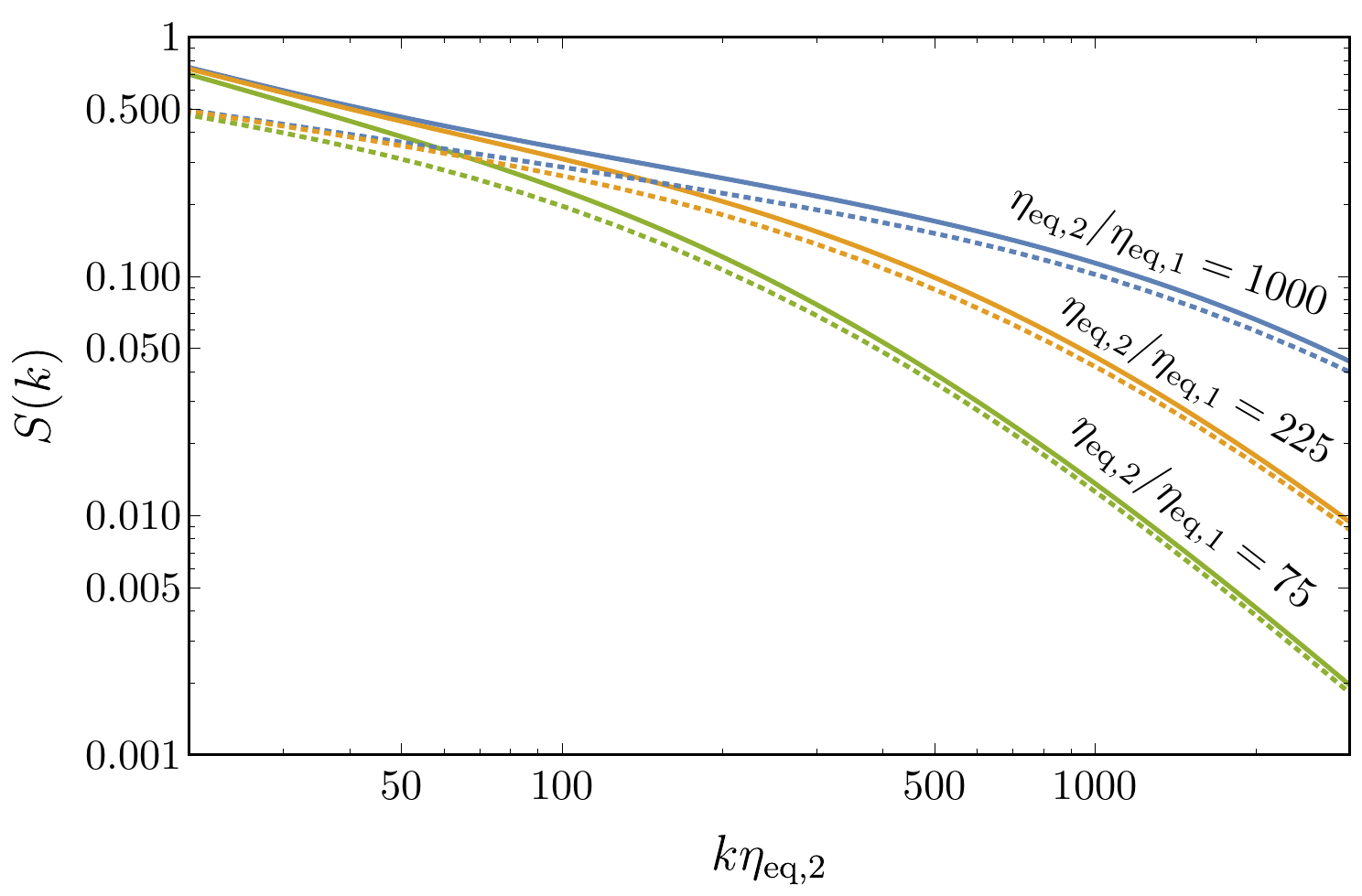}
        \caption{\justifying
        The wavenumber dependence of the normalization factor. 
        The solid and dashed lines are the numerical results and the approximate lower bound Eq.~(\ref{eq:s_low}), respectively.
      From \rref{Inomata:2020lmk}.
        }
        \label{fig:s_low}
\end{figure}

Figure~\ref{fig:s_low} compares the numerical result of $S$ and $S_\text{low}$.
We can see that $S_\text{low}$ fits the numerical result well especially for the small scale perturbations, which dominantly contribute to the poltergeist mechanism.

%%%%%%%%%%%%%%%%%
\para{Non-monochromatic PBH mass function}
%%%%%%%%%%%%%%%%%

Although we have assumed the monochromatic mass function for the PBH distribution so far, a PBH mass spectrum is generally non-monochromatic. 
When it is non-monochromatic, each PBH has a different mass and lifetime. 
This leads to a more gradual transition from the PBH-dominated era to a radiation era, which suppresses the enhancement of the poltergeist mechanism. 
In the following, we describe the outline of the calculation. See \rref{Inomata:2020lmk} for more details. 

We here parametrize the initial PBH mass function as 
\begin{align}
\rho_{\text{PBH,i,tot}} =& \int \rho_\text{PBH,i}(M_\text{PBH,i}) \text{d} \ln M_\text{PBH,i} 
\nonumber \\
\simeq &  \int \rho_\text{PBH,i}(M_\text{PBH,i}(\eta_\text{eva})) \text{d} \ln \eta_\text{eva},
\end{align}
where $M_\text{PBH,i}$ is the initial PBH mass at its production time ($t_i$ or $\eta_i$), $\rho_{\text{PBH,i,tot}}$ is the total energy density of PBHs and $\rho_{\text{PBH,i}}(M)$ is the energy density for a log bin $\ln M$. 
In the second line, we have used $t_\eva \gg t_i$ and $t_\eva^{1/3} \propto \eta_\eva$, valid during a MD era, which leads to $M_\text{PBH,i} \simeq (3A)^{1/3} t_\eva^{1/3} \propto \eta_\eva$.

We can express the evolution of the PBH energy density as
\begin{align}
\rho_\text{PBH} (\eta) 
\simeq & \int \rho_\text{PBH,i}(M_\text{PBH,i}(\eta_\text{eva})) \left( \frac{a(\eta_\text{i})}{a(\eta)} \right)^3  \nonumber \\
& \times  \left( 1 - \left(\frac{ \eta }{\eta_\text{eva}} \right)^3 \right)^{\frac{1}{3}} \Theta (\eta_\text{eva} - \eta )  \text{d} \ln \eta_\text{eva}, 
\label{rho_PBH_convolution} 
\end{align}
where we have used that the PBH number density is $\propto a^{-3}$ and each PBH mass evolves as Eq.~(\ref{eq:mass-time_relation}) (with $t \propto \eta^3$ during the PBH-dominated era).
From this, we can calculate the time dependence of the decay rate through $\Gamma(\eta) = -a^{-1}(\eta) \dd \ln [a^3(\eta) \rho_\PBH(\eta))]/\dd \eta$.
Using this, we calculate the evolution of the other quantities.

\begin{figure}%[tbh!]
\begin{center}
  \includegraphics[width = 1. \columnwidth]{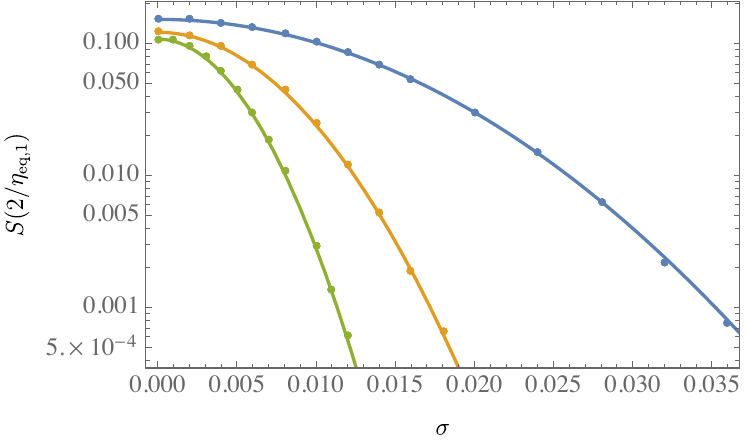}
      \caption{\justifying
        The $\sigma$ dependence of the normalization factor.
        The blue (top), orange (middle), and green (bottom) dots are the numerical results with $\eta_\text{eq,2}/\eta_\text{eq,1} = 75$, 150, and 225, respectively.
        The solid lines are the fitting formula, Eq.~(\ref{eq:sup_factor_with_sigma}), with $c^2 = 0.18$.
        From \rref{Inomata:2020lmk}.
}
\label{fig:sigma_dependence}
      \end{center}
\end{figure}

To characterize the PBH mass spectrum, we assume that the PBH mass spectrum has a log-normal peak. 
Since $M_\text{PBH,i} \propto \eta_\eva$, we can express the PBH mass function with the evaporation time $\eta_\eva$ as 
\begin{align}
\rho_\text{PBH,i}(M_\text{PBH,i}(\eta_\text{eva})) = \frac{\rho_\text{PBH,i} }{\sqrt{2\pi} \sigma } \exp\left( - \frac{\left(\text{ln}(\eta_\text{eva}/\eta_\text{eva,0})\right)^2}{2 \sigma^2} \right),
\label{eq:pbh_mass_func_with_sigma}
\end{align}
where $\eta_{\eva,0}$ corresponds to the evaporation time of the mean mass PBH, and $\sigma$ denotes the dimensionless width of the distribution. 

Figure~\ref{fig:sigma_dependence} shows the $\sigma$-dependence of the normalization factor. 
The dots in the figure show the numerical results, which can be well-fitted with 
\begin{align}
S(k, \sigma) = & S(k) \exp \left( - \left(c \sigma k \eta_\text{eq,2}\right)^2 \right), 
\label{eq:sup_factor_with_sigma}
\end{align}
where $c$ is a fitting parameter. 
In Fig.~\ref{fig:sigma_dependence}, we can see that the large $\sigma$ leads to large suppression.
This is because the broader PBH mass spectrum leads to a more gradual transition, which results in a more suppression of the gravitational potential before its oscillation.

\begin{figure}%[tbh!] 
  \centering \includegraphics[width=0.9\columnwidth]{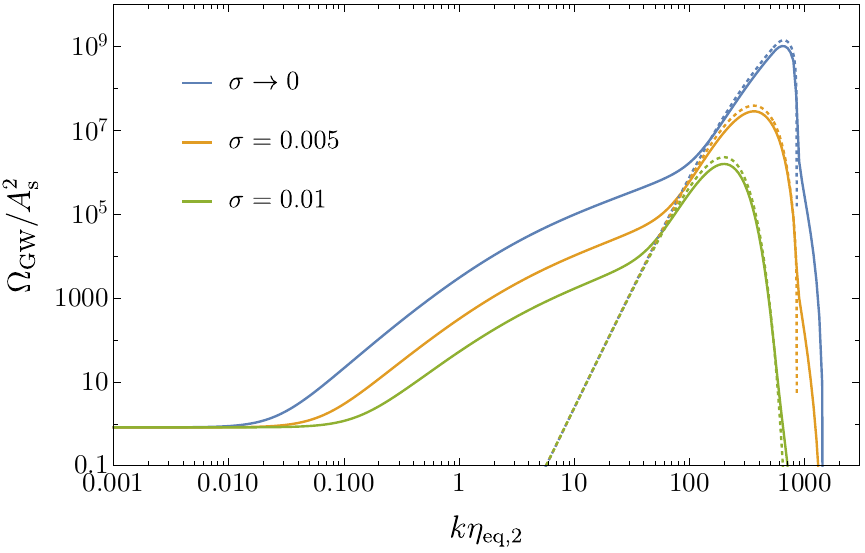}
        \caption{\justifying
        The $\sigma$-dependence of the GW spectrum. 
        We take $\sigma \to 0$ for the top line, $=0.005$ for the middle line, and $=0.01$ for the bottom line. We take $\eta_{\text{eq},2}/\eta_{\text{eq},1} = 225$ for all the lines. 
        The dotted lines are the approximate formula for the resonant peak contribution. 
        From \rref{Inomata:2020lmk}.
        }
        \label{fig:gw_profile_with_sigma}
\end{figure}

Using the fitting formula Eq.~(\ref{eq:sup_factor_with_sigma}), we calculate GWs. 
Figure~\ref{fig:gw_profile_with_sigma} shows the $\sigma$-dependence of the GW spectrum with $\mathcal P_\zeta (k) = A_\ss \Theta(k_\NL-k)$, where $k_\NL$ is the nonlinear scale where the matter (PBH) density perturbation becomes $\delta_\mm = 1$ at the PBH evaporation. 
We numerically obtain $k_\NL$ by solving the equation of motion of the perturbations. 
The solid lines are the numerical results, while the dotted lines are based on an approximate analytical formula in \rref{Inomata:2020lmk} for the resonance part of the GW spectrum. Although we refer the reader to the original reference for the details of this approximation, the basic assumption there is that the product of the smooth power spectrum $\mathcal{P}_\zeta$ and the suppression factor $S(k, \sigma)^2$ constitutes the effective power spectrum that can be locally approximated by power law.  The approximate analytic formula for the GW spectrum for the power-law $\mathcal{P}_\zeta(k)$ (not specific for the PBH application) is presented in Appendix~\ref{app:analytic_formulas}.

\begin{figure}%[ht] 
  \centering \includegraphics[width=1\columnwidth]{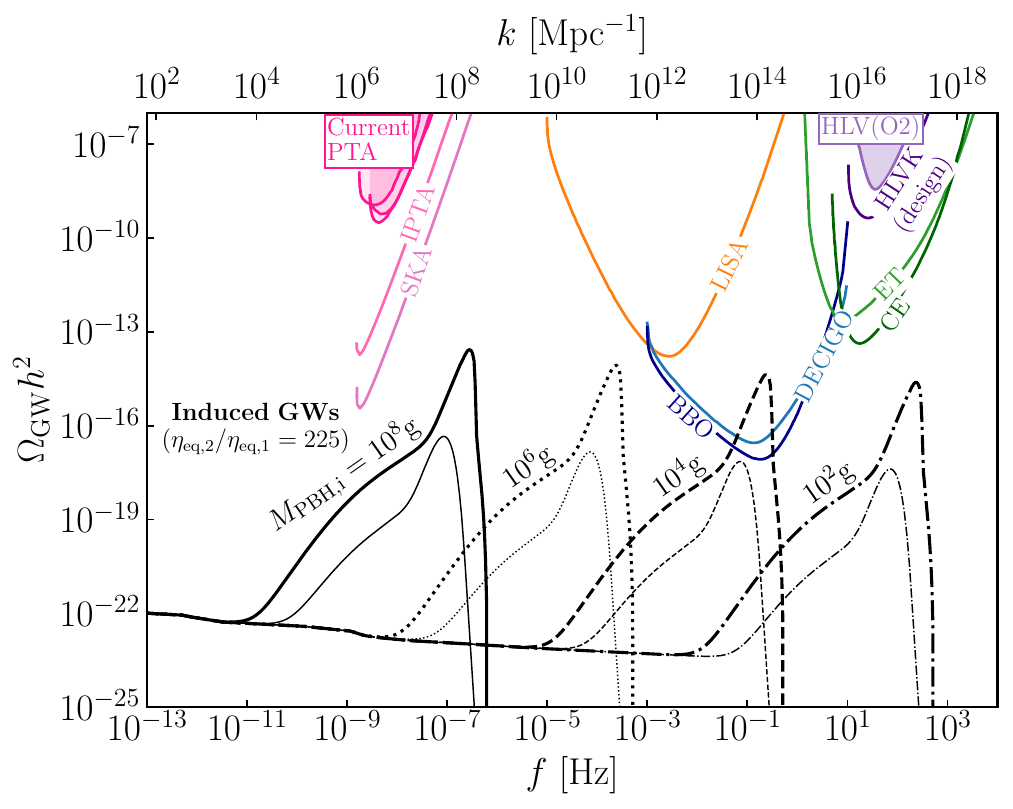} 
        \caption{\justifying
        GW spectrum with different values of the mean PBH mass and the width $\sigma$.
        The curvature power spectrum is given by Eq.~(\ref{eq:pzeta_inst}) with $k_\tmax = k_\NL$.
        The black thick and thin lines show the GW spectrum with $\sigma \rightarrow 0$ and $= 0.01$, respectively.
      From \rref{Inomata:2020lmk} with updated sensitivity curves.
        }
        \label{fig:gw_spectrum_erd}
\end{figure}

Figure~\ref{fig:gw_spectrum_erd} shows the GW spectrum with the sensitivities of the current and future GW experiments.\footnote{
In this scenario, GWs are also induced by the perturbations that produce PBHs during the eRD era, and are also induced through the Hawking evaporation of the PBHs and the mergers of the PBHs before they evaporate, though it is difficult for current GW projects to measure them because of their high frequencies~\cite{Inomata:2020lmk,Cheek:2022mmy}.
}
We consider different values of the mean PBH mass and the width $\sigma$.
In the figure, we take Eq.~(\ref{eq:pzeta_inst}) for the curvature power spectrum with $k_\tmax = k_\NL$.
We can see that the PBH mass spectrum must be sharply peaked ($\sigma < 0.01$) for the GWs to be observed by future experiments, such as BBO~\cite{phinney2003big,Crowder:2005nr,Corbin:2005ny,Harry:2006fi} and DECIGO~\cite{Seto:2001qf,Kawamura:2006up,Yagi:2011wg}. 
Although we here used the approximate formula Eq.~(\ref{eq:sup_factor_with_sigma}) to calculate GWs, \rref{Pearce:2025ywc} numerically discusses the effects of the finite-width PBH mass function without using the approximate formula.

\begin{figure}%[ht] 
  \centering \includegraphics[width=1\columnwidth]{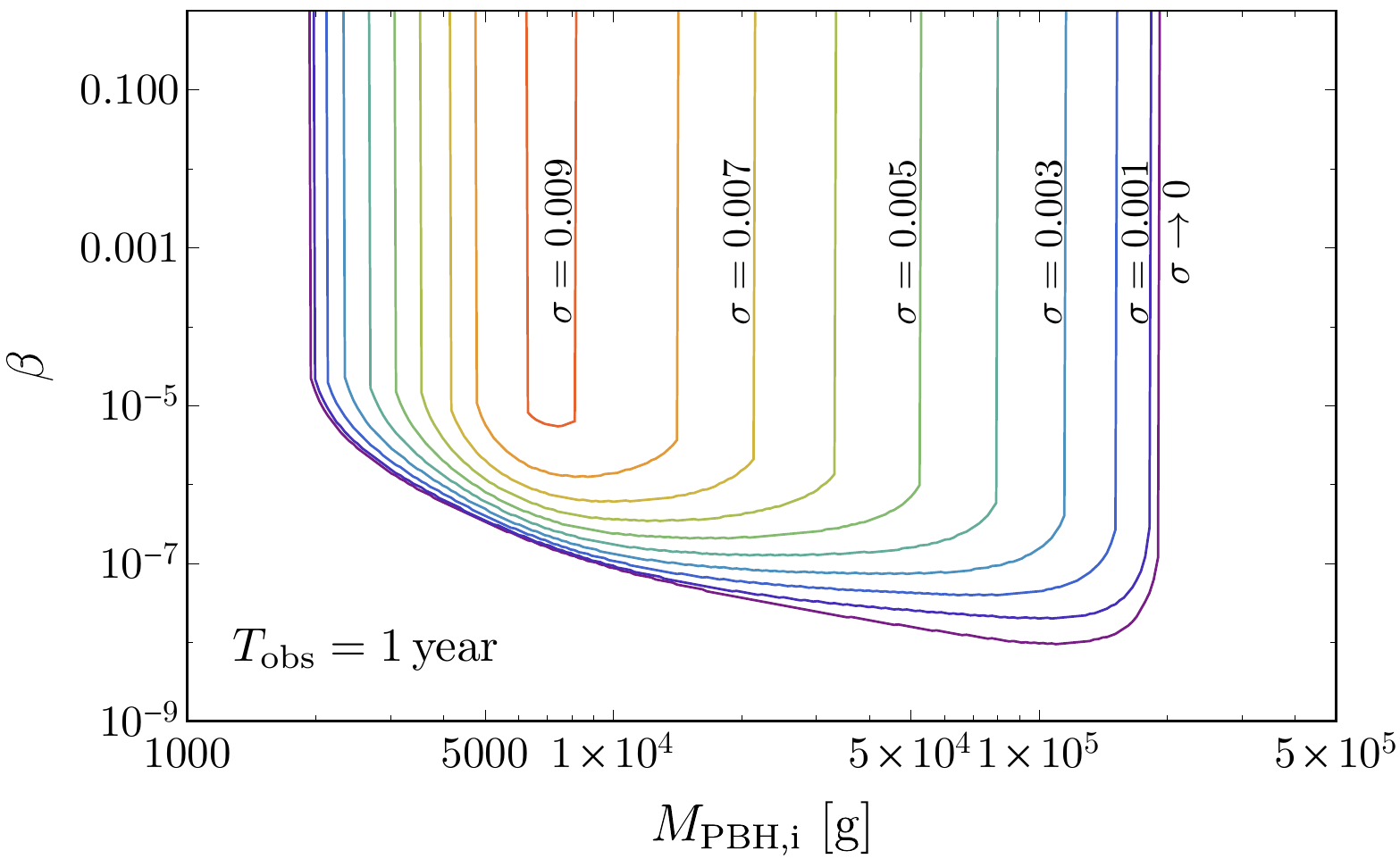} 
        \caption{\justifying
        The PBH abundance that DECIGO can explore in a one-year observation with the signal-to-noise ratio being unity.
        Each line is for a PBH mass spectrum with a different $\sigma$.
        The regions above the lines can be probed by DECIGO.
        The difference of $\sigma$ for adjacent lines is $\Delta\sigma = 0.001$.
       From \rref{Inomata:2020lmk}.
        }
        \label{fig:beta_const}
\end{figure}

This kind of enhanced GWs can be used to probe the abundance of the tiny PBHs that have evaporated by now. 
To characterize the PBH abundance, we here define the initial PBH fraction as 
\begin{align}
    \beta \equiv \frac{\rho_{\PBH,\ii}}{\rho_{\tot,\ii}},
\end{align}
where the subscript i again means the value at the PBH production. 
Note that $\beta$ determines the length of the PBH-dominated era and the PBH mass determines the peak scale of GWs. 
Figure~\ref{fig:beta_const} shows future constraints on the PBH abundance from DECIGO.
See \rref{Inomata:2020lmk} for a more detailed analysis. 

Investigating PBH abundance might also provide insights into the supersymmetric models that predict PBH production via Q-balls~\cite{Flores:2023dgp}.
In addition, the poltergeist GWs could possibly explain the recent PTA results~\cite{Basilakos:2023xof,Zhu:2023gmx,Bhaumik:2023wmw,Basilakos:2024diz}.

%%%%%%%%%%%%%%%%%
\para{Poisson fluctuations of PBH distribution}
%%%%%%%%%%%%%%%%%

We have so far only focused on the adiabatic curvature perturbations.
However, due to the randomness of the PBH formation process, the PBH spatial distribution can be affected by the Poisson fluctuations~\cite{Afshordi:2003zb}.
The energy density of PBHs at their production can be expressed as
\begin{align}
  \rho_{\text{PBH,i}}(\bfx) = M_{\text{PBH,i}} \bar n_{\text{PBH,i}} (1 + \delta_\PBH(\bfx)), 
\end{align}
where $M_{\text{PBH,i}}$ is the PBH mass at the production, $\bar n_{\text{PBH,i}}$ is the spatially averaged PBH number density, and $\delta_\PBH$ is the density (Poisson) fluctuations.
For convenience, we here define $k_\UV$ through $\bar n_{\text{PBH,i}} = 3 k_\UV^{3}/(4\pi)$, where $k_\UV$ corresponds to the inverse of the mean comoving distance between PBHs.
Note that, on the scale $k> k_\UV$, the fluid picture of the PBH energy density breaks down and therefore $k<k_\UV$ is a necessary (but, as we will see below, not a sufficient) condition for the perturbation theory. 
Then, we can express the power spectrum of the Poisson fluctuation $\delta_\PBH$ at the PBH production time as
\begin{align}
    \label{eq:poisson_cutoff_pzeta}
  \mathcal P_{\delta_\PBH}(k) &= \frac{k^3}{2\pi^2} \bar n_{\text{PBH,i}}^{-1} \nonumber \\ 
  &= \frac{2}{3\pi} \left(\frac{k}{k_\UV}\right)^3 \Theta(k_\UV -k),
\end{align}
where we have introduced the cutoff at $k_\UV$ to restrict ourselves to the regime where the fluid picture of PBHs is valid.
The power spectrum of these Poisson fluctuations grows $\propto k^3$ and can be larger than the adiabatic power spectrum on small scales.

The PBH Poisson fluctuations can be regarded as isocurvature perturbations. 
There are several works studying phenomena associated with the PBH isocurvature perturbations in the context of induced GWs.
\rref{Papanikolaou:2020qtd} focused on the second-order tensor perturbations induced by the PBH isocurvature perturbation during the PBH-dominated era and discusses when the backreaction from the induced tensor perturbations cannot be neglected. 
As a side note, the induced tensor perturbations during the PBH-dominated era can be different from GWs after the PBH evaporation, as we can see in Fig.~\ref{fig:gw_sup_result}.
\rref{Domenech:2020ssp} studied the poltergeist mechanism due to the PBH isocurvature perturbations for the first time (with the non-zero PBH spin effects later discussed in \rref{Domenech:2021wkk}).
\rref{Papanikolaou:2022chm} discussed the PBH isocurvature contribution in the extended (non-monochromatic) PBH mass function.
\rref{Bhaumik:2022pil, Bhaumik:2022zdd} showed that the GW spectrum can have double peaks with the poltergeist mechanism--one peak is from the adiabatic perturbations and the other is from the PBH isocurvature perturbations.
The detectability of the double peaks spectrum was studied in \rref{Paul:2025kdd}.
The poltergeist-enhanced GWs from the PBH isocurvature perturbations were also studied in the models for dark matter production from PBH Hawking evaporation~\cite{Samanta:2021mdm,Borah:2022vsu,Barman:2022pdo,Chen:2023lnj,Borah:2024lml} and dark matter as PBH evaporation remnant~\cite{Domenech:2023mqk}. 
Such GWs from the PBH isocurvature perturbations were also studied as a probe of primordial non-Gaussianities that deviate the spatial distribution of PBHs from Poisson statistics through the clustering effect~\cite{Papanikolaou:2024kjb,He:2024luf}.
See also \rref{Gross:2024wkl,Gross:2025hia} for the comparison between GWs from PBH isocurvature perturbations and the other sources, such as Hawking evaporation and the gravitational scattering of the inflaton. 
The poltergeist GWs with the mixed initial condition (adiabatic + isocurvature) were discussed in \rref{Zeng:2025tno}. 
The poltergeist GWs from isocurvature contributions were also discussed in the context of modified gravity~\cite{Papanikolaou:2025ddc}, leptogenesis~\cite{Ghoshal:2023fno,Borah:2024qyo}, the general background (general $w$ era)~\cite{Domenech:2024wao}, runaway inflation models~\cite{Dalianis:2021dbs}, the two-stage inflationary scenario~\cite{Wang:2025lti}, the memory burden effect~\cite{Barman:2024iht, Balaji:2024hpu, Bhaumik:2024qzd}, the PBH reformation~\cite{Kim:2024gqp}, and the combination with cosmic-string-induced GWs~\cite{Borah:2023iqo,Datta:2024bqp}.

However, we need to be careful about the fact that the larger density perturbations become nonlinear $|\delta| = 1$ earlier during the PBH-dominated era. 
We note that many references that focus on the induced GWs from PBH isocurvature perturbations use the linear perturbation theory in the regime of $|\delta| > 1$~\cite{Domenech:2020ssp,Domenech:2021wkk,Bhaumik:2022pil,Bhaumik:2022zdd,Paul:2025kdd,Barman:2022pdo,Borah:2022vsu,Borah:2024lml,Samanta:2021mdm,Chen:2023lnj,Domenech:2023mqk,Papanikolaou:2024kjb,He:2024luf,Gross:2024wkl,Borah:2024qyo,Ghoshal:2023fno,Domenech:2024wao,Dalianis:2021dbs,Wang:2025lti,Balaji:2024hpu,Barman:2024iht,Bhaumik:2024qzd,Kim:2024gqp,Borah:2023iqo,Datta:2024bqp}.
To obtain reliable results in perturbation theory, we must set the UV cutoff as $k_\NL$ (the scale where $|\delta| = 1$) at the PBH evaporation.
Note that, in the presence of a PBH-dominated era, we find $k_\NL < k_\UV$ because of the density perturbation growth during that era.
Unlike the density perturbations, the gravitational potential is constant on subhorizon scales and remain smaller than unity even if we use the linear perturbation theory in the regime of $|\delta|> 1$. 
We however stress that this does not mean that the linear perturbation theory is reliable in $|\delta| > 1$.
Once the density perturbation becomes nonlinear, the linear perturbation theory does not give a reliable prediction on the evolution of all the linear perturbations, including the gravitational potential.
One might think that we can take a time-dependent UV cutoff by imposing $\delta(t)|_{k_\UV(t)} = 1$.
However, once the density perturbations on some scales become $\delta > 1$, the perturbation theory breaks down. 
The evolution of density fluctuations after that can be followed only with the nonperturbative methods. 
Even if the perturbation theory finds $\delta$ decreasing after it becomes $\delta > 1$ and finally becoming $\delta < 1$ on some scales, that behavior is unreliable because the decrease of $\delta$ depends on the perturbation theory when $\delta > 1$ on those scales. 
Furthermore, as we can see in Sec.~\ref{sec:gradual} (on the gradual transition case), the final GW spectrum cannot be calculated correctly unless we follow the tensor perturbations until the eMD era ends (until the source decouples from the tensor perturbations).
This means that, for the scales where the density perturbations become $\delta > 1$ before the end of the eMD era, we cannot get a reliable result of the GW spectrum even if we focus on the contribution of the perturbations before they become $\delta > 1$.

Let us here roughly estimate the upper bound on the induced GWs within the valid linear perturbation theory.
For the adiabatic perturbations, we have considered the scale-invariant power spectrum in $k < k_\NL$. 
We here extend the power spectrum on $k>k_\NL$ scale-dependently with the condition that $\mathcal P_\delta(k) = 1$ at the PBH evaporation.
This is a toy situation, but it gives an estimate of the maximum GWs that can be calculated with the linear perturbation theory.
In this situation, from the Poisson equation (Eq.~(\ref{eq:pertb_eq_00_new2})), we can roughly approximate the power spectrum of the gravitational potential at the PBH evaporation as
\begin{align}
  \label{eq:p_phi_poisso}
  \mathcal P_\Phi(k,\eta_\R) \simeq A_\Phi \left( \Theta(k_\NL - k) + \left(\frac{k}{k_\NL} \right)^{-4} \Theta(k-k_\NL) \right), 
\end{align}
where $A_\Phi$ is the amplitude of $\mathcal P_\Phi$ on $k_\NL$ at the PBH evaporation.
As we have seen in Eq.~(\ref{eq:Omega_GW_rough_behavior}), the enhancement of $\Omega_\GW$ is $\propto k_\tmax^7$ for the scale invariant power spectrum.
By taking into account the scale dependence $\mathcal P_\Phi \propto k^{-4}$ in $k>k_\NL$, the relation $\mathcal P_h \propto \mathcal P_\Phi^2$, and the scale dependence of the suppression factor Eq.~(\ref{eq:s_low}), we can approximate $\Omega_\GW \propto k^{-7/3}$ in $k > k_\NL$.
This means that the peak of $\Omega_\GW$ should be located around $k_\NL$ even for the power spectrum Eq.~(\ref{eq:p_phi_poisso}).
From this, we can conclude that, if we restrict ourselves to the valid linear perturbation theory, the Poisson fluctuation contributions can be important only when we consider a short eMD era at the end of which the adiabatic density perturbations, whose curvature power spectrum is (almost) scale-invariant, are sufficiently smaller than unity at the PBH evaporation.
This means that, in our fiducial setup with the curvature power spectrum of Eq.~(\ref{eq:pzeta_inst}) and $\eta_{\eq,2}/\eta_{\eq,1} = 225$, the Poisson fluctuation contribution cannot surpass the adiabatic perturbation contribution (Fig.~\ref{fig:gw_spectrum_erd}) unless the perturbation theory breaks down.

Imposing a hard cutoff at $k = k_\text{NL}$ and neglecting all the contributions from $k > k_\text{NL}$, one can conservatively estimate the lower bound of the GW spectrum. In reality, however, there are contributions from nonlinearly growing modes $k > k_\text{NL}$. The extension beyond $k_\text{NL}$ with reduced $\mathcal{P}_\Phi (k)$ such that it saturates $\mathcal{P}_\delta (k) = 1$ at the PBH evaporation time (as in Eq.~\eqref{eq:p_phi_poisso}) is a prescription to include some of the additional contributions while still remaining in the linear regime. Ideally, the nonlinear contributions should be included by using nonlinear computational methods such as lattice simulation. See Refs.~\rref{Zeng:2025cer, Zeng:2025tno} for lattice simulations of the induced GWs. 

\subsubsection{Other compact objects}

The poltergeist mechanism could enhance the GW production even for the evaporation of other objects, such as Q-balls and oscillons.

Q-balls are non-topological solitons produced through a coherent rotation of a U(1) complex scalar field in its field space~\cite{Coleman:1985ki}. 
They arise when the power of the scalar potential is less than quadratic, which allows stable and localized field configurations whose phase rotates in time.
The Q-ball configuration minimizes energy at fixed (Q-)charge, making them energetically favorable. 
The coherent rotation, required for Q-ball formation, can be initiated by the Affleck-Dine (AD) mechanism~\cite{Affleck:1984fy}, which explains the origin of the matter-antimatter asymmetry in the Universe.
In particular, the flat directions in supersymmetric theories naturally provide a platform where the AD mechanism works.

Q-balls can form after the inflation, behave as non-relativistic matter, and dominate the Universe before they decay to radiation, similar to PBHs. 
The timescale of the Q-ball evaporation depends on the model.
\rref{White:2021hwi} first discussed the poltergeist mechanism by assuming the instantaneous transition from a Q-ball-dominated era to a RD era. 
\rref{Yu:2025jgx} also discussed the poltergeist mechanism with the same assumption (instantaneous Q-ball evaporation) from the aspect of GW anisotropies associated with non-Gaussian distribution of Q-balls.
\rref{Kasuya:2022cko} revisited the assumption of the instantaneous Q-ball evaporation. 
They focused on the delayed-type Q-ball scenario and found that the suppression during the transition is important.

Later, \rref{Pearce:2025ywc} studied thin-wall, thick-wall, and delayed Q-balls, and compared them with the PBH case.
The mass of the Q-ball evolves as~\cite{Kasuya:2022cko,Pearce:2025ywc}
\begin{align}
  M_Q(t) = M_{Q,i} \left( 1 - \frac{t}{t_\eva} \right)^{n},
\end{align}
where $M_{Q,i}$ is the initial Q-ball mass at its production and $t_\eva$ here means the time at the completion of the Q-ball evaporation.
The $n$ determines the speed of the evaporation and is $n= 3$ for the thin-wall, $=1$ for the thick-wall, and $=3/5$ for the delayed Q-balls.
For PBHs, we find $n=1/3$ from Eq.~(\ref{eq:mass-time_relation}).
Since the lower $n$ leads to more rapid increase of the decay rate around the completion of the evaporation, the above Q-balls lead to more gradual transition than PBHs. 
This means that the amount of the poltergeist GWs in the delayed Q-ball case (the most rapid transition among the above three) gets more suppressed than in the PBH case.
\rref{Pearce:2025ywc} additionally took into account the effects of the finite width of the Q-ball (and PBH) mass distribution.

\rref{Kawasaki:2023rfx} studied the effects of the nonlinear perturbations on the scales where the density perturbations become $|\delta| > 1$ during a (delayed-type) Q-ball-dominated era. 
They partially utilize the results of N-body simulations to obtain the evolution of the density perturbations in the regime of $|\delta| > 1$ during the Q-ball-dominated era.
The complete analysis of the evolution of $|\delta| > 1$ during the transition is left for future study.

Apart from Q-balls, the poltergeist mechanism could be caused by oscillons, which are long-lived and spatially-localized configurations of an oscillating scalar field~\cite{Bogolyubsky:1976yu,Gleiser:1993pt,Copeland:1995fq,Kasuya:2002zs}.
\rref{Lozanov:2022yoy,Sui:2024grm} studied the poltergeist-enhanced GWs by assuming the instantaneous transition from an oscillon-dominated Universe to a RD era.\footnote{
See also \rref{Lozanov:2023aez,Lozanov:2023knf} for GWs induced by soliton isocurvature perturbations without the poltergeist mechanism.
}

\subsubsection{Kinematical blocking}

The quasi-instantaneous change of the equation of state necessary for the poltergeist mechanism can arise not only from a population of localized objects as we have reviewed above but also from homogeneous fields as we discuss in this and the following subsubsections. In the former case, we saw that, practically, there are various impeding sources such as the distribution of mass, spin, and charge, which make the transition slower.  An advantage of the homogeneous mechanism is that it is free of this problem. In this sense, it is a more robust underlying source for the poltergeist mechanism.

We explain a toy model\footnote{
The stability of the mass hierarchy against quantum corrections is to be explained in a complete model. 
} that can realize a sudden transition from an eMD era to a RD era by three real scalar fields: the matter-dominating field $\phi$, the daughter field $\chi$, and the trigger field or triggeron $\tau$.
The Lagrangian is given by 
\begin{align}
\mathcal{L}=& -\frac{1}{2}  \partial^\mu \phi \partial_\mu \phi  -\frac{1}{2}  \partial^\mu \chi \partial_\mu \chi  -\frac{1}{2}  \partial^\mu \tau \partial_\mu \tau  -V, \\
V=& \frac{1}{2} M^2 \phi^2 + \frac{1}{2}m^2 \tau^2 + \frac{\lambda}{4} \tau^2 \chi^2 + \frac{c}{2}M \phi \chi^2,
\end{align}
where $M$ and $m (\ll M)$ are the mass of $\phi$ and $\tau$, respectively, and $\lambda$ and $c$ are dimensionless coupling constants. 
$\phi$ first dominates the Universe, which leads to an eMD era. 
Then, $\phi$ decays to $\chi$ with the decay rate:
\begin{align}
\Gamma = \frac{c^2 M}{32 \pi} \sqrt{1 - \frac{m_{\chi,\text{eff}}^2}{(M/2)^2}}\Theta\left(M^2-4m_{\chi,\text{eff}}^2\right),
\end{align}
where $m_{\chi,\text{eff}}^2 = \langle \lambda \tau^2 /2 \rangle$.
The point is that the mass of $\chi$ depends on the field value of $\tau$.
We consider the situation where $\tau$ initially has a large field value $\tau_0$, which leads to $m_{\chi,\text{eff}} > M/2$. 
Until $m > H$, $\tau$ is still on the potential.
Let us assume a positive initial field value of $\tau$ without loss of generality.
During this phase, the decay of $\phi$ into $\chi$ is kinematically blocked. 
After a while, the Hubble parameter decreases to satisfy $m \gtrsim H$ and $\tau$ starts to roll down the potential.
When $\tau$ crosses the critical value $\tau_c = M/\sqrt{2\lambda}$, the decay channel from $\phi$ to $\chi$ opens.
We consider the case where the initial field value $\tau_0$ is much larger than $\tau_c$, which leads to the sudden opening of the decay channel.
At the same time, $\tau_0$ must be smaller than $M_\Pl$ to be subdominant until it starts to oscillate when $H \sim m$.
In addition, the typical decay rate after the decay channel opens must be larger than the Hubble parameter at the time, which leads to the condition $c^2 M \gg m$.
Once these conditions are satisfied, this model can realize a sudden reheating~\cite{Inomata:2019ivs}.

Let us suppose that $\phi$ is the inflaton. Since the field $\tau$ has fluctuations $\delta \tau$ independent of the dominant field fluctuation $\delta \phi$, we need to be careful about the non-Gaussianity produced by $\delta \tau$. 
We can express the time evolution of $\tau$ as $\tau = \tau_0 \sin(mt)/(mt)$.
In the following, we estimate the magnitude of the non-Gaussianity in this model.
We here assume the instantaneous reheating at $\tau_c$. 
Then, the decay time is given by 
\begin{align}
m t_c= & \pi \left( 1 -  \frac{\tau_{c}}{\tau_{0}} \right).
\end{align}
The e-folds depend on the decay time as~\cite{Kohri:2009ac} 
\begin{align}
e^N  \propto t_c^{1/6}.
\end{align}
From this, we obtain $N' = (1/6) t_c'/t_c$ and $N'' = (1/6) (t_c''/t_c - (t_c'/t_c)^2)$, where the prime denotes the derivative with respect to $\tau_0$. 
We explicitly obtain 
\begin{align}
N' \simeq &  \frac{\tau_\text{c}}{6 \tau_0^2}, 
&  N''\simeq & - \frac{\tau_\text{c}}{3 \tau_0^3},
\end{align}
where we have used $\tau_{c}\ll \tau_0$. 
To be concrete, we assume $\delta \tau_0 \simeq H_\text{inf}/(2\pi)$ with $H_\text{inf}$ being the Hubble parameter during the inflation. 
Then, we can obtain the curvature power spectrum from $\delta \tau_0$ and the local type non-Gaussianity parameter $f_\NL$ as 
\begin{align}
\mathcal{P}_{\zeta^{(\tau)}} =& (N' \delta \tau_{0} )^2 \simeq  \frac{1}{36} 
  \left(  \frac{\tau_{c}}{\tau_{0}} \right)^2 \left( \frac{H_{\text{inf}}}{2 \pi \tau_{0}} \right)^2, \\
f_{\text{NL}} = & \frac{5}{6} \left( \frac{\mathcal{P}_{\zeta^{(\tau)}}}{\mathcal{P}_\zeta} \right)^2 \frac{N''}{(N')^2} \simeq - 10 
   \left( \frac{\mathcal{P}_{\zeta^{(\tau)}}}{\mathcal{P}_\zeta} \right)^2 \frac{\tau_{0}}{\tau_{c}} \nonumber \\ 
   \simeq & - \frac{5}{162} \epsilon_\text{inf}^2 \left(\frac{M_\Pl}{\tau_0}\right)^4 \left(\frac{\tau_c}{\tau_0} \right)^3,
\end{align}
where $\zeta^{(\tau)}$ denotes the curvature perturbation induced by $\delta \tau_0$ and we have used the slow-roll relation $\mathcal P_\zeta \simeq H^2_\text{inf}/(8\pi^2 \epsilon_\text{inf} M_\Pl^2)$ with $\epsilon_\text{inf} \equiv -\dot H_\text{inf}/H^2_\text{inf}$.
We can see that, if $\tau_0$ is large enough, we can make $f_\NL$ small enough to be consistent with the observational bounds. 

As a possible variant, we mention that the sudden decay of $\phi$ could possibly be realized by the first-order phase transition involving the abrupt change of the field value of $\tau$~\cite{Inomata:2019ivs}.

\subsubsection{Rotating axion field}
\label{sssec:axion_poltergeist}

We here discuss the poltergeist mechanism in the axion rotation scenario.
The axion is a particle originally proposed to resolve the strong CP problem in QCD~\cite{Peccei:1977hh,Peccei:1977ur,Weinberg:1977ma, Wilczek:1977pj}. 
The Peccei-Quinn mechanism addresses this issue by introducing a global U(1) symmetry, which, when spontaneously broken, gives rise to a pseudo-Nambu-Goldstone boson~\cite{Nambu:1961tp,Goldstone:1961eq,Goldstone:1962es}—the axion.
Beyond the QCD axion, many extensions of the Standard Model, especially those inspired by string theory, predict the existence of axion-like particles~\cite{Svrcek:2006yi}. 
In the following, axions simply mean any pseudo-Nambu-Goldstone bosons that originate from a spontaneous symmetry breaking of U(1), that is, axion-like particles. 

In cosmology, axion is typically considered to undergo coherent oscillations around the minimum of its potential, giving rise to the well-known misalignment mechanism for dark matter production~\cite{Preskill:1982cy,Abbott:1982af,Dine:1982ah}.
However, in many realistic models, the axion field can carry a nonzero conserved charge density, corresponding to a global U(1) rotation in the complex field space. 
Similarly to Affleck–Dine-like mechanisms~\cite{Affleck:1984fy}, the nonzero U(1) charge can be induced by, \textit{e.g.}, Planck suppressed terms that explicitly break the U(1) symmetry only when the radial field value is large, but finally become negligible as the field value becomes small due to the Hubble friction.
In this case, instead of simple oscillations, the axion field undergoes rotational motion along the angular direction of its potential. 
This rotation plays a central role in a wide range of early Universe phenomena~\cite{Kamada:2019uxp,Co:2019jts,Co:2020xlh,Gouttenoire:2021wzu,Gouttenoire:2021jhk,Co:2022qpr,Eroncel:2022efc,Eroncel:2025qlk}. 
An aspect relevant to the poltergeist mechanism is that the axion rotation can realize a sudden transition from a MD era to a kination era, depending on the parameters of the axion potential. 
Since the Universe has pressure during a kination era similar to a RD era, the sudden transition from a MD era to a kination era triggers the poltergeist mechanism. 

As a concrete example, we consider the supersymmetric two-field model, where, after the explicit U(1)-breaking terms become negligible and the heavy field decouples, we can express its effective Lagrangian as~\cite{Co:2021lkc}
\begin{align}
\label{eq:Leff}
    {\cal L} &=  \! \left( 1 + \frac{f_a^4}{16|P|^4} \right)  |\partial P|^2 -  m^2_S \! \left( |P|  - \frac{(1+d)f_a^2}{4|P|}\right)^{\! 2} \! ,
\end{align}
where $P$ is the complex scalar field containing the axion in its phase component, expressed as $P = S \ee^{i \theta}/\sqrt{2}$, and $|\partial P|^2 \equiv - g^{\mu\nu}\partial_\mu P^\dag \partial_\nu P$.
$f_a$ is the axion decay constant, $m_S$ is the mass parameter of the radial direction field $S$, which originates from supersymmetry breaking, and $1+d$ is a ratio of mass parameters of the two fields that appear in the UV completion. 
The stable axion rotation requires $d \geq 0$ and the perturbativity of the model requires $f_a > m_S$~\cite{Co:2021lkc}.
We note that the potential is nearly quadratic in $S\gg f_a$, which leads to an eMD era.

\begin{figure}
        \centering \includegraphics[width=0.99\columnwidth]{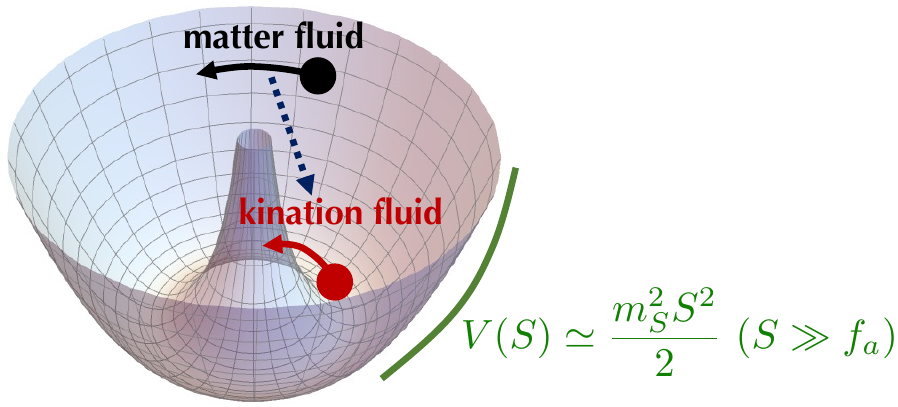}
        \caption{ \justifying
        Schematic picture of the potential for axion rotation Eq.~(\ref{eq:Leff}), which describes the axion rotation after the explicit U(1) breaking terms become negligible. 
        The black and red circles indicate the axion dynamics during its matter-fluid and kination-fluid phases, respectively.
        }
        \label{fig:axion_rot}
\end{figure}

In this scenario, the Universe undergoes the following transition in the equation-of-state parameter: $w=1/3$\,(eRD) $\to w=0$\,(eMD) $\to w=1$\,(kination) $\to w = 1/3$\,(RD).
Figure~\ref{fig:axion_rot} shows a schematic picture of the potential and the axion rotation after the explicit U(1)-breaking terms become negligible.
First, the radial field is large ($S \gg f_a$) and the axion rotation behaves as matter because of the asymptotic form of the potential ($V(S) \simeq m_S^2 S^2/2)$.
At this point, the energy of axion is equally distributed to its potential and kinetic energy.
The Universe is first dominated by radiation, but after some time, the matter-like axion rotation comes to dominate the Universe (at $\eta_{\eq,1}$).
Afterwards, the radial field approaches the potential minimum and the energy of the axion is dominated by its kinetic energy (at $\eta_{\text{kin}}$), which makes the axion rotation behave as kination fluid with $w = 1$.
Finally, radiation dominates the Universe again (at $\eta_{\eq,2}$) because the kination fluid redshifts ($\propto a^{-6}$) faster than radiation ($\propto a^{-4}$).

\begin{figure}
        \centering \includegraphics[width=0.99\columnwidth]{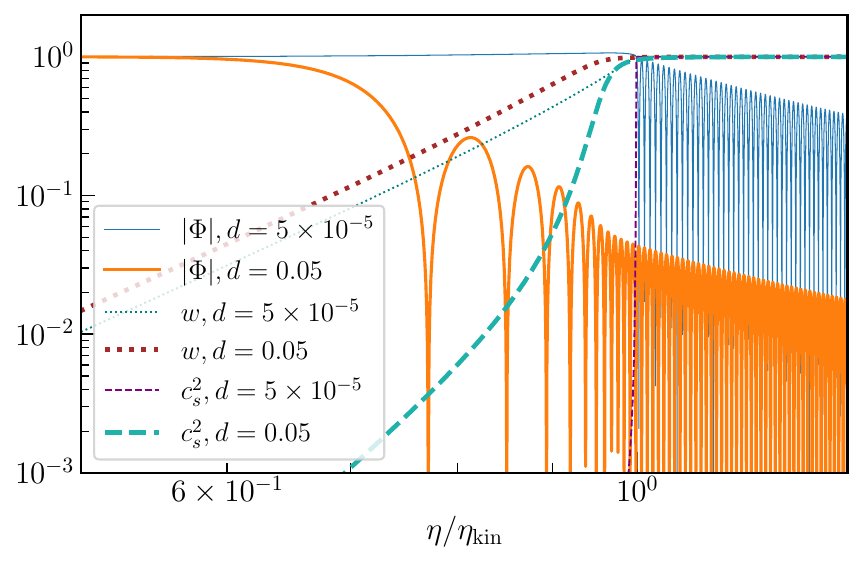}
        \caption{ \justifying 
        The evolution of $c_\ss^2$, $w$, and $\Phi$, in the two-field model, Eq.~(\ref{eq:Leff}).
        The thin (thick) lines are for $d = 5\times 10^{-5} \text{ } (0.05)$.
        $\Phi$ is normalized as $\Phi = 1$ during the MD era.
        $k \eta_\kk = 450$ is taken for $\Phi$. 
        From \rref{Harigaya:2023mhl}.
        }
        \label{fig:phi_ax_rot}
\end{figure}

Figure~\ref{fig:phi_ax_rot} shows the evolution of the sound speed, the equation-of-state parameter, and the gravitational potential.
Note that $\eta_\kk$ is defined as the time when $c_\ss^2 = 0.95$, which we regard as the beginning of the kination era. 
This figure shows that a smaller $d$ leads to a more sudden transition from an eMD era to a kination era. 
The rapid oscillation of the gravitational potential after the sudden transition causes the poltergeist mechanism. Note that a small $d$ is technically natural as a $\mathbb{Z}_2$ symmetry appears in the limit~\cite{Harigaya:2023mhl}.

\begin{figure}
        \centering \includegraphics[width=0.99\columnwidth]{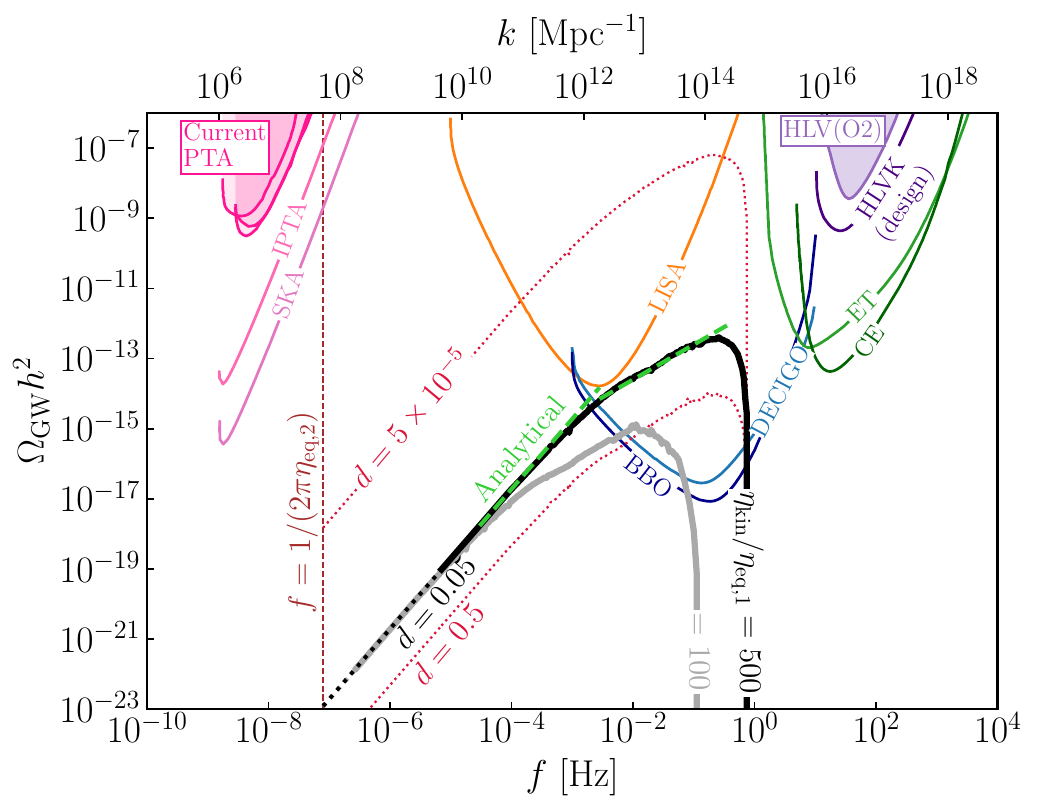} 
        \caption{ \justifying
        The GW spectrum with different $\eta_\kk/\eta_\eqf$ and $d$.
        The curvature power spectrum is given by Eq.~(\ref{eq:pz}) with $A_\ss = 2.1\times 10^{-9}$.
        The black and gray lines are the GWs with $d=0.05$ and $\eta_\eqs = 2\times 10^6\,$s. 
        For comparison purposes, the red dotted lines show the spectra with $\eta_\kk/\eta_\eqf = 500$ and different values of $d$.
        In particular, the case with $d =5 \times 10^{-5}$ is unreliable due to the breakdown of the perturbation theory (see the text). 
        The green dashed line shows the analytical approximation, given by Eq.~(\ref{eq:omega_gw_app}), for the black line.
        The black and gray dotted lines are the GWs in $f < 1/(2\pi \times 100 \eta_\kk)$, which are unreliable because they are based on the linear perturbation theory when the UV cutoff perturbations (at $k_\tmax$) are nonlinear, $|\delta| \simeq \mathcal O(1)$ (see the text).
        From \rref{Harigaya:2023mhl}.
        }
        \label{fig:gw_ax_rot}
\end{figure}

Figure~\ref{fig:gw_ax_rot} shows the GW spectrum with some fiducial parameter sets. 
We take the following curvature power spectrum:
\begin{align}
  \mathcal P_\zeta (k)= A_\ss\, \Theta(k_\tmax-k)\Theta(k-1/\eta_\kk).
  \label{eq:pz}
\end{align}
The UV cutoff is introduced for the same reason as before---to avoid the nonlinear density perturbation during the eMD era. 
In Fig.~\ref{fig:gw_ax_rot}, we basically set $k_\tmax = k_\NL$ under the condition $k_\NL \leq 500/\eta_\kk$. 
When this condition is not satisfied, we reset it to $k_\tmax = 500/\eta_\kk$ to save the computational cost, which can happen if $k_\tmax \gtrsim 1/\eta_\eqf$ because of the eRD era preceding the eMD era.
On the other hand, the IR cutoff is introduced to focus on the poltergeist mechanism, caused by the subhorizon perturbations at the beginning of the kination era. 
We here assume that the axion-direction isocurvature perturbations are negligibly small for simplicity. For the SIGWs with large axion isocurvature perturbations in a gradual transition, see \rref{Bodas:2025wef}.
For the GW calculation, we must generalize the formalism in the previous examples, which focus on the transition from an eMD era to a RD era.
See~\rref{Harigaya:2023mhl} for the general expressions with the transition from an eMD era to an era with a general $w$, which we use for Fig.~\ref{fig:gw_ax_rot}.

We here highlight the two key differences between the case with a RD or a kination era after the eMD era.
First, there is no sharp rise ($\propto k^7$ in RD case) around the UV cutoff scale. This is because the resonant amplification, which makes the $k^7$ rise in the RD case, occurs only when $c_\ss < 1$. In the case of $c_\ss = 1$, the momentum conservation between the tensor and scalar perturbations prohibits the resonant amplification~\cite{Domenech:2019quo}.
Second, in addition to the poltergeist mechanism, the existence of the kination era increases $\Omega_\GW$ as follows.  The GWs are produced by the kination fluid that has a larger energy density than the radiation but finally becomes negligible without entropy production. 
Once GWs are produced during a kination era, $\Omega_\GW$ increases as $\propto a^2$ because $\rho_\GW \propto a^{-4}$, while the background energy density during the kination era, denoted by $\rho_\kk$, behaves as $\rho_\kk \propto a^{-6}$.
This $\propto a^2$ growth of $\Omega_\GW$ during the kination era leads to the additional enhancement of GWs.

The peak scale of the SIGWs is located around $k_* \equiv \min[k_\tmax,1/\eta_{\eq,1}]$.
The analytical expression for the GW spectrum around the peak is given by \cite{Harigaya:2023mhl}
\begin{align}
        \Omega_\GW h^2
        &\simeq 2\times 10^{-11} A_\ss^2 Q^4 B(k) \, \frac{\eta_\kk^2}{\eta_\eqf^2} \,  k_*^5 k \eta_\kk^{6},
        \label{eq:omega_gw_app}
\end{align}
where $Q$ is the normalized value of $\Phi$ on the peak scale $k_*$ at $\eta_\kk$ (up to the oscillations) to take into account the non-instantaneous transition, and $B(k)$ is $1$ for $k>4/\eta_\kk$ and $0.535 \times(k\eta_\kk)$ for $k < 4/\eta_\kk$, showing the spectral break around the MD-to-kination transition.  
The factor $\eta_\kk^2/\eta_\eqf^2 >1$ in Eq.~(\ref{eq:omega_gw_app}) takes into account the $\propto a^2$ growth of $\Omega_\GW$ during the kination era. 
See Appendix~\ref{app:analytic_formulas} for the derivation of Eq.~(\ref{eq:omega_gw_app}).

In this scenario, we need to be careful about the limitations of the perturbation theory from the following two points:

\noindent 
1) A rapid change of the state of the Universe, characterized with $w$ and $c_\ss^2$, generally enhances the contributions from the higher-order perturbations. 
This behavior is often called the strong coupling of the perturbations in the context of the effective field theory of inflation~\cite{Cheung:2007st,Adshead:2014sga,Inomata:2021tpx}. 
In our scenario, a small $d$ could break the perturbation theory, while it enhances the GW production. 
\rref{Harigaya:2023mhl} has numerically confirmed that the case of $d = 0.05$ marginally remains in the valid linear perturbation theory, while $d \lesssim 0.01$ results in the non-negligible higher order contributions.
Because of this, the case of $d = 5\times 10^{-5}$ in Figs.~\ref{fig:phi_ax_rot} and \ref{fig:gw_ax_rot} should not be taken as reliable results. 
Nevertheless, we still plot them to convey the essential idea of the sharp transition.

\noindent
2) Unlike during a RD era, the density perturbation $\delta$ continues to grow on subhorizon scales during a kination era as $\delta \propto \eta^{1/2}$ up to the oscillations. 
This growth starts after horizon entry, and the smallest-scale mode $k_\tmax$ is the first to reach $|\delta| = 1$, beyond which linear perturbation theory breaks down. 
Since the perturbations around $k_\tmax$ are the main source even for the GWs in $k \ll k_\tmax$, this issue needs a careful treatment. 
However, once GWs enter the horizon, they evolve freely, decoupled from the source~\cite{Domenech:2020kqm}. 
Thus, we can reliably compute induced GWs for scales that re-enter before the density perturbations at $k_\tmax$ become $|\delta| = 1$.

Figure~\ref{fig:gw_ax_rot} shows that LISA, BBO, and DECIGO could detect the induced GWs in some parameter space. 
Since the speed of the MD-to-kination transition primarily depends on the shape of the potential in the radial direction, the radial mass parameter as well as the axion decay constant can be probed by the GW observations.  In the supersymmetric two-field model, the radial mass is given by soft supersymmetry breaking. Remarkably, LISA and DECIGO/BBO can probe the scale of soft supersymmetry breaking terms maximally up to about $10^4$ GeV and $10^7$ GeV, respectively, which is impossible to reach by any near-future particle colliders. It is also possible to probe a part of the axion parameter space that supports the ALP cogenesis scenario~\cite{Co:2020xlh}, in which the baryon asymmetry of the Universe and the ALP dark matter abundance are explained simultaneously. See ~\rref{Harigaya:2023mhl} for more detailed discussions.  These highlight the complementarity between GW observations and other experiments/observations.

%%%%%%%%%%%%%%%%%%%%%%%%%%%%%%%%
\section{Summary}
\label{sec:conclusions}
%%%%%%%%%%%%%%%%%%%%%%%%%%%%%%%%

GWs have been attracting significant attention as a probe of the early Universe.
In this review, we have highlighted the potential of GWs induced by scalar perturbations (scalar-induced GWs: SIGWs) in the early Universe.
In particular, we have focused on SIGWs associated with an eMD era, which can be realized through the coherent oscillation or rotation of a field, or through the domination of compact objects such as PBHs, Q-balls, and oscillons. 
A key point is that the amplitude of SIGWs depends on the timescale of the transition from the eMD era to the subsequent era. 
If this transition occurs on a timescale much shorter than the Hubble timescale at that time, the production of SIGWs is significantly enhanced.
This enhancement mechanism is called the poltergeist mechanism, which has been the main focus of this review.

Throughout this review, we have explained both the conceptual essence and the technical details of the poltergeist mechanism. 
During an eMD era, density perturbations on subhorizon scales grow proportionally to the scale factor, and they begin to oscillate after the transition from an eMD era to an era with $w \neq 0$.
The oscillation timescale is $\sim 1/k$, and hence density perturbations on deeply subhorizon scales at the end of the eMD era oscillate much faster than the Hubble timescale. 
These rapid oscillations correspond to large velocity perturbations, which are the primary source of the poltergeist mechanism.
A sudden transition is essential, since a gradual transition significantly suppresses the amplitude of the subhorizon density perturbations.

We have also discussed several concrete examples.
In particular, the poltergeist GWs from PBH evaporation offer a unique probe of the abundance of tiny PBHs that have evaporated by now.
In this case, we need to be careful about the finite duration of each PBH evaporation and the finite width of the PBH mass spectrum, both of which suppress the poltergeist GWs.
We also emphasize that, compared to the contribution from adiabatic perturbations, the contribution arising from Poisson fluctuations in the PBH spatial distribution becomes nonlinear over a shorter duration of the eMD era, rendering the linear perturbation theory unreliable.
A careful treatment of these Poisson fluctuations is therefore required, and we leave this for future work. 
Beyond PBHs, evaporation of other compact objects such as Q-balls and oscillons could also trigger the poltergeist mechanism.

Apart from compact objects, the poltergeist mechanism can arise in scenarios involving kinetic blocking, which can realize sudden reheating through the dynamics of a scalar field rolling down its potential. 
The axion rotation can also lead to the poltergeist mechanism through a transition from an eMD era to a kination era. 
In these scenarios, the poltergeist GWs provide a valuable probe of parameter regions inaccessible to other (non-GW) experiments.

Since the poltergeist mechanism can occur whenever an eMD era ends suddenly and is followed by an era with $w \neq 0$, its full potential may not be fully uncovered.
Exploring new scenarios that naturally realize such sudden transitions is an intriguing direction for future research.

Another important future direction is understanding the GW spectrum in the presence of nonlinear density perturbations.
While our current treatment is conservative, the introduction of the UV cutoff scale is artificial and affects the spectral shape. 
Refining this aspect will be crucial for meaningful comparison with observations.
If the resonance peak persists even after properly accounting for the nonlinearity, the peak feature could play a key role in distinguishing between different GW sources. 

Although the poltergeist mechanism typically arises in scenarios involving rapid cosmic transitions, its occurrence offers an exceptional opportunity to uncover hidden aspects of the early Universe. A detection of GWs consistent with this mechanism would provide invaluable clues about the cosmic history and the underlying physics. Such a signal would represent a rare yet powerful window into the dynamics of the early Universe.

\vspace{5pt}
%%%%%%%%%%%%%%%%%%%%%%%%%%%%%%%%
\acknowledgments
%%%%%%%%%%%%%%%%%%%%%%%%%%%%%%%%
We thank Keisuke Harigaya, Masahiro Kawasaki, Kyohei Mukaida, Tomohiro Nakama, and Tsutomu T. Yanagida for their collaborations on our previous works~\cite{Inomata:2019zqy,Inomata:2019ivs,Inomata:2020lmk,Harigaya:2023mhl}.
KI was supported by JSPS Postdoctoral Fellowships for Research Abroad.
KK was supported by KAKENHI Grants No.~JP23KF0289, No.~JP24H01825, and No.~JP24K07027.
TT was supported by the 34th (FY 2024) Academic research grant (Natural Science) No.~9284 from DAIKO FOUNDATION.
%%%%%%%%%%%%%%%%%%%%%%%%%%%%%%%%%%%
%%%%%%%%%%%%%%%%%%%%%%%%%%%%%%%%%%%

\appendix

%%%%%%%%%%%%%%%%%%%%%%%%%%%%%%%%
\section{Concrete expressions up to first order}
\label{app:einstein_em_tensor}
%%%%%%%%%%%%%%%%%%%%%%%%%%%%%%%%

In this appendix, we summarize the concrete expressions of the background and the first-order perturbations of important quantities in a general gauge, focusing on the scalar perturbations. 

\subsection{Metric}

The metric at the first order is given by 
\begin{align}
  &g_{00} = -a^2(1+2\Phi), \ g_{0i} = a^2 B_{,i}, \nonumber \\
  &g_{ij} = a^2[ (1-2 \Psi) \delta_{ij} + 2E_{,ij}], \\ 
  &g^{00} = -\frac{1}{a^2}(1-2\Phi), \ g^{0i} = \frac{1}{a^2} B_{,i}, \nonumber \\ 
  &g^{ij} = \frac{1}{a^2}[ (1+2\Psi)\delta_{ij} - 2E_{,ij}],
\end{align}
where we have neglected the first-order vector and tensor perturbations.
Note that the transformation between the superscript and subscript in the 3-dimensional space is performed with $\delta_{ij}$ because we consider the flat Universe.
These expressions satisfy $g^{\mu\lambda}g_{\lambda \nu} = \delta^\mu_{\ \nu}$ up to the first order in perturbations.

\subsection{Christoffel symbols}
The Christoffel symbol is given by 
\begin{align}
  \Gamma^\lambda_{\ \mu\nu} \equiv \frac{1}{2} g^{\lambda \rho} (g_{\rho \mu, \nu} + g_{\rho \nu, \mu} - g_{\mu \nu, \rho}).
\end{align}
In the flat FLRW metric, the background Christoffel symbols are given as
\begin{align}
  &\bar \Gamma^0_{\ 00} = \mathcal H, \ \ \bar \Gamma^i_{\ 0j} = \mathcal H \delta^i_{\ j}, \ \ \bar \Gamma^0_{\ ij} = \mathcal H \delta_{ij}, \nonumber \\ 
  &\bar \Gamma^i_{\ 00} = \bar \Gamma^0_{\ 0i} = \bar \Gamma^i_{\ jk} = 0,
  \label{eq:chris_exp}
\end{align}
where the bar denotes the background value.
The perturbations of the Christoffel symbols are given by 
\begin{align}
  \label{eq:gamma_pertb_000}
  \delta \Gamma^{0}_{\ 00} &= \Phi', \\
   \delta \Gamma^0_{\ 0i}&= \Phi_{,i} + \mathcal H B_{,i}, \\
  \delta \Gamma^i_{\ 00} &= a^{-1}(a B_{,i})' + \Phi_{,i}, \\
  \delta \Gamma^{0}_{\ ij} &= -2 \mathcal H \Phi \delta_{ij} - B_{,ij} - a^{-2} [a^2 (\Psi \delta_{ij} - E_{,ij})]', \\
  \delta \Gamma^i_{\ 0j} &= - \Psi' \delta_{ij} + E_{,ij}', \\
  \label{eq:gamma_pertb_ijk}  
  \delta \Gamma^i_{\ jk} &= - \Psi_{,j}\delta_{ik} - \Psi_{,k} \delta_{ij} + \Psi_{,i} \delta_{jk} - \mathcal H B_{,i}  \delta_{jk} + E_{,ijk}.
\end{align}

\subsection{Einstein tensor}

The Einstein tensor is given by 
\begin{align}
  G_{\mu \nu} &\equiv R_{\mu \nu} - \frac{1}{2} R g_{\mu \nu}, \\
  R_{\mu \nu} &\equiv \Gamma^{\lambda}_{\ \mu \nu, \lambda} - \Gamma^\lambda_{\ \mu \lambda, \nu} + \Gamma^\kappa_{\ \mu \nu} \Gamma^\lambda_{\ \kappa \lambda} - \Gamma^\kappa_{\ \mu \lambda} \Gamma^{\lambda}_{\ \nu \kappa}, 
\end{align}
where $R = g^{\mu \nu} R_{\mu\nu}$.
Then, the background Einstein tensor is given by 
\begin{align}
  &\bar G^0_{\ 0} = -\frac{3}{a^2}\mathcal H^2, \ 
  \bar G^i_ {\ j} = -\frac{1}{a^2} (2\mathcal H' + \mathcal H^2) \delta^i_{\ j}, \nonumber \\ 
  &\bar G^i_{\ 0} = \bar G^0_{\ i} = 0.
\end{align}
The perturbed Einstein tensor is given by
\begin{align}
  \label{eq:einstein_prtb_00}
  \delta G^0_{\ 0} =&\, \frac{2}{a^2} \left[ 3 \mathcal H^2 \Phi + 3\mathcal H \Psi' - \Psi^{,i}_{\ ,i} + \mathcal H (B^{,i}_{\ ,i} - {E^{,i}_{\ ,i}}') \right], \\
  \label{eq:einstein_prtb_0i}  
  \delta G^0_{\ i} =&\, \frac{1}{a^2}\left( -2 \mathcal H \Phi_{,i} - 2 \Psi'_{, i} \right), \\
  \label{eq:einstein_prtb_i0}  
  \delta G^i_{\ 0} =&\, \frac{1}{a^2}\left[ 2 \mathcal H \Phi_{,i} + 2 \Psi'_{, i} + 2(\mathcal H^2 -\mathcal H') B_{,i}\right], \\
  \label{eq:einstein_prtb_ij}  
  \delta G^i_{\ j} =&\, \frac{1}{a^2} \left[ \left( (2 \mathcal H^2 + 4 \mathcal H') \Phi + 2 \mathcal H \Phi' + \Phi^{,i}_{\ ,i} + 4 \mathcal H \Psi' + 2 \Psi''  \right.  \phantom{ \left( \frac{1}{2} \right)^{i}_{j} }  \right. \nonumber \\
  & \left. - \Psi^{,i}_{\ ,i} + 2 \mathcal H B^{,i}_{\ ,i} + {B^{,i}_{\ ,i}}' -  2\mathcal H {E^{,i}_{\ ,i}}' - {E^{,i}_{\ ,i}}'' \right) \delta_{ij} \nonumber \\
 &+ \left(- \Phi + \Psi - 2\mathcal H B - B' + 2\mathcal H E' + E'' \right)^{,i}_{\ ,j} \bigg].
\end{align}

\subsection{Energy-momentum tensor}

To summarize the expressions of the energy-momentum tensor, we first summarize the expressions of $u^\mu$ and $u_\mu$, which we express as $u^\mu = (1/a + \delta u^0, \delta u^i)$ and $u_\mu=(-a + \delta u_0, \delta u_i)$.
From the normalization condition $g^{\mu \nu} u_\mu u_\nu = g_{\mu\nu} u^\mu u^\nu = -1$, we obtain
\begin{align}
  \label{eq:u0_express}
  \delta u_0 = - a\Phi, \ \delta u^0 = - \frac{\Phi}{a}.
\end{align}
From $u^i = g^{i\mu} u_\mu$, we obtain 
\begin{align}
  \delta u^i = -\frac{B_{,i}}{a} + \frac{\delta u_{,i}}{a^2}.
\end{align}
Using these expressions, we can express the background of the energy-momentum tensor as 
\begin{align}
  \label{eq:t_mn_back}
  \bar T^0_{\ 0} &= - \bar \rho,\ \bar T^i_{\ j} = \bar P \delta^{i}_{\ j},\  \bar T^{0}_{\ i} = \bar T^{i}_{\ 0} = 0, 
\end{align}
and its perturbations as 
\begin{align}
  \label{eq:t_mn_00}
  \delta T^0_{\ 0} &= - \delta \rho,\\
  \label{eq:t_mn_0i}  
  \delta T^{0}_{\ i} &= \frac{1}{a} (\bar \rho + \bar P)\delta u_{,i}, \\
  \label{eq:t_mn_i0}  
  \delta T^{i}_{\ 0} &= \frac{1}{a} (\bar \rho + \bar P) \left( a B_{,i} - \delta u_{,i} \right), \\
  \label{eq:t_mn_ij}  
  \delta T^i_{\ j} &= \delta P \delta^{i}_{\ j}.
\end{align}
%%

%%%%%%%%%%%%%%%%%%%%%%%%%%%%%%%%
\section{Linear perturbations during transitions}
\label{app:perturbation_during_trans}
%%%%%%%%%%%%%%%%%%%%%%%%%%%%%%%%

In this appendix, we derive the equations for the perturbations that experience the transition from a MD era to a RD era in a general gauge (Eqs.~(\ref{eq:delta_m_evo})-(\ref{eq:theta_r_evo})).
We begin with the perturbed energy momentum conservation, Eqs.~(\ref{eq:d_t_m_munu}) and (\ref{eq:d_t_r_munu}):
\begin{align}
  \delta(T^\mu_{\text{m}\, \nu; \mu}) &= \delta(\Gamma\, T^\mu_{\text{m}\, \nu} u_{\text{m}\,\mu}), \tag{\ref{eq:d_t_m_munu}} \\
  \delta(T^\mu_{\text{r}\, \nu; \mu}) &= -\delta(\Gamma\, T^\mu_{\text{m}\, \nu} u_{\text{m}\,\mu}). \tag{\ref{eq:d_t_r_munu}}
\end{align}
The perturbation of the covariant derivative of the energy-momentum tensor is given by
\begin{align}
  \delta(T^\mu_{\ \nu;\mu}) &=  \delta T^{\mu}_{\ \nu,\mu} + \bar \Gamma^{\mu}_{\ \mu \lambda} \delta T^{\lambda}_{\ \nu} - \bar \Gamma^{\lambda}_{\ \mu \nu} \delta T^{\mu}_{\ \lambda} \nonumber \\
  &\quad
  + \delta\Gamma^{\mu}_{\ \mu \lambda} \bar T^{\lambda}_{\ \nu} - \delta \Gamma^{\lambda}_{\ \mu \nu} \bar T^{\mu}_{\ \lambda}.
\end{align}
Substituting Eqs.~(\ref{eq:gamma_pertb_000})-(\ref{eq:gamma_pertb_ijk}) and (\ref{eq:t_mn_00})-(\ref{eq:t_mn_ij}) into this expression, we get
\begin{align}
  \label{eq:t_mu0_pertb} 
  \delta (T^{\mu}_{\ 0;\mu}) &=
  \delta T^\mu_{\ 0,\mu} + 3 \mathcal H \delta T^0_{\ 0} -\mathcal H \delta T^i_{\ i} - \delta \Gamma^i_{\ i0} (\bar\rho + \bar P) \nonumber \\
  &= -\delta \rho' + \frac{1}{a} (\bar \rho + \bar P) (a \Delta B - \Delta \delta u) \nonumber \\
  &\quad - 3 \mathcal H \delta \rho  - 3 \mathcal H \delta P + (3\Psi' - \Delta E')(\bar \rho + \bar P), \\
  \label{eq:t_mui_pertb} 
  \delta (T^{\mu}_{\ i;\mu}) &= 
  \delta T^\mu_{\ i,\mu} + 3 \mathcal H \delta T^0_{\ i} - \mathcal H \delta T^i_{\ 0} + \delta \Gamma^0_{\ 0i} (\bar \rho + \bar P) \nonumber \\
  &= \left( (\bar \rho + \bar P) \frac{\delta u_{,i}}{a} \right)'  + \delta P_{,i}\nonumber \\ 
  &\quad  + (\bar \rho + \bar P) \left( 4\mathcal H\frac{\delta u_{,i}}{a} + \Phi_{,i} \right).
\end{align}
For later convenience, we write down the expressions for $(\delta (T^{\mu}_{i;\mu}))^{,i}$,
\begin{align}
\label{eq:t_muii_pertb}  
(\delta (T^{\mu}_{\ i;\mu}))^{,i} &=  \left( (\bar \rho + \bar P) \frac{\Delta \delta u}{a} \right)'  + \Delta \delta P \nonumber \\ 
 &\quad + (\bar \rho + \bar P) \left( 4\mathcal H\frac{\Delta \delta u}{a} + \Delta \Phi \right).
\end{align}
The perturbations of the right-hand side in Eqs.~(\ref{eq:d_t_m_munu}) and (\ref{eq:d_t_r_munu}) are given by 
\begin{align}
  \label{eq:delta_gamma_0}
  \delta(\Gamma\, T^{\mu}_{\mm \,\,0} u_{\text{m}\, \mu} ) &= a \delta \Gamma \bar \rho_\mm + a \bar \Gamma \delta \rho_\mm + a \bar \Gamma \bar \rho_\mm \Phi, \\
  \label{eq:delta_gamma_i}  
  \delta(\Gamma\, T^{\mu}_{\mm \,\,i} u_{\text{m}\, \mu} ) & = -\bar \Gamma \bar \rho_\mm \delta u_{\mm,i}, \\
  \label{eq:delta_gamma_ii}
  (\delta(\Gamma\, T^{\mu}_{\mm \,\,i} u_{\text{m}\, \mu} ))^{,i} & = -\bar \Gamma \bar \rho_\mm \Delta \delta u_\mm,
\end{align}
where we have used the relation $\delta u_0 = -a\Phi$, given in Eq.~(\ref{eq:u0_express}).
Note again that we take into account the perturbation of the decay rate as $\Gamma = \bar \Gamma + \delta \Gamma$.

%%%%%%%%%%%%%%%%%
\subsection{Equations for matter perturbations}
%%%%%%%%%%%%%%%%%

Let us here focus on the equations for matter perturbations ($\bar P_\mm = \delta P_\mm =0$).
Combining Eqs.~(\ref{eq:t_mu0_pertb}) and (\ref{eq:delta_gamma_0}), we obtain 
\begin{align}
&-\delta \rho_\mm' + \frac{\bar \rho_\mm}{a} (a \Delta B - \Delta \delta u_\mm) - 3 \mathcal H \delta \rho_\mm + (3\Psi' - \Delta E')\bar \rho_\mm \nonumber \\ 
&= a \delta \Gamma \bar \rho_\mm + a \Gamma \delta \rho_\mm + a\Gamma \bar\rho_\mm \Phi.
\end{align}
Dividing both sides by $\bar \rho_\mm$, we obtain the expression of Eq.~(\ref{eq:delta_m_evo}) in a general gauge:
\begin{align}
\label{eq:delta_m_evo_app}
\delta_\mm' - (\Delta B - \theta_\mm) + (-3\Psi' + \Delta E') = -a \delta \Gamma - a\Gamma \Phi,
\end{align}
where $\delta_\mm \equiv \delta \rho_\mm/\bar \rho_\mm$ and $\theta_\mm \equiv \Delta \delta u_\mm/a$ and we have used $\bar \rho_\mm' + 3 \mathcal H \bar \rho_\mm = - a \bar \Gamma \bar \rho_\mm$ (Eq.~(\ref{eq:rho_m_with_decay})).
Combining Eqs.~(\ref{eq:t_muii_pertb}) and (\ref{eq:delta_gamma_ii}), we obtain 
\begin{align}
 ( \bar \rho_\mm \theta_\mm )' + \bar \rho_\mm  \left( 4 \mathcal H \theta_\mm + \Delta \Phi \right) = -a \Gamma \bar \rho_\mm \theta_\mm.
\end{align}
Dividing both sides by $\bar \rho_\mm$, we obtain
\begin{align}
\label{eq:theta_m_evo_app}  
 \theta_\mm' + \mathcal H \theta_\mm + \Delta \Phi = 0.
\end{align}
Although this equation is for a general gauge, it is the same as that in the Newtonian gauge, Eq.~(\ref{eq:theta_m_evo}).
We note that Eqs.~(\ref{eq:delta_m_evo_app}) and (\ref{eq:theta_m_evo_app}) are consistent with those obtained in \rref{Poulin:2016nat}.

%%%%%%%%%%%%%%%%%
\subsection{Equations for radiation perturbations}
%%%%%%%%%%%%%%%%%

Next, let us obtain the equations for the radiation perturbations ($\bar P_\rr = \bar \rho_\rr/3,\, \delta P_\rr = \delta \rho_\rr/3$).
Combining Eqs.~(\ref{eq:t_mu0_pertb}) and (\ref{eq:delta_gamma_0}), we obtain
\begin{align}
&-\delta \rho_\rr' + \frac{4\bar \rho_\rr}{3a} (a \Delta B - \Delta \delta u_\rr) - 4 \mathcal H \delta \rho_\rr  + \frac{4\bar \rho_\rr}{3}(3\Psi' - \Delta E') \nonumber \\
&= - (a \delta \Gamma \bar \rho_\mm + a \bar \Gamma \delta \rho_\mm + a\bar \Gamma \bar \rho_\mm \Phi).
\end{align}
Dividing both sides by $\bar \rho_\rr$, we obtain the expression of Eq.~(\ref{eq:delta_r_evo}) in a general gauge:
\begin{align}
\label{eq:delta_r_evo_app}  
&\delta_\rr' - \frac{4}{3} ( \Delta B - \theta_\rr) - \frac{4}{3}(3\Psi' - \Delta E') \nonumber \\ 
&=  a \frac{\bar \rho_\mm}{\bar \rho_\rr} \bar \Gamma \left( \frac{\delta \Gamma}{\bar \Gamma} + \delta_\mm - \delta_\rr + \Phi \right),
\end{align}
where $\delta_\rr \equiv \delta \rho_\rr/\bar \rho_\rr$ and $\theta_\rr \equiv \Delta \delta u_\rr/a$ and we have used $\bar\rho_\rr' + 4 \mathcal H \bar \rho_\rr = a \bar \Gamma \bar\rho_\mm$.
Combining Eqs.~(\ref{eq:t_muii_pertb}) and (\ref{eq:delta_gamma_ii}), we obtain
\begin{align}
 \left( \frac{4}{3} \rho_\rr \theta_\rr \right)'  + \frac{1}{3} \Delta \delta \rho_\rr + \frac{4}{3} \bar \rho_\rr \left(4 \mathcal H \theta_\rr
  + \Delta \Phi\right) = a \Gamma \bar \rho_\mm \theta_\mm.
\end{align}
Dividing both sides by $4\bar \rho_\rr/3$, we obtain the expression of Eq.~(\ref{eq:theta_r_evo}) in a general gauge:
\begin{align}
\label{eq:theta_r_evo_app}  
 \theta_\rr'  + \frac{1}{4} \Delta \delta_\rr + \Delta \Phi = a \Gamma \frac{3\bar \rho_\mm}{4\bar \rho_\rr} \left( \theta_\mm - \frac{4}{3} \theta_\rr \right).
\end{align}
Eqs.~(\ref{eq:delta_r_evo_app}) and (\ref{eq:theta_r_evo_app}) are consistent with those obtained in \rref{Poulin:2016nat}.

%%%%%%%%%%%%%%%%%%%%%%%%%%%%%%%%
\section{GW spectrum formula for the poltergeist mechanism in the instantaneous-limit transition}
\label{app:analytic_formulas}
%%%%%%%%%%%%%%%%%%%%%%%%%%%%%%%%

In this appendix, we collect the approximate analytic formulas for $\Omega_\text{GW}$ induced by the poltergeist mechanism in the instantaneous-limit transition, for which the approximate analytic calculation is possible. For the derivation, see the original references~\cite{Inomata:2019ivs} for subsection~\ref{ssec:analytic_transition_RD} and \cite{Harigaya:2023mhl} for subsection~\ref{ssec:analytic_transition_KD}. 

\subsection{Transition into RD era}\label{ssec:analytic_transition_RD}
Since the oscillation-averaged integration kernel has been computed both in a transient MD era and a transient RD era~\cite{Kohri:2018awv, Terada:2025cto}, the remaining task is just a matter of computation. As explained in the main text, the dominant contribution to the induced GWs is produced just after the transition in the RD era.  

Except for the GW mode close to the cutoff scale $k \sim 2 k_\text{max}$, $\Omega_\text{GW}$ is dominantly induced by the smallest-scale modes (see Fig.~\ref{fig:gw_prof}). This is realized in two ways.  For generic low-frequency modes $k \ll 2 k_\text{max}$, the $(s,t)$-integral of $\Omega_\text{GW}$ in Eq.~(\ref{eq:gw_cosmo_para_general_ts}) is dominated in the large $t$ region.  The large $t$ region physically corresponds to the squeezed limit configuration of the momentum vectors.  Another type is the resonance contribution satisfying the on-shell dispersion relation of GWs, which enables sustained production of GWs on subhorizon scales. Therefore, it is relevant around $k\sim 2 k_\text{max} / \sqrt{3}$. Mathematically, this originates from the logarithmic singularity of the cosine integral function.  One can extract these two contributions separately by different approximations~\cite{Inomata:2019ivs}.  

Before discussing $\Omega_\text{GW}(k)$, we summarize the integration kernel $\overline{I_\text{RD}(s, t, x)}^2$ in these two approximations. It is convenient to extract the time dependence as $I_\text{RD} = \frac{1}{x-x_\text{R}/2} \mathcal{I}_\text{RD}$.  Then, the large-scale contribution (with $t\gg 1$) is 
\begin{widetext}
\begin{align}
    \overline{\mathcal{I}_\text{RD}^2}|_{t \gg 1}\simeq & \frac{9 t^4 x_\text{R}^8}{163840000} \left( \pi^2 + \pi^2 \cos y_\text{R} + 2 \mathrm{Ci} ( y_\text{R}^-)^2 + 2 \mathrm{Ci} (y_\text{R}^+)^2 + 4 \cos y_\text{R} \mathrm{Ci}(y_\text{R}^+)\mathrm{Ci}(y_\text{R}^-) \right. + 2 \pi \sin y_\text{R} \left(\mathrm{Ci}(y_\text{R}^+) - \mathrm{Ci}(y_\text{R}^-) \right) \nonumber \\
    & - 2 \pi \left( 1 + \cos y_\text{R}\right) \left( \mathrm{Si}(y_\text{R}^-) + \mathrm{Si}(y_\text{R}^+) \right) + 4 \sin y_\text{R} \left( \mathrm{Ci}(y_\text{R}^-) \mathrm{Si}(y_\text{R}^+) - \mathrm{Ci}(y_\text{R}^+) \mathrm{Si}(y_\text{R}^-)\right)  \nonumber \\
    & \left.  + 2 \mathrm{Si}(y_\text{R}^-)^2 + 2 \mathrm{Si}(y_\text{R}^+)^2 + 4 \cos y_\text{R} \mathrm{Si}(y_\text{R}^+) \mathrm{Si}(y_\text{R}^-)  \right),
\end{align}
where we have introduced $y_\text{R} \equiv \frac{x_\text{R}}{\sqrt{3}}$ and $y_\text{R}^\pm \equiv \frac{x_\text{R}}{2} \pm \frac{s x_\text{R}}{2\sqrt{3}}$ to simplify the expression, and $\mathrm{Si}(z) \equiv \int_0^z \mathrm{d}\zeta \, \sin(\zeta)/\zeta $ and $\mathrm{Ci}(z) \equiv - \int_z^\infty \mathrm{d}\zeta \, \cos(\zeta)/\zeta$ are the sine integral and cosine integral functions, respectively.  Since the $s$-dependence is weak, one may further approximate it by setting $s=0$, leading to
\begin{align}
    \overline{\mathcal{I}_\text{RD}^2}|_{t \gg 1, \, s=0}\simeq & \frac{9t^4 x_\text{R}^8 \left(4 \mathrm{Ci}\left(\frac{x_\text{R}}{2}\right)^2 + \left( \pi - 2 \mathrm{Si}\left(\frac{x_\text{R}}{2}\right)\right)^2 \right)}{81920000}.
\end{align}

The resonance component is obtained as
\begin{align}
    \overline{\mathcal{I}_\text{RD}^2}|_{t \approx \sqrt{3}-1}\simeq & \frac{9(t^2 + 2 t + s^2 - 5)x_\text{R}^8}{81920000(1-s+t)^2 (1+s+t)^2} \mathrm{Ci}\left(\left|\frac{(t-\sqrt{3}+1) x_\text{R}}{2\sqrt{3}}\right|\right)^2.
\end{align}

To obtain explicit formulas of $\Omega_\text{GW}(k)$, we consider the power-law power spectrum for the curvature perturbations with a UV cutoff
\begin{align}
    \mathcal{P}_\zeta(k) = A_\ss \left(\frac{k}{k_0} \right)^{n_\ss - 1} \Theta (k_\text{max} - k),
\end{align}
where $k_0$ is the pivot scale, $A_\ss$ is the overall amplitude, $n_\ss-1$ is the spectral index, and $k_\text{max}$ is the cutoff scale to avoid the nonlinearity. 

$\Omega_\text{GW}(k)$ is given by the sum of the large-scale contribution $\Omega_\text{GW}^\text{(LS)}(k)$ and the resonance contribution $\Omega_\text{GW}^\text{(res)}(k)$, 
\begin{align}
    \Omega_\text{GW}(k) = & \Omega_\text{GW}^\text{(LS)}(k) + \Omega_\text{GW}^\text{(res)}(k).
\end{align}
Assuming $n_\text{s} > - 3/2$, the large-scale component is 
\begin{align}
    \Omega_\text{GW}^{\text{(LS)}} \simeq & \frac{3 \left(4 \mathrm{Ci}\left(\frac{x_\text{R}}{2}\right)^2 + \left( \pi - 2 \mathrm{Si}\left(\frac{x_\text{R}}{2}\right)\right)^2 \right) A_\ss^2 x_\text{max,R}^8 }{625 (3 + 2 n_\text{s}) 2^{17+2 n_\ss}} \left( \frac{2 x_\text{max,R}}{x_\text{R}} - 1 \right)^{2 n_\ss} \left( \frac{x_\text{R}}{x_{*\text{, R}}} \right)^{2 (n_\ss - 1)}  \nonumber \\
    & \times \left( \widetilde{\Omega}_\text{GW}^{\text{(LS,1)}} \Theta (x_\text{max,R}-x_\text{R}) + \widetilde{\Omega}_\text{GW}^{\text{(LS,2)}} \Theta(x_\text{R} - x_\text{max,R}) \right) \Theta (2 x_\text{max,R} - x_\text{R} ),
\end{align}
where $x_\text{R} = k \eta_\text{R}$ and $x_\text{max,R} = k_\text{max} \eta_\text{R}$, and
\begin{align}
    \widetilde{\Omega}_\text{GW}^{\text{(LS,1)}}(k)=& \frac{1}{(2+n_\ss)(3+n_\ss)(4+n_\ss)(5+2n_\ss)(7+2n_\ss)} \left( 1536 - 6144 \tilde{k} + (7168-1920 n_\ss -256 n_\ss^2 )\tilde{k}^2 \right. \nonumber \\
    & + (5760 n_\ss + 768 n_\ss^2) \tilde{k}^3 + (1328 n_\ss + 3056 n_\ss^2 + 832 n_\ss^3 + 64 n_\ss^4) \tilde{k}^4 \nonumber \\
    & - (7168 + 12256 n_\ss + 7392 n_\ss^2 + 1664 n_\ss^3 + 128 n_\ss^4 ) \tilde{k}^5 + (7392 + 10992 n_\ss + 5784 n_\ss^2 + 1248 n_\ss^3 + 96 n_\ss^4) \tilde{k}^6 \nonumber \\
    & - (2784 + 3904 n_\ss + 1960 n_\ss^2 + 416 n_\ss^3 + 32 n_\ss^4) \tilde{k}^7 + (370 + 503 n_\ss + 247 n_\ss^2 + 52 n_\ss^3 + 4 n_\ss^4 )\tilde{k}^8 \nonumber \\
    & \left. -256 (1 - \tilde{k})^6 \left(6 + 6(2 + n_\ss) \tilde{k} + (2+ n_\ss)(5 + 2 n_\ss) \tilde{k}^2 \right) \left( 1 - \frac{\tilde k}{2 - \tilde k} \right)^{2 n_\ss} \right), \\
    \widetilde{\Omega}_\text{GW}^{\text{(LS,2)}}(k) =& 2 (2 - \tilde{k})^4 \Gamma (4 + 2 n_\ss) \left( \frac{\tilde{k}^4}{\Gamma(5 + 2 n_\ss)} - \frac{4 \tilde{k}^2 (2 - \tilde{k})^2}{\Gamma (7 + 2 n_\ss)} + \frac{24(2 - \tilde{k})^4}{\Gamma (9 + 2 n_\ss)} \right),
\end{align}
with $\Gamma(\cdot)$ being the gamma function and $\tilde{k}=k/k_\text{max}(= x_\text{R}/x_\text{max,R})$. Note that the factor $\mathrm{Ci}(z) \sim \log (z) $ (for $z \ll 1$) includes the logarithmic dependence, see~\cite{Yuan:2019wwo}. 
The resonance component is 
\begin{align}
    \Omega_\text{GW}^{\text{(res)}}(k) \simeq & \frac{2.3 \times \sqrt{3} 3^{n_\ss} Y}{2^{13+2n_\ss} \times 625} A_\ss^2 x_\text{R}^7 \left( \frac{x_\text{R}}{x_{*\text{,R}}} \right)^{2(n_\ss-1)} s_0 (x_\text{R})  \nonumber \\
    & \times \left( 4 \HGF{\frac{1}{2}}{1-n_\ss}{\frac{3}{2}}{\frac{s_0^2(x_\text{R})}{3}} - 3 \HGF{\frac{1}{2}}{-n_\ss}{\frac{3}{2}}{\frac{s_0^2(x_\text{R})}{3}} - s_0^2(x_\text{R}) \HGF{\frac{3}{2}}{-n_\ss}{\frac{5}{2}}{\frac{s_0^2(x_\text{R})}{3}} \right),
\end{align}
where $\HGF{a}{b}{c}{z}$ is the hypergeometric function, $Y\simeq 2.3$ is a fudge factor to match the numerical result (in the case of $n_\ss = 1$), and 
\begin{align}
    s_0(x_\text{R}) = \begin{cases}
        1 & \left( x_\text{R} \leq \frac{2 x_\text{max,R}}{1 + \sqrt{3}} \right)\\
        \frac{2 x_\text{max,R}}{x_\text{R}} - \sqrt{3} & \left(\frac{2 x_\text{max,R}}{1+\sqrt{3}} < x_\text{R} \leq \frac{2 x_\text{max,R}}{\sqrt{3}} \right) \\
        0 & \left(\frac{2 x_\text{max,R}}{\sqrt{3}} < x_\text{R} \right)
    \end{cases}
\end{align}
takes care of the suppression close to the cutoff scale. 

Taking the limit of scale invariance ($n_\ss = 1$), we obtain
\begin{align}
    \Omega_\text{GW}^\text{(LS)}(k) \simeq & \frac{4 \mathrm{Ci}(\frac{x_\text{R}}{2})^2 + \left(\pi - 2 \mathrm{Si}(\frac{x_\text{R}}{2}) \right)^2}{86016000000} A_\ss^2 x_\text{R}^3 x_\text{max,R}^5 \nonumber \\
    & \times \left( \Theta (x_\text{max,R} - x_\text{R}) \left(5376 - 17640 \tilde{k} + 23760 \tilde{k}^2 - 16425 \tilde{k}^3 + 5825 \tilde{k}^4 - 847 \tilde{k}^5 \right) \right. \nonumber \\
    & \left. + \Theta (x_\text{R} - x_\text{max,R}) \left(2 - \tilde{k}\right)^6 \left( 4 - 8\tilde{k} - 9 \tilde{k}^2 + 13 \tilde{k}^3 + 49 \tilde{k}^4 \right) \right), \label{Omega_GW_LS_SI}
\end{align}
and 
\begin{align}
    \Omega_\text{GW}^\text{(res)} (k) \simeq \frac{2.3 \sqrt{3} Y}{102400000} A_\ss^2 x_\text{R}^7 s_0(x_\text{R}) \left( 15-10s_0^2 (x_\text{R}) + 3 s_0^4 (x_\text{R}) \right). \label{Omega_GW_res_SI}
\end{align}
\end{widetext}
Based on Eqs.~\eqref{Omega_GW_LS_SI} and \eqref{Omega_GW_res_SI}, the approximate spectral shape of the induced GWs is summarized in Eq.~\eqref{eq:Omega_GW_rough_behavior} as a broken power-law formula. 

\subsection{Transition into kination era}\label{ssec:analytic_transition_KD}

For the transient MD era transitioning into a transient kination era, we add one more layer of an approximation.  This is because we do not have an analytic formula for the integration kernel $I$ of the induced GWs for a transient kination era although for the pure kination case as well as a cosmic era characterized by the generic constant value of the equation-of-state parameter $w$, it was derived in ~\rref{Domenech:2019quo}.  

In a kination era, the two independent solutions of the tensor mode are given by the Bessel functions $J_0 (y)$ and $Y_0 (y)$, where we define
\begin{align}
y \equiv k \left(\eta -  \frac{3}{4} \eta_\text{kin}\right),
\end{align}
with $\eta_\text{kin}$ denoting the conformal transition time from the MD era to the kination era.  Therefore, it is convenient to decompose the integration kernel in the kination era in the form
\begin{align}
    I(u, v, x) \simeq & \int_{x_\text{kin}}^x \mathrm{d} \bar{x} \, k G_k (\eta, \bar {\eta}) f(u, v, \bar{x}, x_\text{kin}) \nonumber \\
    & = \frac{\pi}{2}  \left( Y_0 (y) \mathcal{I}_J (u, v, x) - J_0 (y) \mathcal{I}_Y (u, v, x) \right),
\end{align}
where $\mathcal{I}_{X} (u, v, x) \equiv \int_{y_\text{k}}^y \mathrm{d}\bar{y}\, \bar{y} X_0 (\bar{y}) f(u, v, \bar{x}) $ ($X= J$ or $Y$ denoting the Bessel function), $x_\text{kin}=k \eta_\text{kin}$, and $y_\text{kin} = \frac{1}{4} k \eta_\text{kin}$.  In particular, on the subhorizon scales of observational interest, 
\begin{align}
    \overline{I^2(u,v,x(\gg1))} \simeq \frac{\pi}{4 y} \left( \mathcal{I}_J^2(u,v,x) + \mathcal{I}_Y^2(u,v,x) \right).
\end{align}
How this quantity relates to the observable $\Omega_\text{GW}(k)$ in the present time depends on the assumption about the cosmic history after the kination era.  Here, we mainly focus on the model-independent part $\mathcal{I}_{J/Y}^2$. 

The idea of the approximation is simple. We divide the integration region $[y_\text{kin}, \infty)$ into $[\min[y_\text{kin}, 1], \max[y_\text{kin}, 1]]$ and $[\max[y_\text{kin}, 1], \infty)$, and approximate the Bessel function $J_0 (\bar y)$ or $Y_0 (\bar y)$ with $\bar y \ll 1$ or $\bar y \gg 1$ in the respective integration domain. Of course, this introduces some error around $\bar y \approx 1$.  Also, this approximation artificially produces a pole for a generic value of $w$~\cite{Harigaya:2023mhl}, so the estimate on the resonance contribution does not work.  However, in the kination era, the resonance condition is satisfied only at the kinematic endpoint ($k = 2k_\tmax$), so it is known that the resonance does not appear in the kination case~\cite{Domenech:2019quo}. Therefore, the induced GW spectrum can be obtained by the large $t$ approximation valid on large scales, \textit{i.e.}, except $k \sim 2 k _\text{max}$. We obtain
\begin{widetext}
    \begin{equation}
    \mathcal{I}_J^2 + \mathcal{I}_Y^2 \simeq
    \begin{dcases}
        \begin{split}%\MoveEqLeft
        &\frac{2}{\pi} \left( \frac{3}{50} \right)^2 y_\text{kin}^6 t^4 \left[\left(\frac{\sin (1+s- \pi/4)}{1+s} + \frac{\sin (1-s-\pi/4)}{1-s} - \frac{\sqrt{2\pi}\sin s}{s} \right)^2 \right. \\
         & \qquad \left. + \left(\frac{\cos(1+s-\pi/4)}{1+s} + \frac{\cos (1-s-\pi/4)}{1-s} + 6 \sqrt{\frac{2}{\pi}} \frac{(\gamma - \log 2 ) \sin s - \mathrm{Si}(s)}{s} \right)^2 \right]
        \end{split}   & (y_\text{kin} \ll 1)\\
        \frac{2}{\pi} \left(\frac{3}{25}\right)^2 y_\text{kin}^5 \frac{t^4}{(1-s^2)^2} & (y_\text{kin} \gg 1) 
    \end{dcases},
\end{equation}
where $\gamma = 0.577215\dots$ is the Euler-Mascheroni constant. 

Now, we consider the power-law power spectrum $\mathcal{P}_\zeta$ (with the cutoff $k_\text{max}$) and neglect the $s$-dependence in its argument (either $u k = (t+s+1)k/2$ or $v k = (t-s+1)k/2$). Then, we can integrate over $s$ (numerically for $y_\text{kin} \ll 1$) and also over $t$.  The power spectrum of the tensor mode is 
\begin{align}
    \overline{\mathcal{P}_h (\eta, k)} \simeq & \frac{9 \cdot 2^{3 + 2 n_\ss} }{625 (3 + 2 n_\ss)} \frac{1}{y} y_\text{kin}^5 \left( \frac{2}{\tilde{k}} - 1 \right)^{3+2 n_\ss} A_\ss^2  \times 
    \begin{cases}
        2.14 \, y_\text{kin}  & (y_\text{kin} \ll 1) \\
        1 & (y_\text{kin} \gg 1 \text{ and } k \ll k_\text{max})
    \end{cases}. \label{P_h_axion_rotation}
\end{align}

Finally, we relate this formula to $\Omega_\text{GW}$ in the axion rotation scenario discussed in the main text. In the scenario, an early RD era is followed by the MD era, the kination era, and the late (or standard) RD era. Remember that the radiation just redshifts as $a^{-4}$ from the early RD era to the late RD era without entropy production in this scenario and that the matter and kination in this scenario are two limiting behaviors of the same axion field.  Taking into account the redshift factors, $\Omega_\text{GW}$ well after the kination-to-radiation transition is
\begin{align}
    \Omega_\text{GW}(\eta, k)h^2 \simeq \frac{1}{24} \left( \frac{(1+3w)y_\text{c}}{2} \right)^2 \frac{a_\text{kin}}{a_\text{eq,1}} \left( \frac{a_\text{kin}}{a(\eta_\text{c})} \right)^2 \overline{\mathcal{P}_h (\eta_\text{c}, k)},
\end{align}
where the subscripts ``eq,1'' and ``eq,2'' denote the first equality (between the early RD era and the MD era) and the second equality (between the kination era and the late RD era), respectively.  The time dependence cancels at $\eta \gg \eta _\text{eq,2}$ and $\eta_\text{c}$ in the above expression is arbitrary as long as $\eta_\text{kin}\ll \eta_\text{c} \ll \eta_\text{eq,2}$ (Note $a \propto \eta^{1/2}$ during a kination era). 
Substituting Eq.~\eqref{P_h_axion_rotation}, setting $n_\ss = 1$, and effectively taking into account the suppression factor $Q$ (which can be numerically obtained) for the gravitational potential $\Phi$ due to the non-instantaneous transition, we obtain Eq.~\eqref{eq:omega_gw_app}.  In particular, the power of $k$ is $2$ and $1$ for $k \eta_\text{kin} \ll 1 $ and $k \eta_\text{kin} \gg 1$, respectively, consistent with the known IR behavior \cite{Domenech:2020kqm}. 

\end{widetext}

\small
\bibliographystyle{apsrmp4-2}
\bibliography{poltergeist_review}

\end{document}